\documentclass[journal,letterpaper, draftclsnofoot, 12pt, onecolumn]{IEEEtran}
\usepackage{cite}
\usepackage{babel}
\usepackage{refstyle}
\usepackage{tablefootnote}
\usepackage{amsmath}
\usepackage{amsthm}
\usepackage{amssymb}
\usepackage{stmaryrd}
\usepackage{graphicx}
\usepackage{subfig}
\usepackage{tikz}
\usepackage{pgfplots}
\usepackage{pgfgantt}
\usepackage{pdflscape}
\usepackage{bbm}
\usepackage{verbatim}
\pgfplotsset{compat=1.11} 
\pgfplotsset{plot coordinates/math parser=false}
\newlength\fwidth
\newref{eq}{name=Eq.~}
\newref{subsec}{name=Subsection~}
\newref{sec}{name=Section~}
\newref{thm}{name=Theorem~}
\newref{lem}{name=Lemma~}
\newref{cor}{name=Corollary~}
\newref{fig}{name=Fig.~}
\newref{def}{name=Definition~}
\newref{exa}{name=Example~}
\newref{eq}{refcmd=(\ref{#1})}
\newref{app}{name=Appendix~}

\theoremstyle{definition}
\newtheorem{defn}{\protect\definitionname}
\theoremstyle{definition}
\newtheorem{example}{\protect\examplename}
\theoremstyle{plain}
\newtheorem{thm}{\protect\theoremname}
\theoremstyle{remark}
\newtheorem{rem}{\protect\remarkname}
\theoremstyle{plain}
\newtheorem{cor}{\protect\corollaryname}
\theoremstyle{plain}
\newtheorem{lem}{\protect\lemmaname}

\providecommand{\corollaryname}{Corollary}
\providecommand{\definitionname}{Definition}
\providecommand{\examplename}{Example}
\providecommand{\lemmaname}{Lemma}
\providecommand{\remarkname}{Remark}
\providecommand{\theoremname}{Theorem}

\newcommand{\revisioncolor}{black}
\newcommand{\minorrevisioncolor}{black}

\begin{document}
%
\title{Bivariate Polynomial Coding for Efficient Distributed Matrix Multiplication}
%
%
%

\author{Burak Hasırcıoğlu,
        Jesús Gómez-Vilardebó,
        and~Deniz Gündüz
\thanks{Burak Hasırcıoğlu and Deniz Gündüz are with the Department
of Electrical and Electronic Engineering, Imperial College London, UK. E-mail: \{b.hasircioglu18, d.gunduz\}@imperial.ac.uk}
\thanks{Jesús Gómez-Vilardebó is with Centre Tecnològic de Telecomunicacions
de Catalunya (CTTC/CERCA), Barcelona, Spain. E-mail: jesus.gomez@cttc.es}
\thanks{Parts of this paper were presented in 2020 IEEE International Symposium on Information Theory (ISIT) \cite{hasirciouglu2020bivariate} and 2020 IEEE Global Communication Conference (Globecom) \cite{hasirciouglu2020globecom}. }
}

\maketitle

\begin{abstract}
Coded computing is an effective technique to mitigate “stragglers” in large-scale and distributed matrix multiplication. In particular, univariate polynomial codes have been shown to be effective in straggler mitigation by making the computation time depend only on the fastest workers. However, these schemes completely ignore the work done by the straggling workers resulting in a waste of computational resources. To reduce the amount of work left unfinished at workers, one can further decompose the matrix multiplication task into smaller sub-tasks, and assign multiple sub-tasks to each worker, possibly heterogeneously, to better fit their particular storage and computation capacities. In this work, \color{\revisioncolor}we propose a novel family of \color{black} \textit{bivariate polynomial codes} to efficiently exploit the work carried out by straggling workers. We show that bivariate polynomial codes bring significant advantages in terms of upload communication costs and storage efficiency, measured in terms of the number of sub-tasks that can be computed per worker. We propose two bivariate polynomial coding schemes. The first one exploits the fact that bivariate interpolation is always possible on a rectangular grid of evaluation points. We obtain such points at the cost of adding some redundant computations. For the second scheme, we relax the decoding constraints and require decodability for almost all choices of the evaluation points. We present interpolation sets satisfying such decodability conditions for certain storage configurations of workers. Our numerical results show that bivariate polynomial coding considerably reduces the average computation time of distributed matrix multiplication. We believe this work opens up a new class of previously unexplored coding schemes for efficient coded distributed computation.
\end{abstract}


%
\IEEEpeerreviewmaketitle

\section{Introduction} 
The availability of massive datasets and model sizes makes computation tasks for machine learning applications so demanding that they cannot be carried out on a single machine within a reasonable time frame. To speed up learning, most demanding computation tasks, e.g., matrix multiplication, are distributed to multiple dedicated servers, called \emph{workers}. However, due to unpredictable delays in their service time, some workers, called \emph{stragglers}, may
become a bottleneck for the overall computation task. 
One can mitigate the effects of stragglers
by assigning redundant computations. 
In particular, one can treat stragglers as random erasures, and improve the computation time by creating redundant computations similarly to channel coding for erasure channels. \color{\revisioncolor} Assuming all the workers start computing simultaneously, we define the \textit{computation time} as the time from the start until sufficiently many computations that allow decoding $AB$ at the master are received. It excludes the communication time as well as the encoding and decoding times. \color{black}
For the matrix multiplication task, the authors in \cite{lee2017speeding} propose to partition one of the matrices, encode its partitions by using an MDS code, and send coded partitions to the workers together with the other matrix (which is not partitioned or coded). It is then shown that the full matrix multiplication can be decoded by using only a subset of the multiplications between the coded partitions of the first matrix and the second matrix. 

In \cite{yu2017polynomial}, polynomial codes are proposed to speed up the multiplication of matrices $A$ and $B$. In this scheme, a \emph{master} partitions $A$ row-wise and $B$ column-wise. Then, two separate encoding polynomials, whose coefficients are the partitions of $A$ and $B$, respectively, are generated. The master evaluates both polynomials at the same point and sends the evaluations to the workers, which multiply them and return the result to the master. Using a subset of the responses from the fastest workers, the full multiplication can be recovered. This scheme is
optimal in terms of the \emph{download rate}, which is defined as the ratio between the total number of bits needed to be downloaded from the workers
and the number of bits needed to represent the result of the multiplication.
In \cite{dutta2019optimal}, MatDot codes are proposed, which use
an alternative partitioning scheme for matrices; that is, $A$ is partitioned column-wise
and $B$ is partitioned row-wise. The authors show that, compared to \cite{yu2017polynomial},
their approach improves the \emph{recovery threshold}, which is defined
as the minimum number of responses the master must receive from the workers to guarantee decoding the product $AB$. However, 
both the amount of computation each worker should carry out, referred to as the
\emph{computation cost}, and the download rate are higher than in \cite{yu2017polynomial}.
Also in \cite{dutta2019optimal}, PolyDot codes are proposed as an interpolation between polynomial codes in \cite{yu2017polynomial},
and MatDot codes by trading off between the recovery threshold and the computation
and download costs. In \cite{yu2020straggler}, the same problem
is studied, and entangled polynomial codes
are proposed, which improve the recovery threshold in \cite{dutta2019optimal}
under a  fixed  computation cost and a fixed download rate. Generalized PolyDot
codes are proposed in \cite{dutta2018unified} achieving the same recovery threshold in \cite{yu2020straggler}.
In \cite{jia2019cross}, batch multiplication of matrices, i.e.,
$A_{i}B_{i},$ $i\in[L]$ where $L>1$, is studied and cross subspace
alignment (CSA) codes are proposed. It is shown that, in the batch
multiplication setting, CSA codes improve the upload-download cost
trade-off compared to applying entangled polynomial codes separately
for each multiplication task in the batch. \color{\revisioncolor}
Since the decoding process in the polynomial coding approaches is based on polynomial interpolation, numerical stability becomes an important research problem for practical implementations. In \cite{fahim2021numerically, ramamoorthy2019numerically, subramaniam2019random, das2020efficient}, numerically stable coding schemes are proposed for distributed coded matrix multiplication problem. 
\color{black}


In all of these approaches, the result of all the work assigned to a worker is communicated to the master only if it is finished completely. Workers 
that fail to complete all their assignments by the time the recovery threshold is reached are treated as erasures, which implies
ignoring completely the work done by them. Such an approach is sub-optimal, especially if the workers' speeds are close to each other, in which case, the ignored workers have probably
completed a significant portion of the work assigned to them \cite{ferdinand2018hierarchical, amiri2019computation}.
To exploit the partially completed work done by stragglers, a multi-message
approach is considered in \cite{amiri2019computation, kiani2018exploitation, ozfatura2020straggler},
where workers' tasks are divided into smaller sub-tasks, and the result
of each sub-task is communicated to the master as soon as it is completed.
The approaches in \cite{amiri2019computation,ozfatura2020straggler} are based on uncoded computation and a hybrid of uncoded and coded computation,
respectively, and it is shown that uncoded computation may be more
beneficial if the workers' computation speeds are similar. In our work, we allow the workers
to be heterogeneous, as encountered in serverless computing, peer-to-peer
applications, or edge computing, and consider coded computation
with multi-message communication. A similar setting
is considered in \cite{kiani2018exploitation}, and product codes are employed. 

In \cite{ferdinand2018hierarchical}, a hierarchical coding framework
for the straggler exploitation problem is proposed, also taking into account the decoding complexity. 
This work is extended to matrix-vector
and matrix-matrix multiplications in \cite{kiani2019hierarchical}. It is shown that while gaining in terms of the decoding complexity, the computation time of hierarchical coding is only slightly larger than \cite{kiani2018exploitation}
with univariate polynomial coding. Thus, the benefits of hierarchical coding are significant mainly if the decoding time is comparable to the computation time.

In all of the aforementioned polynomial-type coding approaches \cite{lee2017speeding, yu2017polynomial, dutta2019optimal, yu2020straggler, dutta2018unified,jia2019cross}, univariate polynomials are used. As we will show in this paper, under fixed storage capacities at the workers, in
univariate polynomial coding, dividing a task into sub-tasks by a given
factor reduces the \color{\revisioncolor} fraction of work that can be done by \color{black} the workers by the same
factor, resulting in inefficient use of workers' storage capacity and upload costs. Product codes proposed
in \cite{kiani2018exploitation}, which are basically a combination
of two MDS codes, partially address this issue. However, in product
codes, computations at workers are not one-to-any replaceable, \color{\revisioncolor} i.e., some might be redundant, and hence, not useful, \color{black} which results in poor performance in various scenarios.
Moreover, univariate polynomial codes, as well as product codes, \color{\revisioncolor} impose certain constraints preventing fully heterogeneous workloads across workers. \color{black}

\color{\revisioncolor}In this work, we propose bivariate polynomial codes to improve the computation time of distributed matrix-matrix multiplication under limited storage at the workers. \color{\revisioncolor}The main contributions of this work can be summarized as follows:
\begin{itemize} 
    \item 
    We first show the limitation of univariate polynomial codes in terms of both computational and storage efficiency when extended to the multi-message setting. 
    \item We introduce bivariate polynomial coding schemes to address these limitations. Interpolation of bivariate polynomials cannot be guaranteed by simply requiring all evaluation points to be distinct. Here, we introduce the concepts of regular (always invertible), and almost regular (almost always invertible) interpolation matrices. 
    
    \item We first extend the product coding scheme of \cite{kiani2018exploitation} to bivariate polynomial coding, which leads to a regular interpolation matrix by imposing a particular rectangular grid structure on the interpolation points. This strategy attains maximum storage efficiency, but the computation efficiency can be limited due to redundant computations.
    
    \item Next, we propose two novel bivariate coding schemes. \color{\minorrevisioncolor} We demonstrate that unlike univariate schemes, for bivariate coding, the order by which the computations are done at the workers has a non-trivial impact on decodability; and hence, we impose a special computation order for the tasks assigned to each worker. \color{\revisioncolor} These schemes achieve maximum computation efficiency by completely avoiding redundant computations. Their storage efficiency is limited, yet higher than that of univariate schemes. We further propose two alternative bivariate polynomial codes with higher storage efficiency at the cost of a slight decrease in computation efficiency. 
       
    \item We numerically validate our findings assuming a shifted exponential model for computation speeds, and show the superiority of the proposed bivariate schemes compared to univariate alternatives and product codes. 
    
    \item While polynomial codes have been extensively studied with numerous applications in practice, to the best of our knowledge, our work provides the first examples of bivariate polynomial code constructions with superior performance compared to their univariate counterparts.
    
\end{itemize}

\color{black}

\section{System Model and Problem Formulation}\label{sec:System-model}

\begin{figure}
\begin{centering}
\includegraphics[scale=0.4,draft=false]{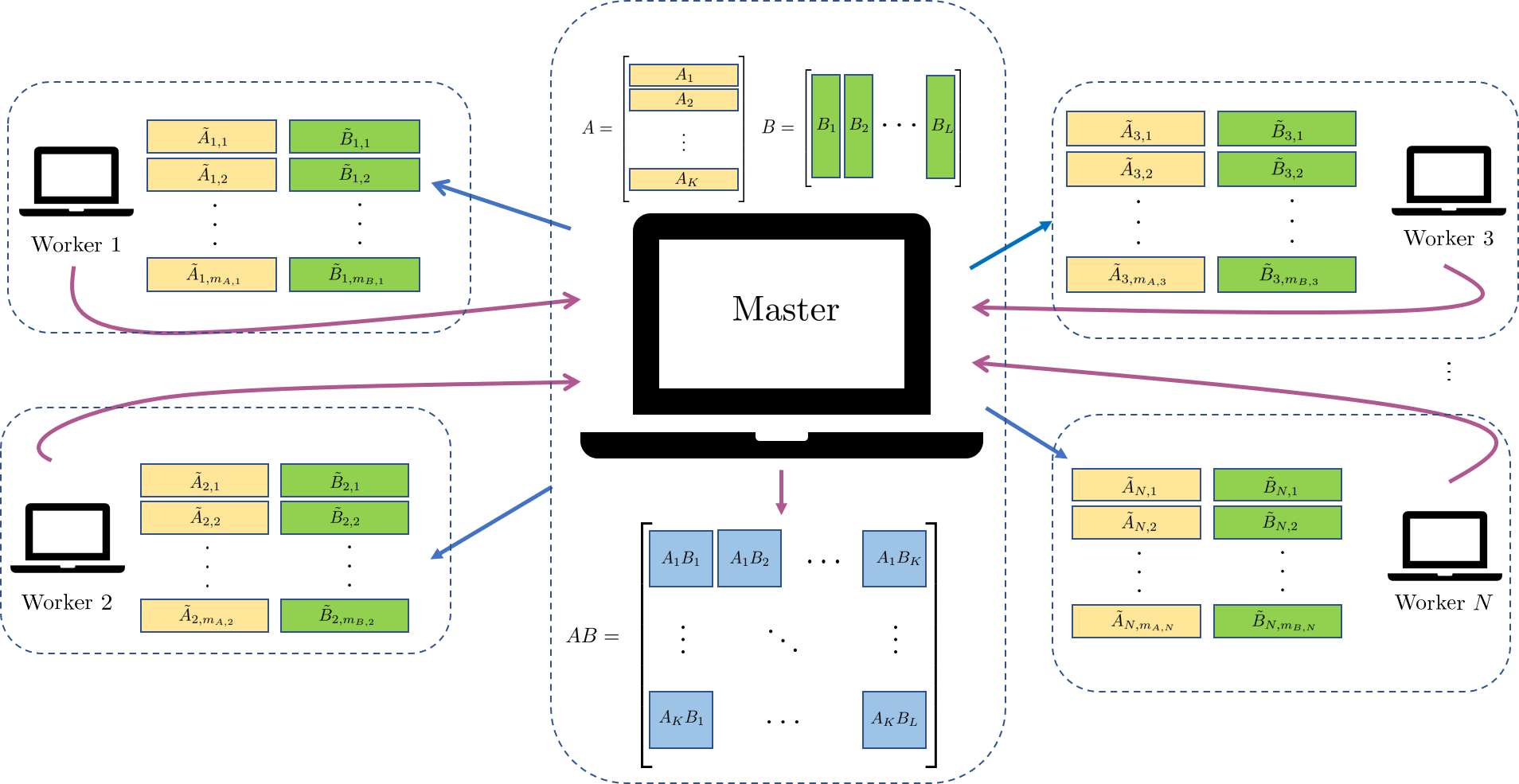} 
\par\end{centering}
\caption{\label{fig:setup}The master computes $AB$ by offloading partial
computations to $N$ workers.}
\end{figure}

\sloppy In our system, illustrated in \figref{setup}, a master server wants to multiply two matrices $A\in\mathbb{R}^{r\times s}$ and $B\in\mathbb{R}^{s\times c}$, $r$, $s$, $c\in\mathbb{Z}^{+}$, by offloading partial computations to $N$ workers with heterogeneous storage capacities and computation speeds. The master divides $A$ horizontally and $B$ vertically into $K$ and $L$ partitions, respectively, such that $A=\begin{bmatrix}A_{1}^{T} & A_{2}^{T} & \cdots & A_{K}^{T}\end{bmatrix}^{T}$ and $B=\begin{bmatrix}B_{1} & B_{2} & \cdots & B_{L}\end{bmatrix}$, where $A_{i}\in\mathbb{R}^{\frac{r}{K}\times s}$, $\forall i\in[1:K]$\footnote{Given $a<b$, we define $[a:b]\triangleq\{a,a+1,a+2,\ldots,b-1,b\}$} and $B_{j}\in\mathbb{R}^{s\times\frac{c}{L}}$, $\forall j\in[1:L]$. The master generates and sends to worker $i\in[1:N]$, $m_{A,i}$ and $m_{B,i}$ coded matrix partitions $\tilde{A}_{i,k}$ and $\tilde{B}_{i,l}$ based on $A$ and $B$, respectively, for $k\in[1:m_{A,i}]$ and $l\in[1:m_{B,i}]$, where $m_{A,i}$ and $m_{B,i}\in\mathbb{Z}^{+}$, and $\tilde{A}_{i,k}\in\mathbb{R}^{\frac{r}{K}\times s},$ $\tilde{B}_{i,l}\in\mathbb{R}^{s\times\frac{c}{L}}$. Thus, worker $i\in[1:N]$ is assumed to store a fraction $M_{A,i}=\frac{m_{A,i}}{K}$ of $A$ and $M_{B,i}=\frac{m_{B,i}}{L}$ of $B$. How these coded matrix partitions are generated depends on the specific coding scheme employed. In this work, they will be obtained as linear combinations of the original matrix partitions. 

Depending on the coding scheme employed, worker $i$ can compute all, or a subset of the products of  coded matrix partitions assigned to it, i.e., $\tilde{A}_{i,k}\tilde{B}_{i,l}$, $k\in[1:m_{A,i}]$, $l\in[1:m_{B,i}]$ in a prescribed order, which is also specific to the coding scheme. \color{\revisioncolor} We denote by $\eta_{i}$ the maximum number of computations worker $i$ can provide, which can be possibly used by the master for decoding $AB$. Thus, $\eta_{i}\leq m_{A,i}m_{B,i}$, \color{black} and the specific value of $\eta_i$  depends on the coding scheme.
In order to exploit the partial work done by straggling workers, the results of these individual products are sent to the master as soon as they are finished. The master collects the responses from the workers until the received set of computations allow the master to uniquely recover $AB$. \color{\revisioncolor} Then, the master instructs all the workers to stop computing and \color{black} decodes $AB$.
\color{\revisioncolor}Note that the recovery threshold, which is defined as the minimum number of computations that guarantee the decodability of $AB$, does not have to be a fixed quantity in our setting. Depending on the coding scheme, $R_{th}$ can be a function of the collected computations by the master.\color{black}

As is common in the related literature, we specify the storage capacity at workers separately for each of the two matrices, i.e., $M_{A,i}$ and $M_{B,i}$. 
However, in practice, it is more appropriate to assume a total storage capacity at each worker, which can be freely allocated between the partitions of the two matrices. Assume that the rows of $A$ and the columns of $B$ require the same amount of storage. We define the storage capacity of worker $i$, denoted by $s_{i}\in\mathbb{N^{+}}$, as the sum of the total number of rows of $A$ and the total number of columns of $B$ that the $i^{th}$ worker can store. Accordingly, for a given $K$, $L$, and $s_i$, we allocate $m_{A,i}$ and $m_{B,i}$ to maximize $\eta_i$ subject to $M_{A,i}r+M_{B,i}c=s_{i}$. Defining $C_{\text{part}}\triangleq \frac{1}{KL}$ as the fraction of work corresponding to a single partial product $\tilde{A}_{i,j}\tilde{B}_{i,l}$, the maximum fraction of work that can be done by  worker $i$ is given by
\begin{equation}
C_{\max,i} \triangleq\eta_{i}C_{\text{part}}=\frac{\eta_{i}}{m_{A,i}m_{B,i}}M_{A,i}M_{B,i}.
\end{equation}

\color{\revisioncolor}Under the same storage constraints, a code that can provide more fraction of work uses its storage more efficiently; hence, $C_{\max,i}$ will be used to measure the \textit{storage efficiency}. \color{black}

\color{\minorrevisioncolor} We define $C_{\text{wasted}}$ as the worst-case fraction of wasted computations with respect to the full product, $AB$. There are two sources of wasted computations. Firstly, depending on the coding scheme, some of the computations completed by the workers may not be used in decoding $AB$. Secondly, when $R_{th}$ is reached, the master instructs all the workers to stop their computations and the ongoing computations of the workers are wasted. We assume that the communication time for the stop signal to reach from the master to the workers is short enough that the workers receive this instruction before finishing their ongoing computations. In the following sections, we compute the fraction of the wasted computations of the second type based on this assumption. If this assumption does not hold, the wasted computations of the second type may increase. While it is out of the scope of this work, designing coding schemes that minimize wasted computations of the second type when the relative speed of communication is comparable to the speed of a unit computation can be an interesting challenge for a follow-up study.
\color{black}


\color{\revisioncolor}For a fixed $N$ and a total storage capacity at worker $i$, $s_i$, our objective is to minimize the average computation time of $AB$. This depends on the statistics of the computation speeds of the workers and is difficult to obtain in closed form. Instead, we use $C_{max,i}$ and $C_{wasted}$ as proxies for the performance of a code. These metrics do not depend on the worker's speeds and provide general indicators on the code performance. Note that, especially in heterogeneous settings, in which some workers
may be much faster than the others, the higher fraction of work provided
by faster workers helps to finish the task earlier. Therefore, storage
efficiency, or $C_{\max,i}$, is a factor to be optimized to improve
the average computation time. Moreover, low $C_{\text{wasted}}$ implies that more of the available computation capacity across the workers is exploited towards completing the desired computation. Therefore, to minimize the average computation time,
we are interested in maximizing $C_{\max,i}$ and minimizing $C_{\text{wasted}}$. \color{black}  Table \ref{tab:schemes_comparison} summarizes the key code parameters $C_{\text{max,i}}$, $C_{\text{wasted}}$ and system constraints for the schemes considered in this work. A detailed discussion on these parameters is postponed to the later sections.


\begin{table}
\begin{centering}
\begin{tabular}{|l|c|c|c|}
\hline 
Scheme  & $C_{\max,i}$  & $C_{\text{wasted}}$ & System constraints\tabularnewline
\hline 
\hline 
UPC & $M_{A}$$M_{B}$  & $NM_{A}M_{B}-1$  & $m_{A,i}=m_{B,i}=1$\tabularnewline
\hline 
UPC-PC  & $\frac{M_{A,i}M_{B,i}}{m_i}$  & $\sum_{i=1}^{N-1}\frac{M_{A,i}M_{B,i}}{m_i^{2}}$  & %

$m_{A,i}=m_{B,i}=m_{i}\in[1:\min(K,L)]$

\tabularnewline
\hline 
B-PROC  & $M_{A}M_{B}$  & %
\begin{tabular}{l}
$\sum_{i=1}^{N-1}\frac{M_{A,i}M_{B,i}}{m_{A,i}m_{B,i}}$\tabularnewline$+(n_{A}M_{B}-1)(1-\frac{M_{A}}{m_{A}})$\tabularnewline
$+(n_{B}M_{A}-1)(1-\frac{M_{B}}{m_{B}})$\tabularnewline
\end{tabular} & %
\begin{tabular}{l}
{\small{}{}$N=n_{A}n_{B}$}\tabularnewline
{\small{}{}$m_{A,i}=m_{A}$, $m_{B,i}=m_{B}$}\tabularnewline
{\small{}{}$K\leq n_{A}m_{B}$, $L\leq n_{B}m_{A}$}\tabularnewline
\end{tabular}\tabularnewline
\hline 
BPC-VO & $M_{A,i}M_{B,i}$  & $\sum_{i=1}^{N-1}\frac{M_{A,i}M_{B,i}}{m_{A,i}m_{B,i}}$  & %
\begin{tabular}{l}
$m_{A,i}=1$ and $m_{B,i}\leq L$ \textbf{or}\tabularnewline
$m_{A,i}\geq1$ and $m_{B,i}=L$ \tabularnewline
\end{tabular}\tabularnewline
\hline 
BPC-HO & $M_{A,i}M_{B,i}$  & $\sum_{i=1}^{N-1}\frac{M_{A,i}M_{B,i}}{m_{A,i}m_{B,i}}$  & %
\begin{tabular}{l}
$m_{B,i}=1$ and $m_{A,i}\leq K$ \textbf{or}\tabularnewline
$m_{B,i}\geq1$ and $m_{A,i}=K$\tabularnewline
\end{tabular}\tabularnewline
\hline 
BPC-NZO  & $M_{A,i}M_{B,i}$  & \begin{tabular}{l}
$\sum_{i=1}^{N-1}\frac{M_{A,i}M_{B,i}}{m_{A,i}m_{B,i}}$ \tabularnewline $+(\mu_{B}-2)(\frac{L}{\mu_{B}}-1)\frac{1}{KL}$
\end{tabular}   & 
\begin{tabular}{l}
$\mu_{B}\mid L$, $\mu_{B}\mid m_{B,i}$, $m_{A,i}=K$ and $m_{B,i}\leq L$
\textbf{or}\tabularnewline
$m_{B,i}=\mu_{B},$$\mu_{B}\mid L$, $m_{B,i}\leq L$ and $m_{A,i}\leq K$
\textbf{or}\tabularnewline
$m_{A,i}=1$, $m_{B,i}<\mu_{B}$ and $\mu_{B}\mid L$\tabularnewline
\end{tabular}\tabularnewline
\hline 
BPC-ZZO & $M_{A,i}M_{B,i}$  & \begin{tabular}{l}
$\sum_{i=1}^{N-1}\frac{M_{A,i}M_{B,i}}{m_{A,i}m_{B,i}}$\tabularnewline $+(\mu_{A}-2)(\frac{K}{\mu_{A}}-1)\frac{1}{KL}$ 
\end{tabular} & %
\begin{tabular}{l}
$\mu_{A}\mid K$, $\mu_{A}\mid m_{A,i}$, $m_{A,i}\leq K$ and $m_{B,i}=L$
\textbf{or}\tabularnewline
$m_{A,i}=\mu_{A},$$\mu_{A}\mid K$, $m_{A,i}\leq K$ and $m_{B,i}\leq L$
\textbf{or}\tabularnewline
$\mu_{A}\mid K$, $m_{A,i}<\mu_{A}$ and $m_{B,i}=1$\tabularnewline
\end{tabular}\tabularnewline
\hline 
\end{tabular}
\par\end{centering}
\caption{\label{tab:schemes_comparison}Comparison of the key parameters and system constraints.}
\end{table}

\section{Univariate Schemes}\label{sec:Univariate}

We first review the codes based on univariate polynomial interpolation.

\subsubsection*{Univariate Polynomial Codes (UPC)}

With the univariate polynomial codes presented in \cite{yu2017polynomial}, the master encodes the matrix partitions using the polynomials 
\begin{eqnarray}
A(x) & = & A_{1}+A_{2}x+\cdots+A_{K}x^{K-1},\\
B(x) & = & B_{1}+B_{2}x^{K}+\cdots+B_{i}x^{(i-1)K}+\cdots+B_{L}x^{(L-1)K}.
\end{eqnarray}
The master sends $A(x_{i})$ and $B(x_{i})$ to worker $i$, $i\in[1:N]$, for some distinct $x_{i}\in\mathbb{R}$. Thus, every worker receives one coded partition of $A$ and one partition of $B$, i.e., $m_{A,i}=m_{B,i}=1$ and $M_{A,i}=M_A=1/K$, $M_{B,i}=M_B=1/L$, $\forall i\in[1:N]$. Worker $i$ computes ${A}(x_{i}){B}(x_{i})$ and returns the result. No other computations are done at the workers, and thus $\eta_{i}=m_{A,i}m_{B,i}=1$. Once the master receives any $R_{th}=KL$ results, it can interpolate the polynomial
\begin{equation}
A(x)B(x)=\sum_{i=1}^{K}\sum_{j=1}^{L}A_{i}B_{j}x^{i-1+K(j-1)}    
\end{equation}
of degree $KL-1$. Observe that the coefficients of the interpolated polynomial correspond to the $KL$ sub-products $A_{i}B_{j},$ $\forall i\in[1:K]$, $\forall j\in[1:L]$ of $AB$. Finally, notice that 
\begin{equation}
C_{\max,i}=C_{\text{part}}=\frac{1}{KL}=M_{A}M_{B}.    
\end{equation}

Observe that with $N>R_{th},$ this scheme can tolerate up to $N-R_{th}$ stragglers. It helps to reduce the average computation time thanks to the parallelization afforded by redundant workers. However, all the work done by the $N-R_{\text{th}}$ slowest workers are ignored. In the worst case, where the $N-R_{\text{th}}+1$ slowest workers finish simultaneously, we have 
\begin{equation}
\label{eq:upc-c-wasted}
C_{\text{wasted}}=(N-R_{\text{th}})C_{\text{part}}=\color{\revisioncolor} (N-R_{\text{th}})\frac{1}{KL}=\color{black} NM_{A}M_{B}-1.
\end{equation}
\color{\revisioncolor}
We observe from \eqref{upc-c-wasted} that without changing the number of workers $N$ or the storage capacities of workers $M_A$, $M_B$, it is not possible to improve $C_{\text{wasted}}$, and thus reduce the amount of work lost at workers. Next, we provide an extension of UPC such that $C_{\text{wasted}}$ can be improved by increasing $K$ and $L$, exploiting the partial work done at the workers.
\color{black}

\subsubsection*{Univariate Polynomial Codes with Partial Computations (UPC-PC)}

To exploit the partial work done at slower workers, we present an extension of UPC, which is based on the multi-message approach and also allow heterogeneous storage capacities at workers. The main idea is to divide the task assigned to a worker into smaller sub-tasks, i.e., larger $K$ and $L$, and allowing the workers to store multiple partitions. Specifically, we allow worker $i$ to store $m_{i}=m_{A,i}=m_{B,i}$ coded partitions of $A$ and $B$, i.e., $M_{A,i}=m_{i}/K$ and $M_{B,i}=m_{i}/L$. For worker $i$, the master evaluates $A(x)$ and $B(x)$ at $m_{i}$ different points $\{x_{i,1},\ldots,x_{i,m_{i}}\}$ such that $x_{i,k}\neq x_{j,l}$ if $(i,k)\neq(j,l),\forall i,j\in[1:N]$ and $\forall k,l\in[1:m_{i}]$. Worker $i$ computes $A(x_{i,j})B(x_{i,j})$ consecutively for $j\in[1:m_{i}]$ and sends each result as soon as completed. Observe that multiplications are only allowed between $A(x)$ and $B(x)$ evaluated at the same $x_{i,k}$ values, and thus $\eta_{i}=m_{i},$ $\forall i\in[1:N]$. The master is able to interpolate $A(x)B(x)$ as soon as it receives $R_{th}=KL$ responses from the workers. Thus 
\begin{eqnarray}
C_{\text{part}} & = & \frac{1}{KL}=\frac{M_{A,i}M_{B,i}}{m_{i}^{2}},\\
C_{\max,i} & = & m_{i}C_{\text{part}}=\frac{M_{A,i}M_{B,i}}{m_{i}}.
\end{eqnarray}
The total fraction of wasted work in the worst case, in which all the workers were up to finish its ongoing partial multiplication once the $R_{th}$-th result is received by the master, is given by
\begin{equation}
\label{eq:upc-pc-c-wasted}
C_{\text{wasted}}=\left(N-1\right)C_{\text{part}}=\color{\revisioncolor}\left(N-1\right)\frac{1}{KL}= \color{black}\sum_{i=1}^{N-1}\frac{M_{A,i}M_{B,i}}{m_i^{2}}.
\end{equation}

As opposed to UPC, \color{\revisioncolor} according to \eqref{upc-pc-c-wasted}, UPC-PC can improve $C_{\text{wasted}}$ by increasing $K$ and $L$. Stragglers might be unable to complete the full task assigned to them, but they might complete a part of it. Clearly, the smaller are the sub-tasks executed at workers, the smaller is the work that can be lost at a straggler. \color{black} On the other hand, UPC-PC makes quite an inefficient use of the storage capacity at workers. Observe that even if a worker has enough storage to fully store $A$ and $B$, i.e., $M_{A,i}\geq1$ and $M_{B,i}\geq1$, it, alone, can only provide $\min\{K,L\}$ partial computations. Indeed, for a fixed storage capacity at the workers, i.e., $M_{A,i}$ and $M_{B,i}$ are kept constant, the maximum fraction of work done at a worker, $C_{\max,i}$, decreases while \color{\revisioncolor} $K$ and $L$ increases to improve $C_{\text{wasted}}$, which results in less efficient use of the storage capacities of the workers. \color{black} The bivariate schemes presented in the next section address this problem.

\section{Bivariate Polynomial Coding }\label{sec:Bivariate}

For bivariate polynomial coding schemes, we encode the matrix partitions of $A$ and $B$ by using the following encoding polynomials 
\begin{equation}
A(x)=A_{1}+A_{2}x+\cdots+A_{K}x^{K-1},\label{eq:encoding_poly_A}
\end{equation}
\begin{equation}
B(y)=B_{1}+B_{2}y+\cdots+B_{L}y^{L-1}.\label{eq:encoding_poly_B}
\end{equation}
Depending on the coding scheme, coded matrix partitions $\tilde{A}_{i,k}$ and $\tilde{B}_{i,l}$ are either direct evaluations of the encoding polynomials $A(x)$ and $B(y)$, respectively, or the evaluations of their derivatives. Hence, the workers obtain evaluations of the bivariate polynomial
\begin{equation}
A(x)B(y)=\sum_{i=1}^{K}\sum_{j=1}^{L}A_{i}B_{j}x^{i-1}y^{j-1}
\end{equation}
or of its derivatives, by multiplying the coded matrix partitions $\tilde{A}_{i,k}$ and $\tilde{B}_{i,l}$'s. Finally, the  master interpolates the bivariate polynomial $A(x)B(y)$ by making use of these products.

In addition to allowing heterogeneous storage capacities across workers, bivariate coding schemes allow different numbers of stored coded partitions of $A$ and $B$ for each worker, i.e., $m_{A,i}\neq m_{B,i}$ in general. The maximum number of computations a worker can generate is $\eta_{i}=m_{A,i}m_{B,i}$, resulting in
$
C_{\max,i}=m_{A,i}m_{B,i}C_{\text{part}}=M_{A,i}M_{B,i}.
$

Observe that, unlike univariate polynomial coding schemes, for a given storage capacity $M_{A,i}$ and $M_{B,i}$, the maximum amount of work done at worker $i$, $C_{\max,i}$, does not decrease with $m_{A,i}$ and $m_{B,i}$. In univariate schemes, the reason behind storage inefficiency is that the workers can use each evaluation of $A(x)$ and $B(x)$ only for one partial computation. For example, ${A}(x_{i,k}){B}(x_{i,l})$, for $k\neq l$, cannot be used to interpolate $A(x)B(x)$ in a univariate scheme. Bivariate polynomial coding eliminates this limitation and \color{\revisioncolor} allows the workers to provide additional
useful computations at no additional storage cost. \color{black} Moreover, like UPC-PC, bivariate polynomial codes can exploit the computational power of the stragglers. 

\color{\revisioncolor}
Since $A(x)B(y)$ has $KL$ coefficients, we need $KL$ partial computations to interpolate it. However, in some cases, the set of first $KL$ computations may not be enough to guarantee decodability and more computations may be required. Thus, the number of computations needed to guarantee decodability satisfies $R_{th}\geq KL$. Moreover, at the instant when the $R_{th}$-th computation is completed by a worker, all the ongoing computations become unnecessary. Therefore, in this setting, we have two sources of wasted computations: $R_{th}-KL$ redundant computations that have been received by the master but not used for the actual interpolation, and the ongoing computations at all the workers except the worker providing $R_{th}$-th computation. We consider the worst-case scenario and assume all these ongoing computations at the remaining $N-1$ workers are close to end. Thus, we count them as wasted computations.
\color{black} 
As a result, the maximum wasted fraction of computations is given by

\begin{equation}
C_{\text{wasted}}=\left(N-1\right)C_{\text{part}}+(R_{th}-KL)C_{\text{part}}=\sum_{i=1}^{N-1}\frac{M_{A,i}M_{B,i}}{m_{A,i}m_{B,i}}+\frac{R_{th}}{KL}-1.\label{eq:c_wasted_bivariate}
\end{equation}

Before presenting the bivariate schemes, we introduce some basic concepts and definitions from polynomial interpolation theory. 
\begin{defn}
\label{def:interpolation_matrix}
The interpolation of a bivariate polynomial of the form $A(x)B(y)$ can be formulated as a system of linear equations. The unknowns of these equations are the coefficients of $A(x)B(y)$. We define the \textbf{interpolation matrix} as the coefficient matrix of this linear system, denoted by $M$.
\end{defn}
\color{\revisioncolor}Recall that the interpolation matrices we consider result from evaluations of $A(x)B(y)$ or their derivatives at different points. The rules the coding schemes impose on the computations, e.g., computation orders, types of computations assigned to the workers, etc., \color{\minorrevisioncolor} may result in $\det(M)$ to become an identically zero polynomial \color{\revisioncolor} no matter which points are chosen. This is an undesirable situation, and we should show that this does not happen for a proposed scheme. Next, we define two notions in which such undesirable structures are not imposed on the interpolation matrix.\color{black}

\begin{defn}
\cite[Definition 3.1.3]{lorentz2006multivariate} An interpolation scheme is called \textbf{regular} if $\det(M)\neq0$ for every set of \color{\revisioncolor} distinct and non-zero \color{black}evaluation points. On the other hand, if $\det(M)\neq0$ for almost all choices of the evaluation points, then the interpolation scheme is called \textbf{almost regular}. Almost regularity implies that \color{\minorrevisioncolor}$\det(M)$ is not the zero polynomial in general \color{black} and the Lebesgue measure of the evaluation points making $\{\det(M)=0\}$ is zero in $\mathbb{R}^2$.

\color{\revisioncolor}To understand the practical meaning of almost regularity, let us assume that we sample our evaluation points uniform randomly from the interval $[l,u]$, where $l,u\in\mathbb{R}$ and $l<u$. Since the Lebesgue measure of the evaluation points making $\det(M)=0$ is zero, the probability of sampling such evaluation points is exactly zero. Note that this is due to using an uncountable set to sample our evaluation points and there are infinitely many possible choices of evaluation points. Even if we have countably many bad choices of evaluation points, the invertibility of $M$ is guaranteed almost surely.

\color{black}
Univariate polynomial interpolation is regular if the evaluation points are distinct \color{\revisioncolor}and non-zero \color{black} since the corresponding interpolation matrix is a Vandermonde matrix, which is known to be invertible. However, for bivariate interpolation, there are very few cases for which sufficient conditions for regularity are known. Next, we consider one such case.
\end{defn}

\subsection{Bivariate Polynomial Interpolation on Rectangular Grids}\label{subsec:Rectangular}

It is well known that interpolation of $A(x)B(y)$ such that $A(x)$ and $B(y)$ have degrees $K-1$ and $L-1$, respectively, is regular for any rectangular grid of points $\{x_{1},x_{2},\ldots,x_{K}\}\times\{y_{1},y_{2},\ldots,y_{L}\}$ \color{\revisioncolor} provided all $x_i$'s and $y_i$'s are distinct. \color{black}The interpolation scheme we propose next exploits this fact. It was originally proposed in \cite{kiani2018exploitation} using product codes. Here, we present it using bivariate polynomial codes, which is equivalent to the product-code form in terms of all the performance metrics considered in this paper. We further generalize it to allow $m_{A}\neq m_{B}$ and $n_{A}\neq n_{B}$, where $N=n_{A}n_{B}$.

\subsubsection*{Bivariate Product Coding (B-PROC)}

Assume that all the workers can store $m_{A}$ partitions of $A$ and $m_{B}$ partitions of $B$, and $N$ can be factored as $N=n_{A}n_{B}$ such that $K\leq m_{A}n_{A}$ and $L\leq m_{B}n_{\ensuremath{B}}$. The master generates $n_{A}$ disjoint sets $\mathcal{X}_{i}=\{x_{i,1},x_{i,2},\ldots,x_{i,m_{A}}\},i\in[1:n_{A}]$ and $n_{B}$ disjoint sets $\mathcal{Y}_{j}=\{y_{j,1},y_{j,2},\ldots,y_{j,m_{B}}\},j\in[1:n_{B}]$, with distinct elements. Then, it enumerates the workers as $(i,j)$, $i\in [1:n_A]$, $j\in [1:n_B]$ and sends $A(x_{i,k}),k\in [1:m_A]$ and $B(y_{j,l}),l\in [1:m_B]$, to worker $(i,j)$. Worker $(i,j)$ can compute any product $A(x_{i,k})B(y_{j,l})$. Altogether, the set of evaluation points at workers form a rectangular grid of size $m_{A}n_{A}\times m_{B}n_{B}$. Observe that, $n_{A}$ workers have the evaluation $B(\hat{y})$ for any $\hat{y}\in \mathcal{Y}_j$, and each of them can compute $m_{A}$ distinct evaluations of the univariate polynomial $A(x)B(\hat{y})$, of degree $K-1$ with respect to $x$. Once the first $K$ of these evaluations are received at the master, $A(x)B(\hat{y})$ can be interpolated. Similarly, for a given $\hat{x}\in \mathcal{X}_i$, $A(\hat{x})B(y)$ can be interpolated from any $L$ evaluations as it is a univariate polynomial in $y$ with degree $L-1$.  As a result, once we have the evaluations of $A(x)B(y)$ on any rectangular grid of size $K\times L$, either directly received from the workers or via univariate interpolation, the bivariate interpolation problem can be solved.

Observe, however, that the computations that were already interpolated from previous results are redundant. To minimize such computations, in \cite{kiani2018exploitation}, for the particular case of $m_{A}=m_{B}$, $n_{A}=n_{B}$, different heuristics to schedule computations at the workers were discussed.  

\begin{figure}
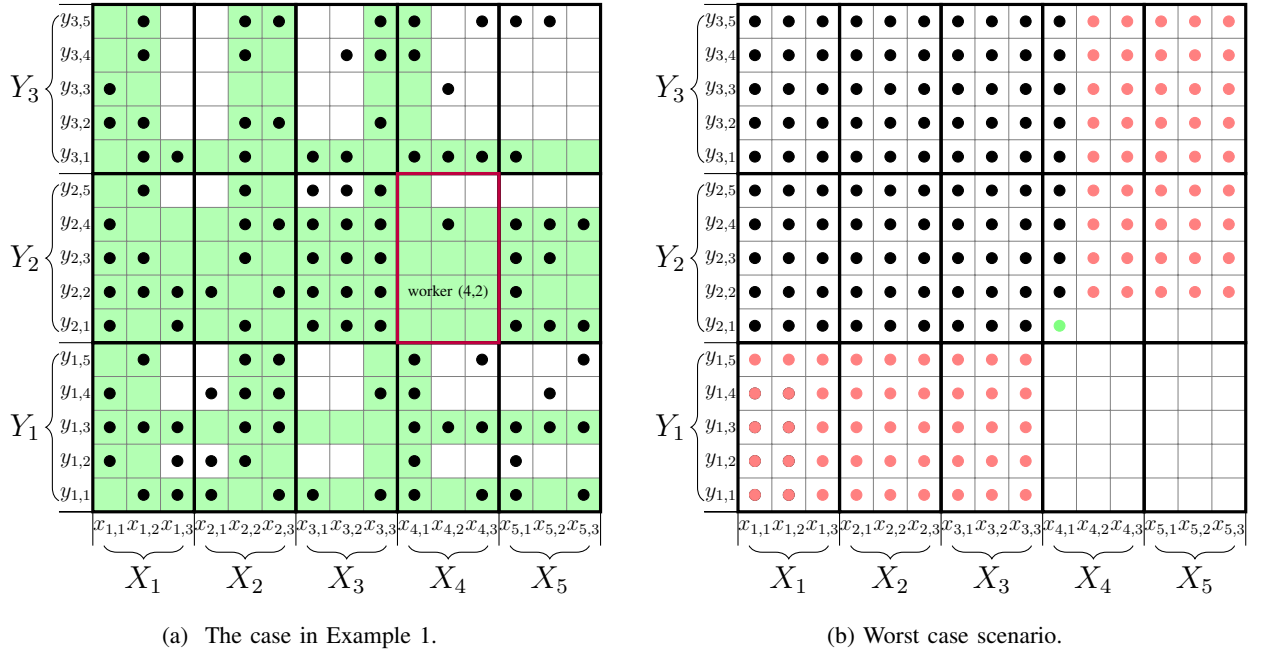

\centering
\subfloat[\
\label{fig:scheme_3_example}The case in \exaref{scheme_3_example}.]{
\centering
\input{figures/product_example}}
\subfloat[\label{fig:c_wasted}Worst case scenario.]{
\centering
\input{figures/product_worst_case}}
\caption{Two different instances of received partial computations at the master
for $K=L=10$, $N=15$, $M_{A}=3/10$ and $M_{B}=5/10$. }

\end{figure}

\begin{example} 
\label{exa:scheme_3_example}
Let us assume that both matrices $A$ and $B$ are divided into $K=L=10$ partitions, and there are $N=15$ workers, while every worker can store $M_{A}=3/10$ of $A$ and $M_{B}=5/10$ of $B$. We take $n_{A}=5$ and $n_{B}=3$. Worker $(i,j)$ stores $\{A(x_{i,1}),A(x_{i,2}),A(x_{i,3})\}$ and $\{B(y_{j,1}),B(y_{j,2}),B(y_{j,3}),B(y_{j,4}),B(y_{j,5})\}$.
Let us assume that the order of computations is random within a worker. \figref{scheme_3_example} shows an instance of the received responses from the workers. Each worker is represented by a $3\times5$ rectangle and each filled circle represents a received computation by the master. To clarify how a worker's finished computations look like, worker $(4,2)$ is emphasized in the figure. All the elements in the columns and the rows coloured by green can be interpolated, i.e., decoded, by using the received responses on the same column or the row. Observe that there are columns/rows coloured by green even if they have less than 10 computations, e.g., the column of $x_{4,1}$. Such rows and columns can be decoded after decoding rows and columns with at least 10 computations, by utilizing all the elements in these columns and rows after decoding. Since there must be at least 10 green columns and 10 green rows in order to decode $A(x)B(y)$, in our example, the received responses are not sufficient, although there are $110>KL=100$ responses received by the master.
\end{example}

The total fraction of the work wasted in the worst case depends on the heuristics employed. If we consider uniform random computation order at the workers, which is reported to perform well in \cite{kiani2018exploitation}, then the computations can be received at any order by the master. In the worst case, there may be $K-1$ fully computed columns, that is, in every column there are exactly $n_{B}m_{B}$ computations, and one column with exactly $L$ computations. Thus, in this case, $n_{B}m_{B}-L$ computations in each of the $K-1$ fully computed columns are wasted. On the other hand, there may be $L-1$ fully computed rows and one row with exactly $K$ computations. In this case, $n_{A}m_{A}-K$ computations in each of the $L-1$ fully computed rows are wasted. Thus, in total, we have $(n_{B}m_{B}-L)(K-1)+(n_{A}m_{A}-K)(L-1)$ wasted computations. \color{\revisioncolor}Therefore, the worst-case $R_{th}$ of B-PROC is
\begin{equation}
\label{eq:b-proc-r-th}
R^{B-PROC}_{th} = KL+(n_{B}m_{B}-L)(K-1)+(n_{A}m_{A}-K)(L-1).
\end{equation}
Note that this expression is a worst-case value and depending on the received responses, the actual number of computations that guarantee decodability may be much lower. \color{black}
If we plug \eqref{b-proc-r-th} into \eqref{c_wasted_bivariate}, the fraction of wasted computations for B-PROC in the worst case becomes 
\begin{eqnarray}
C_{\text{wasted}} & = & \sum_{i=1}^{N-1}\frac{M_{A}M_{B}}{m_{A}m_{B}}+[(n_{B}m_{B}-L)(K-1)+(n_{A}m_{A}-K)(L-1)]\frac{1}{KL}\nonumber\\
 & = & \sum_{i=1}^{N-1}\frac{M_{A}M_{B}}{m_{A}m_{B}}+(n_{B}M_{B}-1)\left(1-\frac{M_{A}}{m_{A}}\right)+(n_{A}M_{A}-1)\left(1-\frac{M_{B}}{m_{B}}\right).
\end{eqnarray}
The expression is highly dependent on how $n_{A}$ and $n_{B}$ are allocated. Increasing the memory, i.e., $M_{A}$ and $M_{B}$, while $K$ and $L$ remain constant, or increasing $K$ and $L$ while the memory remains constant both increase $C_{\text{wasted}}$. However, it is worth noting that $C_{\text{wasted}}$ we calculated here is for the worst-case scenario, and the situation may not be that bad most of the time. For the setting in \exaref{scheme_3_example}, we visualize the worst-case situation in \figref{c_wasted}.

B-PROC requires additional constraints on the system, i.e., $N=n_{A}n_{B}$, $K\leq n_{A}m_{A}$, $L\leq n_{B}m_{B}$ and homogeneous storage capacities at workers, and yet it is not possible to ensure that the first $KL$ results arriving at the master form a regular interpolation problem. \color{\revisioncolor}To address these issues, in the next subsection, we propose novel bivariate polynomial codes. However, showing the regularity of these schemes is a hard problem, if not impossible. Therefore, instead, we use the notion of almost regularity, which is a relaxation of regularity and propose almost regular bivariate interpolation schemes.\color{black}

\subsection{Almost Regular Bivariate Interpolation Schemes}
\label{subsec:almost_regular_schemes}

For the almost regular bivariate interpolation schemes we propose, the
polynomial $A(x)B(y)$ is interpolated from the evaluations of it and its derivatives. Such an interpolation is known as Hermite interpolation in
the literature \cite[Chapter 3.6]{atkinsonNumericalAnalysis}.

\subsubsection{Encoding}

The almost regular interpolation schemes we describe in the sequel
have a common encoding procedure. Consider the polynomials
in \eqref{encoding_poly_A} and \eqref{encoding_poly_B}. 
To each worker $i\in[1:N]$ the master assigns a distinct evaluation point $(x_i,y_i)\in\mathbb{R}^2$, and sends the evaluations of $A(x)$ and $B(y)$, and their derivatives up to order $m_{A,i}-1$ and $m_{B,i}-1$, respectively, at ($x_i,y_i$).
Thus, the values generated and sent to the worker $i$ by the master
are $\mathcal{A}_{i}\triangleq\left\{A(x_{i}),\frac{dA(x_{i})}{dx},\ldots,\frac{d^{(m_{A,i}-1)}A(x_{i})}{dx^{(m_{A,i}-1)}}\right\}$
and $\mathcal{B}_{i}\triangleq\left\{B(y_{i}),\frac{dB(y_{i})}{dy},\ldots,\frac{d^{(m_{B,i}-1)}B(y_{i})}{dy^{(m_{B,i}-1)}}\right\}$.
For brevity, in the remaining of the paper, we use $\partial_{k}A(x_{i})$ and $\partial_{l}B(y_{i})$
to denote $\frac{d^{k}}{dx^{k}}A(x_{i})$and $\frac{d^{l}}{dy^{l}}B(y_{i})$, respectively.

\subsubsection{Computations at workers} For
all the coding schemes, after receiving $\mathcal{A}_{i}$ and $\mathcal{B}_{i}$ from the master, each worker $i$ starts computing, one by one, all the cross products between elements in $\mathcal{A}_{i}$ and those in $\mathcal{B}_{i}$,  and
sends the result of each computation to the master as soon
as it is completed. We require each worker to follow a specific computation
order for these multiplications according to the \textbf{priority
score} of each computation, which we will define shortly for each
scheme. We note that a lower priority score gives more priority to a
computation.
Specifically, for any worker $i$, each computation $\partial_{k}A(x_{i})\partial_{l}B(y_{i})$ for $k \in [0:K-1]$ and $l\in [0:L-1]$ is assigned a priority score denoted as $S(k,l)$. Worker $i$ computes $\partial_{k}A(x_{i})\partial_{l}B(y_{i})$ once all the computations $\partial_{\tilde{k}}A(x_{i})\partial_{\tilde{l}}B(y_{i})$, 
$\tilde{k}\in[0:K-1]$ and $\tilde{l}\in[0:L-1]$ 
such that $S(\tilde{k},\tilde{l})<S(k,l)$ are already computed. 
Notice that priority scores $S(k,l)$ are defined for computations that might not be available at worker $i$, i.e., $K>k\geq m_{A,i}$ or $L>l\geq m_{B,i}$. Whenever such a computation has the lowest priority score among all the remaining computations at worker $i$, the worker must discard all the remaining computations and stop.

\begin{defn}\label{def:InterpolationSpace}
\textbf{Derivative Order Space. }The derivative order space of a bivariate
polynomial $A(x)B(y)$ is defined as the 2-dimensional space of all its
possible derivative orders. When $A(x)$ and $B(y)$ have
degrees $K-1$ and $L-1$, respectively, the derivative order space becomes
$\{(k,l):0\leq k<K,0\leq l<L\}$, where the tuple $(k,l)$ represents
the derivative $\partial_{k}A(x)\partial_{l}B(y)$. 
\end{defn}

\subsubsection*{Bivariate Polynomial Coding with Vertical Computation Order (BPC-VO)}

\begin{figure}
\centering
\subfloat[\label{fig:vo}Vertical order]{
\begin{tikzpicture}

\draw [help lines,  step=0.5cm] (-3.5, -3.5) node (v19) {} grid (-0.5,-0.5);  
\node[scale=0.8] at (-3.25,-3.75) {$0$};   
\node[scale=1] at (-0.25,-3.5) {$k$};   
\node[scale=1] at (-3.5,-0.25) {$l$};    
\node[scale=0.8] at (-1.25,-3.75) {$4$};    
\node[scale=0.8] at (-0.75,-3.75) {$5$};   
\node[scale=0.8] at (-1.75,-3.75) {$3$};   
\node[scale=0.8] at (-2.25,-3.75) {$2$};   
\node[scale=0.8] at (-2.75,-3.75) {$1$};    
\node[scale=0.8] at (-3.75,-3.25) {$0$};           
\node[scale=0.8] at (-3.75,-0.75) {$5$};   
\node[scale=0.8] at (-3.75,-1.25) {$4$};   
\node[scale=0.8] at (-3.75,-1.75) {$3$};   
\node[scale=0.8] at (-3.75,-2.25) {$2$};   
\node[scale=0.8] at (-3.75,-2.75) {$1$}; 
\draw[color=black] (-3.25,-3.25) node (v1) {} circle (.08);  
\draw[color=black] (-3.25,-2.75) node (v2) {} circle (.08);  
\draw[color=black] (-3.25,-2.25) node (v3) {} circle (.08);   
\draw[color=black] (-2.75,-3.25) node (v4) {} circle (.08);   
\draw[color=black] (-2.75,-2.75) node (v5) {} circle (.08);   
\draw[color=black] (-2.75,-2.25) node (v6) {} circle (.08);   
\draw[color=black] (-2.25,-3.25) node (v7) {} circle (.08);  
\draw[color=black] (-2.25,-2.75) node (v8) {} circle (.08);   
\draw[color=black] (-2.25,-2.25) node (v9) {} circle (.08);   
\draw[color=black] (-1.75,-3.25) node (v10) {} circle (.08);   
\draw[color=black] (-1.75,-2.75) node (v11) {} circle (.08);  
\draw[color=black] (-1.75,-2.25) node (v12) {} circle (.08);   
\draw[color=black] (-1.25,-3.25) node (v13) {} circle (.08);   
\draw[color=black] (-1.25,-2.75) node (v14) {} circle (.08);  
\draw[color=black] (-1.25,-2.25) node (v15) {} circle (.08);   
\draw[color=black] (-0.75,-3.25) node (v16) {} circle (.08);   
\draw[color=black] (-0.75,-2.75) node (v17) {} circle (.08);   
\draw[color=black] (-0.75,-2.25) node (v18) {} circle (.08);   
\draw[color=black] (-3.25,-1.75) node (v19) {} circle (.08);   
\draw[color=black] (-3.25,-1.25) node (v20) {} circle (.08);   
\draw[color=black] (-3.25,-0.75) node (v21) {} circle (.08);   
\draw[color=black] (-2.75,-1.75) node (v22) {} circle (.08);  
\draw[color=black] (-2.75,-1.25) node (v23) {} circle (.08);   
\draw[color=black] (-2.75,-0.75) node (v24) {} circle (.08);   
\draw[color=black] (-2.25,-1.75) node (v25) {} circle (.08);   
\draw[color=black] (-2.25,-1.25) node (v26) {} circle (.08);   
\draw[color=black] (-2.25,-0.75) node (v27) {} circle (.08);   
\draw[color=black] (-1.75,-1.75) node (v28) {} circle (.08);   
\draw[color=black] (-1.75,-1.25) node (v29) {} circle (.08);   
\draw[color=black] (-1.75,-0.75) node (v30) {} circle (.08);   
\draw[color=black] (-1.25,-1.75) node (v31) {} circle (.08);   
\draw[color=black] (-1.25,-1.25) node (v32) {} circle (.08);   
\draw[color=black] (-1.25,-0.75) node (v33) {} circle (.08);   
\draw[color=black] (-0.75,-1.75) node (v34) {} circle (.08);   
\draw[color=black] (-0.75,-1.25) node (v35) {} circle (.08);   
\draw[color=black] (-0.75,-0.75) node (v36) {} circle (.08);  
\draw  (-3.5,-0.5) rectangle (-0.5,-3.5);  
\draw[->]  (v1) edge (v2); 
\draw[->]  (v2) edge (v3); 
\draw[->]  (v4) edge (v5);  
\draw[->]  (v5) edge (v6); 
\draw[->]  (v7) edge (v8);  
\draw[->]  (v8) edge (v9);  
\draw[->]  (v10) edge (v11);  
\draw[->]  (v11) edge (v12);  
\draw[->]  (v13) edge (v14);  
\draw[->]  (v14) edge (v15);  
\draw[->]  (v16) edge (v17);  
\draw[->]  (v17) edge (v18);  
\draw[->]  (v3) edge (v19);  
\draw[->]  (v19) edge (v20);  
\draw[->]  (v20) edge (v21);  
\draw[->]  (v22) edge (v23);  
\draw[->]  (v23) edge (v24);  
\draw[->]  (v25) edge (v26);  
\draw[->]  (v26) edge (v27);  
\draw[->]  (v28) edge (v29);  
\draw[->]  (v29) edge (v30);  
\draw[->]  (v31) edge (v32);  
\draw[->]  (v32) edge (v33);  
\draw[->]  (v34) edge (v35);  
\draw[->]  (v35) edge (v36);

\draw[->]  (v6) edge (v22); 
\draw[->]  (v9) edge (v25); 
\draw[->]  (v12) edge (v28); 
\draw[->]  (v15) edge (v31); 
\draw[->]  (v18) edge (v34);

\draw[->]  (v21) edge (v4); 
\draw[->]  (v24) edge (v7); 
\draw[->]  (v27) edge (v10); 
\draw[->]  (v30) edge (v13); 
\draw[->]  (v33) edge (v16); \end{tikzpicture}}
\subfloat[\label{fig:ho}Horizontal order]{
\begin{tikzpicture}

\draw [help lines,  step=0.5cm] (-3.5, -3.5) node (v19) {} grid (-0.5,-0.5);  
\node[scale=0.8] at (-3.25,-3.75) {$0$};   
\node[scale=1] at (-0.25,-3.5) {$k$};   
\node[scale=1] at (-3.5,-0.25) {$l$}; 
\node[scale=0.8] at (-1.25,-3.75) {$4$};    
\node[scale=0.8] at (-0.75,-3.75) {$5$};    
\node[scale=0.8] at (-1.75,-3.75) {$3$};   
\node[scale=0.8] at (-2.25,-3.75) {$2$};    
\node[scale=0.8] at (-2.75,-3.75) {$1$};    
\node[scale=0.8] at (-3.75,-3.25) {$0$};           
\node[scale=0.8] at (-3.75,-0.75) {$5$};   
\node[scale=0.8] at (-3.75,-1.25) {$4$};   
\node[scale=0.8] at (-3.75,-1.75) {$3$};   
\node[scale=0.8] at (-3.75,-2.25) {$2$};   
\node[scale=0.8] at (-3.75,-2.75) {$1$}; 

\draw[color=black] (-3.25,-3.25) node (v1) {} circle (.08);  
\draw[color=black] (-2.75,-3.25) node (v2) {} circle (.08);  
\draw[color=black] (-2.25,-3.25) node (v3) {} circle (.08);   
\draw[color=black] (-3.25,-2.75) node (v4) {} circle (.08);   
\draw[color=black] (-2.75,-2.75) node (v5) {} circle (.08);   
\draw[color=black] (-2.25,-2.75) node (v6) {} circle (.08);  
\draw[color=black] (-3.25,-2.25) node (v7) {} circle (.08);   
\draw[color=black] (-2.75,-2.25) node (v8) {} circle (.08);   
\draw[color=black] (-2.25,-2.25) node (v9) {} circle (.08);   
\draw[color=black] (-3.25,-1.75) node (v10) {} circle (.08);   
\draw[color=black] (-2.75,-1.75) node (v11) {} circle (.08);  
\draw[color=black] (-2.25,-1.75) node (v12) {} circle (.08);   
\draw[color=black] (-3.25,-1.25) node (v13) {} circle (.08);  
\draw[color=black] (-2.75,-1.25) node (v14) {} circle (.08);  
\draw[color=black] (-2.25,-1.25) node (v15) {} circle (.08);   
\draw[color=black] (-3.25,-0.75) node (v16) {} circle (.08);   
\draw[color=black] (-2.75,-0.75) node (v17) {} circle (.08);   
\draw[color=black] (-2.25,-0.75) node (v18) {} circle (.08);   
\draw[color=black] (-1.75,-3.25) node (v19) {} circle (.08);   
\draw[color=black] (-1.25,-3.25) node (v20) {} circle (.08);   
\draw[color=black] (-0.75,-3.25) node (v21) {} circle (.08);   
\draw[color=black] (-1.75,-2.75) node (v22) {} circle (.08);   
\draw[color=black] (-1.25,-2.75) node (v23) {} circle (.08);   
\draw[color=black] (-0.75,-2.75) node (v24) {} circle (.08);  
\draw[color=black] (-1.75,-2.25) node (v25) {} circle (.08);   
\draw[color=black] (-1.25,-2.25) node (v26) {} circle (.08);   
\draw[color=black] (-0.75,-2.25) node (v27) {} circle (.08);   
\draw[color=black] (-1.75,-1.75) node (v28) {} circle (.08);   
\draw[color=black] (-1.25,-1.75) node (v29) {} circle (.08);   
\draw[color=black] (-0.75,-1.75) node (v30) {} circle (.08);   
\draw[color=black] (-1.75,-1.25) node (v31) {} circle (.08);   
\draw[color=black] (-1.25,-1.25) node (v32) {} circle (.08);   
\draw[color=black] (-0.75,-1.25) node (v33) {} circle (.08);   
\draw[color=black] (-1.75,-0.75) node (v34) {} circle (.08);   
\draw[color=black] (-1.25,-0.75) node (v35) {} circle (.08);  
\draw[color=black] (-0.75,-0.75) node (v36) {} circle (.08);  
\draw  (-3.5,-0.5) rectangle (-0.5,-3.5);  
\draw[->]  (v1) edge (v2);  
\draw[->]  (v2) edge (v3);  
\draw[->]  (v4) edge (v5);  
\draw[->]  (v5) edge (v6);  
\draw[->]  (v7) edge (v8);  
\draw[->]  (v8) edge (v9);  
\draw[->]  (v10) edge (v11);  
\draw[->]  (v11) edge (v12);  
\draw[->]  (v13) edge (v14);  
\draw[->]  (v14) edge (v15);  
\draw[->]  (v16) edge (v17);  
\draw[->]  (v17) edge (v18);  
\draw[->]  (v19) edge (v20);  
\draw[->]  (v20) edge (v21);  
\draw[->]  (v22) edge (v23);  
\draw[->]  (v23) edge (v24);  
\draw[->]  (v25) edge (v26);  
\draw[->]  (v26) edge (v27);  
\draw[->]  (v28) edge (v29);  
\draw[->]  (v29) edge (v30);  
\draw[->]  (v31) edge (v32);  
\draw[->]  (v32) edge (v33);  
\draw[->]  (v34) edge (v35);  
\draw[->]  (v35) edge (v36);

\draw[->]  (v3) edge (v19); 
\draw[->]  (v6) edge (v22); 
\draw[->]  (v9) edge (v25); 
\draw[->]  (v12) edge (v28); 
\draw[->]  (v15) edge (v31); 
\draw[->]  (v18) edge (v34);

\draw[->]  (v21) edge (v4); 
\draw[->]  (v24) edge (v7); 
\draw[->]  (v27) edge (v10); 
\draw[->]  (v30) edge (v13); 
\draw[->]  (v33) edge (v16);
\end{tikzpicture}}
\subfloat[\label{fig:no}N-zig-zag order]{
\begin{tikzpicture}

\draw [help lines,  step=0.5cm] (-3.5, -3.5) node (v19) {} grid (-0.5,-0.5); 
\node[scale=0.8] at (-3.25,-3.75) {$0$};  
\node[scale=1] at (-0.25,-3.5) {$k$};  
\node[scale=1] at (-3.5,-0.25) {$l$};   
\node[scale=0.8] at (-1.25,-3.75) {$4$};   
\node[scale=0.8] at (-0.75,-3.75) {$5$};   
\node[scale=0.8] at (-1.75,-3.75) {$3$};   
\node[scale=0.8] at (-2.25,-3.75) {$2$};   
\node[scale=0.8] at (-2.75,-3.75) {$1$};   
\node[scale=0.8] at (-3.75,-3.25) {$0$};          
\node[scale=0.8] at (-3.75,-0.75) {$5$};  
\node[scale=0.8] at (-3.75,-1.25) {$4$};  
\node[scale=0.8] at (-3.75,-1.75) {$3$};  
\node[scale=0.8] at (-3.75,-2.25) {$2$};  
\node[scale=0.8] at (-3.75,-2.75) {$1$};

\draw[color=black] (-3.25,-3.25) node (v1) {} circle (.08); 
\draw[color=black] (-3.25,-2.75) node (v2) {} circle (.08); 

\draw[color=black] (-3.25,-2.25) node (v3) {} circle (.08);  
\draw[color=black] (-2.75,-3.25) node (v4) {} circle (.08);  
\draw[color=black] (-2.75,-2.75) node (v5) {} circle (.08);  
\draw[color=black] (-2.75,-2.25) node (v6) {} circle (.08);  
\draw[color=black] (-2.25,-3.25) node (v7) {} circle (.08);  
\draw[color=black] (-2.25,-2.75) node (v8) {} circle (.08);  
\draw[color=black] (-2.25,-2.25) node (v9) {} circle (.08);  
\draw[color=black] (-1.75,-3.25) node (v10) {} circle (.08);  
\draw[color=black] (-1.75,-2.75) node (v11) {} circle (.08);  
\draw[color=black] (-1.75,-2.25) node (v12) {} circle (.08);  
\draw[color=black] (-1.25,-3.25) node (v13) {} circle (.08);  
\draw[color=black] (-1.25,-2.75) node (v14) {} circle (.08);  
\draw[color=black] (-1.25,-2.25) node (v15) {} circle (.08);  
\draw[color=black] (-0.75,-3.25) node (v16) {} circle (.08);  
\draw[color=black] (-0.75,-2.75) node (v17) {} circle (.08);  
\draw[color=black] (-0.75,-2.25) node (v18) {} circle (.08);  
\draw[color=black] (-3.25,-1.75) node (v19) {} circle (.08);  
\draw[color=black] (-3.25,-1.25) node (v20) {} circle (.08);  
\draw[color=black] (-3.25,-0.75) node (v21) {} circle (.08);  
\draw[color=black] (-2.75,-1.75) node (v22) {} circle (.08);  
\draw[color=black] (-2.75,-1.25) node (v23) {} circle (.08);  
\draw[color=black] (-2.75,-0.75) node (v24) {} circle (.08);  
\draw[color=black] (-2.25,-1.75) node (v25) {} circle (.08);  
\draw[color=black] (-2.25,-1.25) node (v26) {} circle (.08);  
\draw[color=black] (-2.25,-0.75) node (v27) {} circle (.08);  
\draw[color=black] (-1.75,-1.75) node (v28) {} circle (.08);  
\draw[color=black] (-1.75,-1.25) node (v29) {} circle (.08);  
\draw[color=black] (-1.75,-0.75) node (v30) {} circle (.08);  
\draw[color=black] (-1.25,-1.75) node (v31) {} circle (.08);  
\draw[color=black] (-1.25,-1.25) node (v32) {} circle (.08);  
\draw[color=black] (-1.25,-0.75) node (v33) {} circle (.08);  
\draw[color=black] (-0.75,-1.75) node (v34) {} circle (.08);  
\draw[color=black] (-0.75,-1.25) node (v35) {} circle (.08);  
\draw[color=black] (-0.75,-0.75) node (v36) {} circle (.08); 

\draw  (-3.5,-2) rectangle (-0.5,-3.5); 
\draw  (-3.5,-0.5) rectangle (-0.5,-2);

\draw[->]  (v1) edge (v2); 
\draw[->]  (v2) edge (v3); 
\draw[->]  (v3) edge (v4); 
\draw[->]  (v4) edge (v5); 
\draw[->]  (v5) edge (v6); 
\draw[->]  (v6) edge (v7); 
\draw[->]  (v7) edge (v8); 
\draw[->]  (v8) edge (v9); 
\draw[->]  (v9) edge (v10); 
\draw[->]  (v10) edge (v11); 
\draw[->]  (v11) edge (v12); 
\draw[->]  (v12) edge (v13); 
\draw[->]  (v13) edge (v14); 
\draw[->]  (v14) edge (v15); 
\draw[->]  (v15) edge (v16); 
\draw[->]  (v16) edge (v17); 
\draw[->]  (v17) edge (v18); 
\draw[->]  (v18) edge (v19); 
\draw[->]  (v19) edge (v20); 
\draw[->]  (v20) edge (v21); 
\draw[->]  (v21) edge (v22); 
\draw[->]  (v22) edge (v23); 
\draw[->]  (v23) edge (v24); 
\draw[->]  (v24) edge (v25); 
\draw[->]  (v25) edge (v26); 
\draw[->]  (v26) edge (v27); 
\draw[->]  (v27) edge (v28); 
\draw[->]  (v28) edge (v29); 
\draw[->]  (v29) edge (v30); 
\draw[->]  (v30) edge (v31); 
\draw[->]  (v31) edge (v32); 
\draw[->]  (v32) edge (v33); 
\draw[->]  (v33) edge (v34); 
\draw[->]  (v34) edge (v35); 
\draw[->]  (v35) edge (v36);
\end{tikzpicture}}
\subfloat[\label{fig:zo}Z-zig-zag order]{
\begin{tikzpicture}

\draw [help lines,  step=0.5cm] (-3.5, -3.5) node (v19) {} grid (-0.5,-0.5); 
\node[scale=0.8] at (-3.25,-3.75) {$0$};  
\node[scale=1] at (-0.25,-3.5) {$k$};  
\node[scale=1] at (-3.5,-0.25) {$l$};   
\node[scale=0.8] at (-1.25,-3.75) {$4$};   
\node[scale=0.8] at (-0.75,-3.75) {$5$};   
\node[scale=0.8] at (-1.75,-3.75) {$3$};   
\node[scale=0.8] at (-2.25,-3.75) {$2$};   
\node[scale=0.8] at (-2.75,-3.75) {$1$};   
\node[scale=0.8] at (-3.75,-3.25) {$0$};          
\node[scale=0.8] at (-3.75,-0.75) {$5$};  
\node[scale=0.8] at (-3.75,-1.25) {$4$};  
\node[scale=0.8] at (-3.75,-1.75) {$3$};  
\node[scale=0.8] at (-3.75,-2.25) {$2$};  
\node[scale=0.8] at (-3.75,-2.75) {$1$};

\draw[color=black] (-3.25,-3.25) node (v1) {} circle (.08); 
\draw[color=black] (-2.75,-3.25) node (v2) {} circle (.08); 

\draw[color=black] (-2.25,-3.25) node (v3) {} circle (.08);  
\draw[color=black] (-3.25,-2.75) node (v4) {} circle (.08);  
\draw[color=black] (-2.75,-2.75) node (v5) {} circle (.08);  
\draw[color=black] (-2.25,-2.75) node (v6) {} circle (.08);  
\draw[color=black] (-3.25,-2.25) node (v7) {} circle (.08);  
\draw[color=black] (-2.75,-2.25) node (v8) {} circle (.08);  
\draw[color=black] (-2.25,-2.25) node (v9) {} circle (.08);  
\draw[color=black] (-3.25,-1.75) node (v10) {} circle (.08);  
\draw[color=black] (-2.75,-1.75) node (v11) {} circle (.08);  
\draw[color=black] (-2.25,-1.75) node (v12) {} circle (.08);  
\draw[color=black] (-3.25,-1.25) node (v13) {} circle (.08);  
\draw[color=black] (-2.75,-1.25) node (v14) {} circle (.08);  
\draw[color=black] (-2.25,-1.25) node (v15) {} circle (.08);  
\draw[color=black] (-3.25,-0.75) node (v16) {} circle (.08);  
\draw[color=black] (-2.75,-0.75) node (v17) {} circle (.08);  
\draw[color=black] (-2.25,-0.75) node (v18) {} circle (.08);  
\draw[color=black] (-1.75,-3.25) node (v19) {} circle (.08);  
\draw[color=black] (-1.25,-3.25) node (v20) {} circle (.08);  
\draw[color=black] (-0.75,-3.25) node (v21) {} circle (.08);  
\draw[color=black] (-1.75,-2.75) node (v22) {} circle (.08);  
\draw[color=black] (-1.25,-2.75) node (v23) {} circle (.08);  
\draw[color=black] (-0.75,-2.75) node (v24) {} circle (.08);  
\draw[color=black] (-1.75,-2.25) node (v25) {} circle (.08);  
\draw[color=black] (-1.25,-2.25) node (v26) {} circle (.08);  
\draw[color=black] (-0.75,-2.25) node (v27) {} circle (.08);  
\draw[color=black] (-1.75,-1.75) node (v28) {} circle (.08);  
\draw[color=black] (-1.25,-1.75) node (v29) {} circle (.08);  
\draw[color=black] (-0.75,-1.75) node (v30) {} circle (.08);  
\draw[color=black] (-1.75,-1.25) node (v31) {} circle (.08);  
\draw[color=black] (-1.25,-1.25) node (v32) {} circle (.08);  
\draw[color=black] (-0.75,-1.25) node (v33) {} circle (.08);  
\draw[color=black] (-1.75,-0.75) node (v34) {} circle (.08);  
\draw[color=black] (-1.25,-0.75) node (v35) {} circle (.08);  
\draw[color=black] (-0.75,-0.75) node (v36) {} circle (.08); 

\draw  (-3.5,-0.5) rectangle (-2,-3.5); 
\draw  (-2,-0.5) rectangle (-0.5,-3.5);

\draw[->]  (v1) edge (v2); 
\draw[->]  (v2) edge (v3); 
\draw[->]  (v3) edge (v4); 
\draw[->]  (v4) edge (v5); 
\draw[->]  (v5) edge (v6); 
\draw[->]  (v6) edge (v7); 
\draw[->]  (v7) edge (v8); 
\draw[->]  (v8) edge (v9); 
\draw[->]  (v9) edge (v10); 
\draw[->]  (v10) edge (v11); 
\draw[->]  (v11) edge (v12); 
\draw[->]  (v12) edge (v13); 
\draw[->]  (v13) edge (v14); 
\draw[->]  (v14) edge (v15); 
\draw[->]  (v15) edge (v16); 
\draw[->]  (v16) edge (v17); 
\draw[->]  (v17) edge (v18); 
\draw[->]  (v18) edge (v19); 
\draw[->]  (v19) edge (v20); 
\draw[->]  (v20) edge (v21); 
\draw[->]  (v21) edge (v22); 
\draw[->]  (v22) edge (v23); 
\draw[->]  (v23) edge (v24); 
\draw[->]  (v24) edge (v25); 
\draw[->]  (v25) edge (v26); 
\draw[->]  (v26) edge (v27); 
\draw[->]  (v27) edge (v28); 
\draw[->]  (v28) edge (v29); 
\draw[->]  (v29) edge (v30); 
\draw[->]  (v30) edge (v31); 
\draw[->]  (v31) edge (v32); 
\draw[->]  (v32) edge (v33); 
\draw[->]  (v33) edge (v34); 
\draw[->]  (v34) edge (v35); 
\draw[->]  (v35) edge (v36);
\end{tikzpicture}}
\caption{\label{fig:k_3_l_4}Computation orders at the workers proposed
in this work.}
\end{figure}
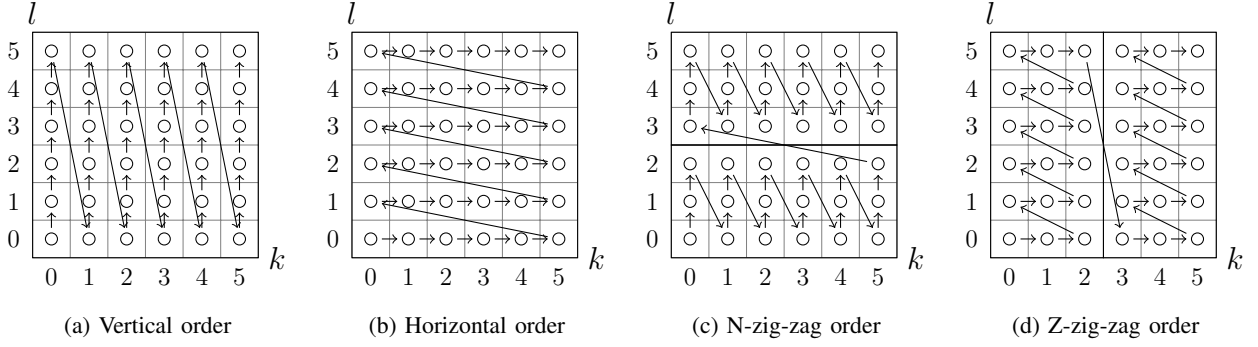

In this scheme, workers follow the \emph{vertical computation order}, illustrated in \figref{vo} for $K=L=6$. In the vertical computation order, a worker first completes a column $k$ in the derivative order space, i.e., all the computations in $\{\partial_{k}A(x_{i})B(y_{i}),\partial_{k}A(x_{i})\partial_{1}B(y_{i}),\dots,\partial_{k}A(x_{i})\partial_{L-1}B(y_{i})\}$ before moving on to the computations from column $k+1$.
Specifically, for any worker $i$, computation $\partial_{k}A(x_{i})\partial_{l}B(y_{i})$ for $k\in[0:K-1]$ and $l\in[0:L-1]$, has a priority score of $\ensuremath{S_{\mathcal{V}}(k,l)\triangleq(K-1)L\left(\left\lceil \frac{l}{L}\right\rceil -1\right)+L(k-1)+l}$.
Because only the computations $\partial_{k}A(x_{i})\partial_{l}B(y_{i})$ $k\in[0:m_{A,i}-1]$, and $l\in[0:m_{B,i}-1]$ can be computed by worker $i$, in order to satisfy the vertical computation order without discarding any computations, worker $i$ can store either:
\begin{enumerate}
\item a single coded partition of $A$, and any number of coded partitions of $B$ not more than $L$, i.e., $m_{A,i}=1$ and $1\leq m_{B,i}\leq L$, or 
\item coded partitions of $B$ equivalent to the full matrix $B$ in size, and not more than $K$ coded partitions of $A$, i.e., $1\leq m_{A,i}\leq K$ and $m_{B,i}=L$. 
\end{enumerate}

\subsubsection*{Bivariate Polynomial Coding with Horizontal Computation Order (BPC-HO)}
In this scheme, workers follow the \emph{horizontal computation order}, illustrated in \figref{ho} for $K=L=6$. In the horizontal computation order, a worker first completes a row $l$ in the derivative order space, i.e., all the computations in $\{A(x_{i})\partial_{l}B(y_{i}),\partial_{1}A(x_{i})\partial_{l}B(y_{i}),\dots,\partial_{K-1}A(x_{i})\partial_{l}B(y_{i})\}$ before moving on to the computations from row $l+1$.
Specifically, for any worker $i$, computation $\partial_{k}A(x_{i})\partial_{l}B(y_{i})$
has a priority score of $\ensuremath{S_{\mathcal{H}}(k,l)\triangleq K(L-1)\left(\left\lceil \frac{k}{K}\right\rceil -1\right)+K(l-1)+k}$. 
As for the vertical computation order, because only computations $\partial_{k}A(x_{i})\partial_{l}B(y_{i})$ $k\in [0:m_{A,i}-1]$, and $l\in[0:m_{B,i}-1]$ can be computed by worker $i$, in order to satisfy the horizontal computation order without discarding any computations, worker $i$ can store either:
\begin{enumerate}
\item a single coded partition of $B$, and any number of coded partitions of $A$ not more than $K$, i.e., $1\leq m_{A,i}\leq K$ and $m_{B,i}=1$, or 
\item coded partitions of $A$ equivalent to the full matrix $A$, and any number of coded partitions of $B$ not more than $L$, i.e., $m_{A,i}=K$ and $1\leq m_{B,i}\leq L$.
\end{enumerate}

\subsubsection*{Bivariate Polynomial Coding with N-zig-zag Computation Order (BPC-NZO) }

For this scheme, we relax the vertical computation order by dividing
the derivative order space into $L/\mu_{B}$ equal horizontal \emph{blocks}, where $\mu_{B}$ is a design parameter such that $\mu_{B}\mid L$. For computation $\partial_{k}A(x_{i})\partial_{l}B(y_{i})$, we assign an \emph{N-zig-zag order} priority score of $S_{\mathcal{N}}(k,l)=(K-1)\mu_{B}\left(\left\lceil \frac{l}{\mu_{B}}\right\rceil -1\right)+\mu_{B}(k-1)+l$. In \figref{no}, we illustrate the N-zig-zag order for $K=L=6$ and $\mu_B=3$. For this computation order, we simply apply vertical computation order inside each horizontal block in the derivative order space starting from the lowermost block. Only when all the computations in a block  are completed, the computations from the next block can start. 
Although they are more relaxed than the vertical computation order, in order to satisfy the N-zig-zag order without discarding any computations at worker $i$, one of the following conditions must be imposed on $m_{A,i}$ and $m_{B,i}$:

\begin{enumerate}
\item $m_{B,i}$ is a positive integer multiple of $\mu_{B}$, and $m_{A,i}=K$, or 
\item $m_{B,i}=\mu_{B}$ and $1\leq m_{A,i}\leq K$, or,
\item $m_{A,i}=1$ and $1\leq m_{B,i}\leq\mu_{B}$ 
\end{enumerate}

Observe that by setting $\mu_B=L$, the N-zig-zag order reduces to the vertical computation order.

\subsubsection*{Bivariate Polynomial Coding with Z-zig-zag Computation Order (BPC-ZZO)}

For this scheme, we relax the horizontal computation order by dividing the derivative order space into $K/\mu_{A}$ equal vertical blocks, where $\mu_{A}$ is a design parameter such that $\mu_{A}\mid L$. For the computation $\partial_{k}A(x_{i})\partial_{l}B(y_{i})$, we define the \emph{Z-zig-zag order} priority score as $S_{\mathcal{Z}}(k,l)=(L-1)\mu_{A}\left(\left\lceil \frac{k}{\mu_{A}}\right\rceil -1\right)+\mu_{A}(l-1)+k$. In \figref{zo}, we visualize the Z-zig-zag computation order when $K=L=6$ and $\mu_A=3$. 
For this computation order, we apply horizontal computation order inside each vertical block starting from the leftmost block. Again, only when all the computations in a block are completed, the computations from the next block can start.
In order to satisfy the Z-zig-zag computation order without discarding any computations at worker $i$, we need to impose one of the following constraints on $m_{A,i}$ and $m_{B,i}$: 
\begin{enumerate}
\item $m_{A,i}$ is a positive integer multiple of $\mu_{A}$, and $m_{B,i}=L$, or 
\item $m_{A,i}=\mu_{A}$ and $1\leq m_{B,i}\leq L$, or 
\item $m_{B,i}=1$ and $1\leq m_{A,i}\leq \mu_{A}$ 
\end{enumerate}
Observe that setting $\mu_A=K$, we recover the horizontal computation order conditions.
\subsubsection{Decoding Procedure of Almost Regular Interpolation Schemes}

For all of the computation orders defined in this section, the master receives responses from the workers and decodes $AB$ by solving a bivariate polynomial interpolation problem. That is, $A(x)B(y)$ is interpolated from the evaluations of $A(x)B(y)$ and its derivatives. Since the degree of $A(x)B(y)$ is $KL$, to solve the interpolation problem, the master needs at least $KL$ computations returned from the workers. In this case, assuming, without loss of generality, the responses are received from all $N$ workers, we have an interpolation matrix as in \eqref{interpol_matrix}.

\begin{equation}
M=\begin{bmatrix}1 & x_{1} & x_{1}^{2} & x_{1}^{3} & \cdots & x_{1}^{K-1} & \cdots & x_{1}^{K-1}y_{1}^{L-1}\\
0 & 1 & 2x_{1} & 3x_{1}^{2} & \cdots & \left(K-1\right)x_{1}^{K-2} & \cdots & \left(K-1\right)x_{1}^{K-2}y_{1}^{L-1}\\
0 & 0 & 2 & 6x_{1} & \cdots & (K-1)(K-2)x_{1}^{K-3} & \cdots & (K-1)(K-2)x_{1}^{K-3}y_{1}^{L-1}\\
\vdots & \vdots & \vdots & \vdots & \ddots & \vdots & \ddots & \vdots\\
1 & x_{N} & x_{N}^{2} & x_{N}^{3} & \cdots & x_{N}^{K-1} & \cdots & x_{N}^{K-1}y_{N}^{L-1}\\
0 & 1 & 2x_{N} & 3x_{N}^{2} & \cdots & \left(K-1\right)x_{N}^{K-2} & \cdots & \left(K-1\right)x_{N}^{K-2}y_{N}^{L-1}
\end{bmatrix}.\label{eq:interpol_matrix}
\end{equation}
In this example, the master received 3 responses from worker 1, and 2 responses from worker $N$. Since we see the derivatives are taken with respect to $x$, we can conclude that this interpolation matrix belongs to BPC-HO or BPC-ZZO.

The next theorem and the corollary characterize the number of computations needed to decode $A(x)B(y)$ in the worst-case scenario by considering the invertibility of the interpolation matrix.

\begin{thm}
\label{thm:main_thm} 
a) For BPC-NZO, the worst-case recovery threshold is $R_{th}^{NZO}\triangleq KL+\max\left\{ 0,(\mu_{B}-2)(\frac{L}{\mu_{B}}-1)\right\} $. 

b) For BPC-ZZO, the worst-case recovery threshold is $R_{th}^{ZZO}\triangleq KL+\max\left\{ 0,(\mu_{A}-2)(\frac{K}{\mu_{A}}-1)\right\} $.

Thus, if the number of computations received by the master is at least $R_{th}^{NZO}$ and $R_{th}^{ZZO}$ for BPC-NZO and BPC-ZZO, respectively, then $\det(M)\neq0$ for almost all choices of the evaluation points.
\end{thm}
The proof of \thmref{main_thm} is given in \secref{proof}. 

\begin{cor}
\label{cor:arbitrary-lower-set} 
\color{\revisioncolor}BPC-VO and BPC-HO can be obtained by setting $\mu_B=L$ and $\mu_A=K$ in BPC-NZO and BPC-ZZO, respectively. Therefore, \color{black} the recovery thresholds of BPC-VO and BPC-HO are $R_{th}^{VO}=R_{th}^{HO}\triangleq KL$, meaning any $KL$ computations received by the master results in $\det(M)\neq0$ for almost all choices of the evaluation points.
\end{cor}

According to \corref{arbitrary-lower-set}, for BPC-VO and BPC-HO, every partial computation sent by the workers is useful at the master, i.e., $R_{th}=KL$. Therefore, for these schemes, the computations are one-to-any replaceable. Thus, according to \eqref{c_wasted_bivariate}, we have
\begin{equation}
C_{\text{wasted, BPC-VO}}=C_{\text{wasted, BPC-HO}}=\sum_{i=1}^{N-1}\frac{M_{A,i}M_{B,i}}{m_{A,i}m_{B,i}}.    
\end{equation}

This is the main advantage of these schemes over B-PROC. 
On the other hand, while in the BPC-HO and BPC-VO, $\eta_i$ is limited by the constraints imposed on $m_{A,i}$ and $m_{B,i}$, in B-PROC, all the available storage can be fully exploited. Therefore, B-PROC has a better storage efficiency $C_{\max,i}$ compared to BPC-VO and BPC-HO.
The main motivation of introducing BPC-NZO and BPC-ZZO is to relax these constraints. According to \thmref{main_thm}, this can be done at the cost of potentially introducing redundant computations; however, the number of redundant computations needed is much less than those needed for B-PROC. Specifically, from \eqref{c_wasted_bivariate}, we have 
\begin{eqnarray}
C_{\text{wasted, BPC-NZO}}&=&\sum_{i=1}^{N-1}\frac{M_{A}M_{B}}{m_{A}m_{B}}+(\mu_{B}-2)\left(\frac{L}{\mu_{B}}-1\right)\frac{1}{KL},\\
C_{\text{wasted, BPC-ZZO}}&=&\sum_{i=1}^{N-1}\frac{M_{A}M_{B}}{m_{A}m_{B}}+(\mu_{A}-2)\left(\frac{K}{\mu_{A}}-1\right)\frac{1}{KL}.
\end{eqnarray}

The following example illustrates the storage efficiency of bivariate polynomial codes.
\begin{example}
Assume $K=L=8$, i.e., the size of partitions of $A$ and $B$ are equal, and each worker can store 8 coded matrix partitions in total, i.e., $m_{A,i}+m_{B,i}=8$. Univariate schemes 
can only carry out $\eta_i=m_{A,i}=m_{B,i}=4$ computations. Instead, B-PROC can set $m_{A,i}=m_{B,i}=4$, resulting in $\eta_{i}=16$ computations. On the other hand, in BPC-VO and BPC-HO, the same worker can generate at most $\eta_i=7$ computations by setting $m_{A,i}=1, m_{B,i}=7$ for BPC-VO, or $m_{A,i}=7, m_{B,i}=1$ for BPC-HO. It is not possible to satisfy condition 2 of BPC-VO and BPC-HO under this storage capacity.
Finally, for BPC-NZO or BPC-ZZO, by setting $\mu_{A}=\mu_{B}=4$ and $m_{A,i}=m_{B,i}=4$, we can also reach $\eta_{i}=16$. Note that BPC-NZO and BPC-ZZO may not always obtain the B-PROC storage efficiency, but they can usually perform very close.
\end{example}

\begin{rem}
\color{\revisioncolor}
Note that $R_{th}^{NZO}$ and $R_{th}^{ZZO}$ provided in \thmref{main_thm} are worst-case values. Depending on the number of computations each worker sends to the master, smaller values, even $KL$ computations may be enough. In \secref{proof}, \lemref{quasi-unique-conds} presents certain conditions under which the computations received from the workers are useful. If the number of computations provided by all workers satisfies these conditions, then all computations are useful and $KL$ computations are enough. Otherwise, we need to discard some computations and since we need to compensate for these discarded computations, the recovery threshold may increase up to the values presented in \thmref{main_thm}. In \secref{proof}, the discussion after \lemref{quasi-unique-conds} explains what kind of computations we discard to guarantee almost regular decodability. 
\color{black}
\end{rem}

\color{\minorrevisioncolor}
\begin{rem}
When the conditions of \thmref{main_thm} are satisfied, the bivariate polynomial interpolation problem has a unique solution. The interpolation problem can be solved simply via inverting the interpolation matrix and multiplying it with the vector of responses collected from the workers. This has a complexity of $\mathcal{O}(rs(KL)^2)$. However, such an interpolation strategy may result in large numerical errors; and hence, more sophisticated methods, such as Newton interpolation, may be needed in practice \cite{sauer1995multivariate, sauer1995computational}. This aspect is worth investigation, but lies beyond the scope of this work. We leave it as a future research direction.
\end{rem}
\color{black}

\subsubsection*{Selecting between computation
orders}

When the partitions of $B$ are smaller than those of $A$, i.e., $c/L<r/K$, under a fixed storage capacity, reducing $m_{A,i}$ by 1 will increase $m_{B,i}$ at least by 1. Since, in this case, the constraints of vertical-type computation orders BPC-VO and BPC-NZO can be satisfied more easily than those of BPC-HO and BPC-ZZO, the schemes having a vertical-type computation order should be chosen. Similarly, when $r/K<c/L$, we should prefer horizontal ordering schemes BPC-HO or BPC-ZZO. Choosing between BPC-HO and BPC-ZZO when $r/K<c/L$, or between BPC-VO and BPC-NNO when $c/L<r/K$, depends on the storage capacity per worker and is discussed further in \secref{Numerical-Results}. 

\subsubsection*{Alternative formulation of almost regular interpolation schemes}

The reason why we formulate almost regular interpolation schemes in terms of Hermite interpolation throughout the paper is to shorten the proof of \thmref{main_thm}. Alternatively, instead of interpolating $A(x)B(y)$ from the evaluations of its derivatives, i.e., Hermite interpolation, almost regular interpolation schemes can also be formulated as the interpolation of $A(x)B(y)$ from its evaluations, as done in B-PROC. Such an approach is equivalent to the Hermite interpolation-based formulation, under the almost regularity condition. We include a more technical discussion about this in the supplementary material. Before reading it, we advise the reader to go through \secref{proof}, since the content in supplementary material is based on the definitions and techniques therein.


\section{\label{sec:Numerical-Results}Numerical Results}

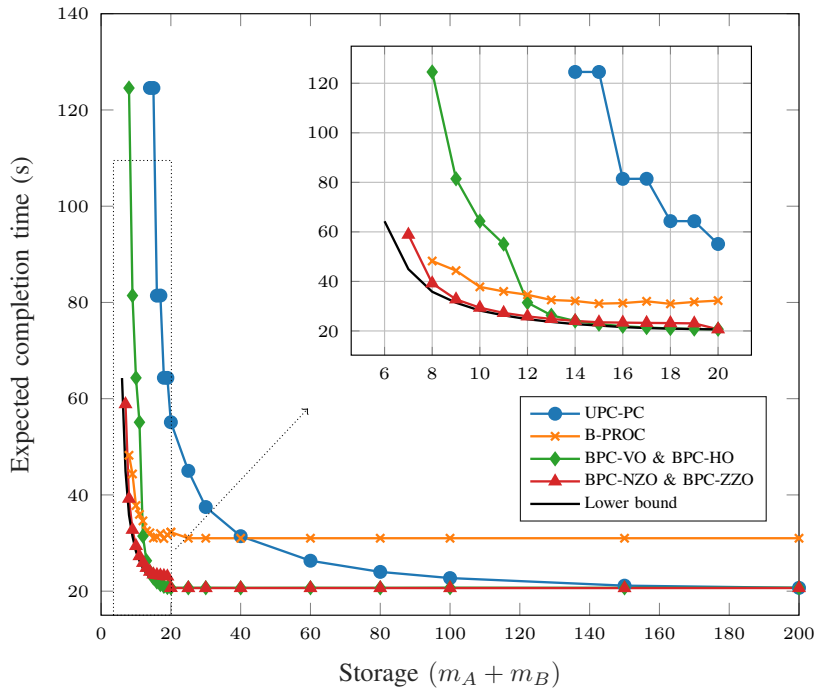
\begin{figure}
\centering \setlength{\fwidth}{0.5\textwidth}
\begin{tikzpicture}[scale=1.1]

\definecolor{color0}{rgb}{0.12156862745098,0.466666666666667,0.705882352941177}
\definecolor{color1}{rgb}{1,0.498039215686275,0.0549019607843137}
\definecolor{color2}{rgb}{0.172549019607843,0.627450980392157,0.172549019607843}
\definecolor{color3}{rgb}{0.83921568627451,0.152941176470588,0.156862745098039}

\begin{axis}[ scale=1,
 every axis plot/.append style={thick},
 scale only axis,
  try min ticks=7,
 yticklabel style={ font=\tiny,},
 xticklabel style={ font=\tiny,},
 xmin=0,
 xmax=200,
 xlabel style={font=\color{white!15!black},
 scale=0.9},
 xlabel={\small $\text{Storage  } (m_A+m_B)$},
 ymin=15,
 ymax=140,
 ylabel style={font=\color{white!15!black},
 scale=0.9},
 ylabel={\small Expected completion time (s)},
 axis background/.style={fill=white},
 legend style={legend cell align=left,
 align=left,
 draw=white!15!black,
 nodes={scale=0.65},
 at={(0.6,0.26)},
anchor=west}] 
\addplot [thick, color0, mark=*, mark size=2, mark options={solid}]  table[row sep=crcr]{ 
 14	124.57371\\ 
 15	124.56499\\ 
 16	81.41024\\ 
 17	81.4322500000002\\ 
 18	64.3429299999997\\ 
 19	64.3236199999999\\ 
 20	55.1115099999997\\ 
 25	45.0099199999998\\ 
 30	37.4829299999999\\ 
 40	31.4254299999999\\ 
 60	26.3239799999999\\ 
 80	24.0045100000001\\ 
 100	22.7211800000001\\ 
 150	21.14579\\ 
 200	20.6861700000001\\ 
 }; \addlegendentry{\footnotesize UPC-PC}
\addplot [thick, color1, mark=x, mark size=2, mark options={solid}]   table[row sep=crcr]{ 8	48.2359999999998\\ 
 9	44.3539999999999\\ 
 10	37.7929999999998\\ 
 11	35.9599999999997\\ 
 12	34.6039999999998\\ 
 13	32.5069999999998\\ 
 14	32.1089999999997\\ 
 15	31.0329999999997\\ 
 16	31.2369999999997\\ 
 17	31.9719999999997\\ 
 18	30.9409999999997\\ 
 19	31.7209999999997\\ 
 20	32.2569999999997\\ 
 20	32.26\\
 25	31\\ 
 30	31\\ 
 40	31\\ 
 60	31\\ 
 80	31\\ 
 100	31\\ 
 150	31\\ 
 200	31\\ 
 }; \addlegendentry{\footnotesize B-PROC}
\addplot [thick, color2, mark=diamond*, mark size=2, mark options={solid, fill}]   table[row sep=crcr]{ 8	124.55474\\ 
 9	81.4294299999998\\ 
 10	64.3284\\ 
 11	55.1022199999997\\ 
 12	31.4412299999999\\ 
 13	26.3228899999998\\ 
 14	24.0233\\ 
 15	22.7498700000001\\ 
 16	21.9094600000001\\ 
 17	21.3747700000001\\ 
 18	21.0104000000001\\ 
 19	20.77353\\ 
 20	20.7019700000001\\ 
 25	20.7041700000001\\ 
 30	20.7130300000001\\ 
 40	20.6952600000001\\ 
 60	20.7072700000001\\ 
 80	20.6982600000001\\ 
 100	20.7225500000001\\ 
 150	20.6959500000001\\ 
 200	20.70971\\ 
 }; \addlegendentry{\footnotesize BPC-VO \& BPC-HO}
\addplot [thick, color3, mark=triangle*, mark size=2, mark options={solid}]
  table[row sep=crcr]{
7   58.8608870000017\\ 
8	39.2202000000027\\ 
9	32.7608000000027\\ 
10	29.3908000000025\\ 
11	27.2849000000026\\ 
12	25.8825000000025\\ 
13	24.8478000000025\\ 
14	24.0454000000026\\ 
15	23.4892000000025\\ 
16	23.3672000000025\\ 
17	23.2476000000024\\ 
18	23.1694000000023\\ 
19	23.0564000000023\\ 
20	20.6659000000013\\ 
25    20.6455\\ 
30    20.6459\\ 
40    20.6496\\ 
60    20.6488\\ 
80    20.6488\\ 
100   20.6440\\ 
150   20.6403\\ 
200   20.6513\\ 
}; 
\addlegendentry{\footnotesize BPC-NZO \& BPC-ZZO}
\addplot [thick, color=black,
 mark options={solid,
 black}]
  table[row sep=crcr]{
6	64.2890490000011\\ 
7	44.9572850000023\\ 
8	35.8271000000027\\ 
9	31.4003000000024\\ 
10	28.2289000000025\\ 
11	26.2943000000024\\ 
12	24.7442000000023\\ 
13	23.6607000000025\\ 
14	22.7654000000024\\ 
15	22.1772000000027\\ 
16	21.5607000000026\\ 
17	21.1642000000025\\ 
18	20.9525000000023\\ 
19	20.7496000000015\\ 
20	20.6241000000011\\ 
};
\addlegendentry{\footnotesize Lower bound}
\coordinate (insetPosition) at (rel axis cs:0.35, 0.3); \end{axis}
\begin{axis}[ tiny,
 scale=2.0,
 at={(105, 105)},
 max space between ticks=10000pt,
  try min ticks=7,
  grid]
\addplot [thick, color2, mark=diamond*, mark size=2, mark options={solid, fill}]  table[row sep=crcr]{ 8	124.55474\\ 
 9	81.4294299999998\\ 
 10	64.3284\\ 
 11	55.1022199999997\\ 
 12	31.4412299999999\\ 
 13	26.3228899999998\\ 
 14	24.0233\\ 
 15	22.7498700000001\\ 
 16	21.9094600000001\\ 
 17	21.3747700000001\\ 
 18	21.0104000000001\\ 
 19	20.77353\\ 
 20	20.7019700000001\\ 
 };
\addplot [thick, color0, mark=*, mark size=2, mark options={solid}]   table[row sep=crcr]{ 
 14	124.57371\\ 
 15	124.56499\\ 
 16	81.41024\\ 
 17	81.4322500000002\\ 
 18	64.3429299999997\\ 
 19	64.3236199999999\\ 
 20	55.1115099999997\\ 
 };
\addplot [thick, color1, mark=x, mark size=2, mark options={solid}]   table[row sep=crcr]{ 8	48.2359999999998\\ 
 9	44.3539999999999\\ 
 10	37.7929999999998\\ 
 11	35.9599999999997\\ 
 12	34.6039999999998\\ 
 13	32.5069999999998\\ 
 14	32.1089999999997\\ 
 15	31.0329999999997\\ 
 16	31.2369999999997\\ 
 17	31.9719999999997\\ 
 18	30.9409999999997\\ 
 19	31.7209999999997\\ 
 20	32.2569999999997\\ 
 };
\addplot [thick, color3, mark=triangle*, mark size=2, mark options={solid}]
  table[row sep=crcr]{
7   58.8608870000017\\ 
8	39.2202000000027\\ 
9	32.7608000000027\\ 
10	29.3908000000025\\ 
11	27.2849000000026\\ 
12	25.8825000000025\\ 
13	24.8478000000025\\ 
14	24.0454000000026\\ 
15	23.4892000000025\\ 
16	23.3672000000025\\ 
17	23.2476000000024\\ 
18	23.1694000000023\\ 
19	23.0564000000023\\ 
20	20.6659000000013\\ 
};
\addplot [thick, color=black,
 mark options={solid,
 black}]
  table[row sep=crcr]{
6	64.2890490000011\\ 
7	44.9572850000023\\ 
8	35.8271000000027\\ 
9	31.4003000000024\\ 
10	28.2289000000025\\ 
11	26.2943000000024\\ 
12	24.7442000000023\\ 
13	23.6607000000025\\ 
14	22.7654000000024\\ 
15	22.1772000000027\\ 
16	21.5607000000026\\ 
17	21.1642000000025\\ 
18	20.9525000000023\\ 
19	20.7496000000015\\ 
20	20.6241000000011\\ 
};
\end{axis}
\draw[densely dotted,
 ->] (.9,.8) -- (2.5,2.5); 
\draw[densely dotted]  (0.15,
0) rectangle (0.85,
5.5);
\end{tikzpicture} 
\caption{\label{fig:simulation_results}Average computation times of univariate
and bivariate polynomial codes \color{\revisioncolor} as a function of available storage when partitions of $A$ and $B$ have equal size. \color{black}}
\end{figure}

\begin{figure}
\centering \setlength{\fwidth}{0.5\textwidth}
\begin{tikzpicture}[scale=1]

\definecolor{color0}{rgb}{0.12156862745098,0.466666666666667,0.705882352941177}
\definecolor{color1}{rgb}{1,0.498039215686275,0.0549019607843137}
\definecolor{color2}{rgb}{0.172549019607843,0.627450980392157,0.172549019607843}
\definecolor{color3}{rgb}{0.83921568627451,0.152941176470588,0.156862745098039}
\definecolor{color4}{rgb}{0.580392156862745,0.403921568627451,0.741176470588235}
\definecolor{color5}{rgb}{0.549019607843137,0.337254901960784,0.294117647058824}

\begin{axis}[ scale=1,
 every axis plot/.append style={thick},
 scale only axis,
  try min ticks=7,
 yticklabel style={ font=\tiny,},
 xticklabel style={ font=\tiny,},
 xmin=6,
 xmax=42,
 xlabel style={font=\color{white!15!black},
 scale=0.9},
 xlabel={\small $\text{Storage  } (m_A+2m_B)$},
 ymin=0,
 ymax=190,
 ylabel style={font=\color{white!15!black},
 scale=0.9},
 ylabel={\small Expected completion time (s)},
 axis background/.style={fill=white},
 legend style={legend cell align=left,
 align=left,
 draw=white!15!black,
 nodes={scale=0.8},
 at={(0.7,0.82)},
anchor=west},
grid] 
\addplot [thick, color0, mark=*, mark size=2, mark options={solid}]
table {%
14 133.026000000000	
15 132.628000000001	
16 81.6579999999996	
17 81.4539999999997	
18 64.2829999999995	
19 64.3819999999996	
20 55.0819999999998	
21 55.0169999999997	
22 49.2369999999998	
23 49.1409999999999	
24 45.0199999999999	
25 45.1209999999998	
26 41.8619999999998	
27 41.9089999999998	
28 39.6139999999998	
29 39.4419999999998	
30 37.4579999999999	
31 37.4389999999998	
32 35.7529999999998	
33 35.7129999999998	
34 34.6099999999999	
35 34.5309999999997	
36 33.3799999999999	
37 33.2719999999998	
38 32.2419999999998	
39 32.3679999999998	
40 31.4329999999998
};
\addlegendentry{\footnotesize UPC-PC}
\addplot [thick, color1, mark=x, mark size=2, mark options={solid}]
table {%
10 152.303000000000	
11 62.2389999999996	
12 62.2389999999996	
13 48.1609999999998	
14 48.1119999999998	
15 44.2529999999998	
16 37.8649999999998	
17 37.8649999999998		
18 36.1219999999997	
19 36.0459999999998	
20 34.9619999999998	
21 32.5829999999998	
22 32.5339999999998	
23 32.2509999999998	
24 31.1099999999999	
25 32.4819999999998	
26 31.4389999999997	
27 31.2469999999997	
28 31.9849999999997	
29 30.9899999999997	
30 32.7739999999997	
31 31.7799999999997	
32 31.8259999999997	
33 32.4209999999997	
34 31.8189999999997	
35 31.7049999999997	
36 32.4529999999997	
37 32.9119999999997	
38 33.0609999999996	
39 33.5539999999996	
40 33.4639999999995
};
\addlegendentry{\footnotesize B-PROC}
\addplot [thick, color5, mark=diamond*, mark size=2, mark options={solid}]
table {%
15 135.283541708204
16 135.272630328349
17 81.0729485451228
18 80.6960279156682
19 63.5066107597795
20 63.069785102794
21 54.2984527055202
22 29.3772342175873
23 23.7633834411042
24 21.1872165348734
25 19.4962335356334
26 18.4715068485804
27 17.5196123312021
28 16.856723496335
29 16.3530211910256
30 15.8088198696273
31 15.4940201940188
32 15.1346839169506
33 14.9311616979368
34 14.683058523078
35 14.3348419125913
36 14.1846752250284
37 14.0546304325115
38 13.9061462079551
39 13.7253611131796
40 13.5296152294937
};
\addlegendentry{\footnotesize BPC-VO}
\addplot [thick, color3, mark=diamond*, mark size=2, mark options={solid}]
table {%
9 134.700687543742
10 80.8026284451043
11 62.9776735970237
12 53.9781833733235
13 35.8521197809046
14 29.4762734154793
15 26.0080534438185
16 23.9808354750925
17 22.4609829654601
18 21.1145824799496
19 20.2084507498733
20 19.5807082356933
21 18.903879945128
22 18.3559293166611
23 17.8315231280331
24 17.4727111266347
25 17.1313398269724
26 16.8082120302327
27 16.6033495928665
28 16.2380897451832
29 16.0134460563775
30 15.892861034008
31 15.6771955680622
32 15.4243980085687
33 15.3177875150811
34 15.1571729771417
35 14.975374441132
36 14.833143326782
37 14.7420529846394
38 14.5983737889714
39 14.4812610715423
40 14.3645731949241
};
\addlegendentry{\footnotesize BPC-HO}
\addplot [thick, color4, mark=triangle*, mark size=2, mark options={solid}]
table {%
8 181.331543980353
9 58.1753752354478
10 46.1756411001613
11 37.5637397310045
12 34.129966504246
13 30.8329708292215
14 28.9787137073382
15 27.1630491085299
16 26.1435236443794
17 24.9736257584482
18 24.1493616481648
19 23.2875769733288
20 22.7619034056514
21 22.0056179993218
22 21.6184580231074
23 21.0988550235384
24 20.8075651038123
25 20.2319579012731
26 20.0479173364096
27 20.0718670414642
28 19.9198757631142
29 19.76793280412
30 16.4440268684943
31 16.3468350918887
32 16.4298627127952
33 16.2789046965433
34 16.2944136945238
35 14.9287018160111
36 14.9127173771777
37 14.897157785779
38 14.7817433644587
39 14.8158045854089
40 14.0391516016208
};
\addlegendentry{\footnotesize BPC-ZZO}
\addplot [thick, color2, mark=triangle*, mark size=2, mark options={solid}]
table {%
12 58.1549091061739
13 37.7243742237464
14 30.8053424401355
15 27.1252432872476
16 24.85844129291
17 23.1167053672201
18 21.8080679197167
19 21.0659886835218
20 20.1998056225954
21 20.2236695078269
22 20.0098877492798
23 20.0945698136197
24 19.8998432925285
25 19.9559579479529
26 19.7566616709562
27 19.9494173530304
28 19.7187176579051
29 19.7134480818421
30 16.4050645054305
31 16.4875375477139
32 16.4773959061601
33 16.3799614890027
34 16.2933601583296
35 16.3994976764765
36 16.225691989663
37 16.3097887573529
38 16.2578274367166
39 16.2930968235749
40 14.8484345456628
};
\addlegendentry{\footnotesize BPC-NZO}

\end{axis}

\end{tikzpicture}
\caption{\label{fig:simulation_results_diff_size}\color{\revisioncolor}Average computation times of univariate
and bivariate polynomial codes as a function of storage capacity, when partitions of $B$ are twice larger than partitions of $A$. \color{black}}
\end{figure}
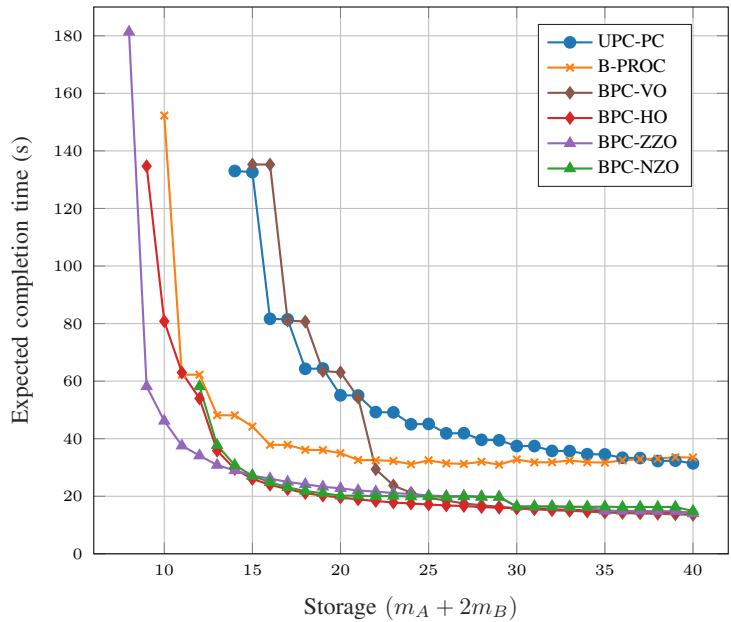

In this section, we compare the schemes presented throughout the paper in terms of the average computation time.
We only focus
on the computation time since the bivariate polynomials to be interpolated
in B-PROC, BPC-VO, BPC-HO, BPC-NZO and BPC-ZZO schemes have the same number of coefficients,
and thus, the variations in their encoding and decoding times are considered negligible.
We also assume that
the communication time is negligible. We model the computation speed
of the workers by the shifted exponential model \cite{liang2014tofec,lee2017speeding},
which is commonly used in the literature to analyze coded computation
schemes. In this model, the probability that a worker finishes at
least $p$ computations by time $t$ is $F(p,t)=1-e^{-\lambda(\frac{t}{p}-\nu)}$, if $t\geq p\nu$, and 0, otherwise. Thus, the probability of completing
exactly $p$ computations by time $t$ is given by $P(p,t)=F(p,t)-F(p+1,t)$
assuming $F(0,t)=1$, and $F(p_{max}+1,t)=0$, where $p_{max}$ is
the maximum number of computations a worker can complete. In $F(p,t)$,
$\nu$ is the minimum duration 
of one computation.
The scale parameter $\lambda$ controls the variance of computation times. The smaller is $\lambda$ the more variance, and thus more heterogeneous
computation speeds among the workers. To cover more heterogeneous
cases, we choose $\nu=0.01$ and $\lambda=0.1$.

We run Monte Carlo simulations to compute the expected computation time
for each scheme under different memory availability. \color{\revisioncolor} We consider two scenarios in which the sizes of the partitions of $A$ and $B$ are equal, i.e., $\frac{c}{L}=\frac{r}{K}$, and the size of the partitions of $B$ is twice larger than those of $A$, i.e., $\frac{c}{L}=\frac{2r}{K}$.  In both cases, \color{black} we assume that the workers have the same storage
capacity, as required by B-PROC. Thus,
$M_{A,i}=M_{A}$ and $M_{B,i}=M_{B}$, $\forall i \in [1:N]$. We set $K=L=10$ and assume $N=15$. In both scenarios, we set $\mu_B=5$ and $\mu_A=5$ for BPC-NZO and BPC-ZZO, respectively. For each memory value, we run $10^{4}$ experiments. \color{\revisioncolor}The results of the first scenario and the second scenario are given in \figref{simulation_results} and \figref{simulation_results_diff_size}, respectively. \color{black} For
each scheme, the minimum memory required to complete $KL=100$ computations
with $N=15$ workers is different. Thus, we plot each scheme starting
from a different minimum memory value.

\color{\revisioncolor}Let us first consider the scenario in which the partitions of $A$ and $B$ have equal size. In this case, since we also have $K=L$ and $\mu_A=\mu_B=5$, \color{black} there is no difference between BPC-HO
and BPC-VO, and also no difference between BPC-NZO and BPC-ZZO. 
In \figref{simulation_results}, we observe that BPC-NZO and BPC-ZZO result in a much lower expected computation
time than the other schemes for low storage capacities. Even though we
allow partial computations, the univariate polynomial coding, which
is UPC-PC, performs far worse than all the others due to inefficient
use of the memory resulting in a much less fraction of work done per worker compared to other schemes.
In B-PROC, despite the optimality in the memory allocation between $m_{A,i}$
and $m_{B,i}$, we see that the higher number of useless computations aggravates
the average computation time. For the same reason, increasing
the storage capacity does not improve the average computation time
beyond a certain point. While simulating B-PROC, we use
a random computation order at the workers, which is reported to perform
well in \cite{kiani2018exploitation} and stop the computation as soon as the master is able to decode. 
On the other hand, for BPC-NZO and BPC-ZZO,
we consider the worst-case scenario, in which the master starts decoding only after $(\mu_{B}-2)(\frac{L}{\mu_{B}}-1)$
computations for BPC-NZO or $(\mu_{A}-2)(\frac{K}{\mu_{A}}-1)$ computations for
BPC-ZZO are collected. Thus, we can expect the performance of BPC-NZO and BPC-ZZO to be even better than what we observe in \figref{simulation_results}. 
We also observe that BPC-VO
and BPC-HO performs significantly better than B-PROC and UPC-PC for the intermediate and large memory values. \color{\revisioncolor}For this storage regime, we also observe that BPC-HO and BPC-VO perform slightly better than BPC-ZZO and BPC-NZO due to the first constraint of these schemes. For instance, in BPC-NZO, when $m_{A,i}=K$, we need $m_{B,i}$ to be a multiple of $\mu_B$. Therefore, increasing storage capacity while keeping $m_{A,i}=K$ improves the expected computation time at some specific memory values. This is the reason for the improvement we observe in the expected computation time of the BPC-NZO in \figref{simulation_results} when the storage is 20. On the other hand, in the low storage regime, there is a significant performance
degradation of BPC-VO and BPC-HO due to the restrictive constraints of these schemes at small storage values. Since in BPC-NZO and BPC-ZZO, the constraints of BPC-VO and BPC-HO are relaxed especially for small storage
values, we observe that BPC-NZO and BPC-ZZO are superior in low storage regimes. To evaluate the performance of our schemes, we also plot a lower bound on the average
computation time of any bivariate polynomial-based coding scheme,
assuming there are no redundant computations and the memory allocation between
$m_{A,i}$ and $m_{B,i}$ is optimal. We observe that the average computation
time of the proposed bivariate schemes is quite close to this lower bound especially for the intermediate and high storage regimes. In a low storage regime, we observe that BPC-NZO and BPC-ZZO perform close to the lower bound although for very low values of storage, the gap between BPC-NZO/ZZO and the lower bound increases. This is due to the third constraint of these schemes which forbids optimal memory allocation between $m_{A,i}$ and $m_{B,i}$. This suggests that there might be still room for improvement in the trade-off between the expected computation time and the storage capacities of the workers. \color{black}

\color{\revisioncolor}
On the other hand, when we consider the scenario in which the partitions of $B$ is twice larger than those of $A$, in \figref{simulation_results_diff_size}, we observe that neither BPC-NZO and BPC-ZZO nor BPC-VO and BPC-HO are equivalent. Recall the discussion at the end of \subsecref{almost_regular_schemes} discussing how we select between computation orders. Since the partitions of $B$ is larger in our case, decreasing $m_{B,i}$ by one may increase $m_{A,i}$ more than one. Therefore, satisfying the constraints of horizontal-type schemes, i.e., BPC-ZZO and BPC-HO, is easier in this case. Therefore, we expect they perform better than the schemes with vertical-type order. We verify this in \figref{simulation_results_diff_size}, in which the performances of BPC-ZZO and BPC-HO are superior especially in the low storage regime. We also observe that for low and intermediate values of the storage, the performance of BPC-VO degrades close to that of UPC-PC. That is because the computations are done column-by-column in BPC-VO, see \figref{vo}, and to assign one more computation twice more storage availability is needed compared to BPC-HO and BPC-ZZO. BPC-NZO suffers from the same problem, but since in the zig-zag order, the constraints are relaxed, i.e., the computation grid is divided into blocks, its performance stays reasonable. We observe that it performs similarly to the BPC-HO in the intermediate and large storage values. Finally, similar to \figref{simulation_results}, we observe that for the large storage regime, almost regular schemes perform close to each other and much better than B-PROC and UPC-PC.
\color{black}

\section{Proof of \thmref{main_thm}\label{sec:proof}}
\color{\revisioncolor}

Without loss of generality, we assume $[1:n],1\leq n \leq N$ is the set of workers which provide at least one computation by the time the master collects sufficient responses 
to decode $AB$. Consider the interpolation matrix $M$ as defined in \defref{interpolation_matrix}. To prove the invertibility of an interpolation matrix $M$, we use Taylor series expansion
of $\det(M)$. Note that $\det(M)$ is a polynomial in the evaluation points $z_i\triangleq(x_{i},y_{i}),i\in [1:n]$. 
We can write the Taylor series expansion of $\det(M)$ around $(x_{i},y_{i})$
by taking the evaluation point $(x_{j},y_{j})$ as the variable, as: 
\begin{equation}
\det(M)=\sum_{(\alpha_1,\alpha_2)\in\mathbb{N}^{2}}\frac{1}{\alpha_{1}!\alpha_{2}!}(x_{j}-x_{i})^{\alpha_{1}}(y_{j}-y_{i})^{\alpha_{2}}D_{\alpha_{1},\alpha_{2}}(\tilde{Z}),\label{eq:taylor_expansion}
\end{equation}
where $\tilde{Z}\triangleq \{(x_k,y_k),k\in[1:n]\}\setminus \{(x_j,y_j)\}$, and
\begin{equation}
D_{\alpha_{1},\alpha_{2}}(\tilde{Z})\triangleq\frac{\partial^{\alpha_{1}+\alpha_{2}}}{\partial x_{j}^{\alpha_{1}}\partial y_{j}^{\alpha_{2}}}\det(M)\biggl|_{x_{j}=x_{i},y_{j}=y_{i}}.
\label{eq:d_a1_a2_y_first}    
\end{equation} 

We call $(x_{i},y_{i})$ the \textbf{pivot} node and $(x_{j},y_{j})$ the
\textbf{variable }node in this expansion. 
If the monomials in the set $\{x^{\alpha_1} y^{\alpha_2}\mid (\alpha_1,\alpha_2)\in\mathbb{N}^{2}\}$ cannot be written as a linear combination of the other monomials in the set, then, they are said to be \emph{linearly independent}. In this sense, the monomials $(x_{j}-x_{i})^{\alpha_{1}}(y_{j}-y_{i})^{\alpha_{2}}$ in \eqref{taylor_expansion} are linearly independent for different $(\alpha_{1},\alpha_{2})$ pairs, as long as, 
there is no dependence between $x_{i},x_{j}$ and $y_{i},y_{j}$. Consequently, $\det(M)=0$ for all values of  $(x_{j},y_{j})\in \mathbb{R}^2$,
if and only if $D_{\alpha_{1},\alpha_{2}}(\tilde{Z})=0,\forall(\alpha_{1},\alpha_{2})\in\text{\ensuremath{\mathbb{N}^{2}}}.$ That is, to show that $M$
is non-singular, it suffices to show that there exists an $(\alpha_{1},\alpha_{2})$ pair such that $D_{\alpha_{1},\alpha_{2}}(\tilde{Z})$ is nonzero. 

Let us choose some $(\alpha_{1},\alpha_{2})$ pair, and analyse $D_{\alpha_{1},\alpha_{2}}(\tilde{Z})$. 
Notice that, $D_{\alpha_{1},\alpha_{2}}(\tilde{Z})$ is a polynomial in the evaluation points, now, in $\tilde{Z}$. Specifically, it does not depend on $x_{j}$ and $y_{j}$ since the derivatives were taken with respect to these variables, and then evaluated
at $x_{j}=x_{i},y_{j}=y_{i}$. We call this procedure the \emph{coalescence} of the evaluation points $(x_i,y_i)$ and $(x_j,y_j)$ into $(x_i,y_i)$.
Next, to show 
$D_{\alpha_{1},\alpha_{2}}(\tilde{Z})\neq 0$, we do a new coalescence, i.e. we write the Taylor series expansion of  $D_{\alpha_{1},\alpha_{2}}(\tilde{Z})$ on a new variable point, choose a new $(\alpha_1,\alpha_2)$ pair, and coalescence them into $(x_i,y_i)$. Our proof technique is based on such recursive Taylor series expansions until all evaluation points are coalesced into one. We will present a technique to choose $(\alpha_1,\alpha_2)$ pairs at each step, which guarantees to obtain a non-zero polynomial at the final coalescence step.

In the following, we first present some preliminaries which we will need while presenting our technique for choosing $(\alpha_1,\alpha_2)$ at each step.

\subsection{Preliminaries}

In order to choose an $(\alpha_1,\alpha_2)$ pair at each step, we will need to analyze $D_{\alpha_{1},\alpha_{2}}(\tilde{Z})$. 
Since $D_{\alpha_{1},\alpha_{2}}(\tilde{Z})$ is derived from the Taylor series expansion of a determinant, in some cases, we can write it, again, in terms of the determinants of other matrices, which turns out to be more insightful than using its polynomial form.
Before showing this, we introduce the notions of \textbf{derivative set} and \textbf{shift}, 
which will be useful in the rest of the proof.

\begin{defn}
Associated to  every evaluation point $z_{i}\triangleq(x_{i},y_{i}),i\in[1:n]$,
there may be one or more rows in $M$ each corresponding to a different derivative order of $A(x)B(y)$ evaluated at $z_i$. We define the \textbf{derivative set}, $\mathcal{U}_{z_{i},M}$, of node $z_{i}$ as the multiset\footnote{Here, we use multiset instead of set as we allow multiple instances for each of its elements. The number of instances of a given element is called the multiplicity of that element in the multiset.} of derivative orders associated to $z_i$ in $M$, i.e., we say $(d_x,d_y)\in \mathcal{U}_{z_{i},M}$, if $M$ has a row corresponding the evaluation $\partial_{d_x} A(x_i)\partial_{d_y} B(y_i)$ or equivalently the master received the evaluation $\partial_{d_x} A(x_i)\partial_{d_y} B(y_i)$ from worker $i$.
\end{defn}

\begin{defn}
\textbf{\label{def:operator}} Let $M\in\mathbb{R}^{KL\times KL}$
be an interpolation matrix such that at least one of its rows depends on $(x_{j},y_{j})$, and let $r_{i}$ denote the $i^{th}$ row in $M$. We
define a \textbf{simple shift}\footnote{The term \textbf{shift} to refer to derivatives of interpolation matrices highlight the fact that derivatives applied to interpolation matrices correspond to shifts in their derivative sets when depicted in the derivative order space, as shown in \figref{example_N_zig_zag}} as
\begin{equation}
\partial_{i,x_j}M\triangleq\left[r_{1}^{T},\dots,\frac{\partial}{\partial x_{j}}r_{i}^{T},\dots,r_{KL}^{T}\right]^{T} \text{ and  } \partial_{i,y_j}M\triangleq\left[r_{1}^{T},\dots,\frac{\partial}{\partial y_{j}}r_{i}^{T},\dots,r_{KL}^{T}\right]^{T}.
\end{equation}

Assume that the $i^{th}$ row of $M$ corresponds to $\partial_{d_{i,x}} A(x_j)\partial_{d_{i,y}} B(y_j)$. Then, the derivative sets of node $z_i$ associated to matrices $\partial_{i,x_j}M$ and $\partial_{i,y_j}M$ are shifted versions of the ones associated to $M$, in the sense that, $\mathcal{U}_{z_j,\partial_{i,x_j}M}=\{({d_{i,x}}+1,d_{i,y})\}\cup \mathcal{U}_{z_j,M} \setminus \{({d_{i,x}},{d_{i,y}})\}$ and $\mathcal{U}_{z_j,\partial_{i,y_j}M}=\{({d_{i,x}},{d_{i,y}}+1)\}\cup \mathcal{U}_{z_j,M} \setminus \{({d_{i,x}},{d_{i,y}})\}$. Note that if the multiplicity of any element in a derivative set is greater than one, then the corresponding interpolation matrix has at least two identical rows making the matrix singular.  $\partial_{i,x_j}$ is called a \textbf{regular} simple shift, if all elements in $\mathcal{U}_{z_j,\partial_{i,x_j}M}$ have a multiplicity of one. Similarly, $\partial_{i,y_j}$ is a regular simple shift if all elements in $\mathcal{U}_{z_j,\partial_{i,y_j}M}$ have a multiplicity of one. Finally, for the \textbf{composition of simple shifts}, we introduce the notation $\nabla^{x_j,y_j}_{\boldsymbol{k},\boldsymbol{l}}M$, where the $i^{th}$ entries in $\boldsymbol{k}$ and $\boldsymbol{l}$
are the total order of the derivatives taken on the $i^{th}$ row of $M$
with respect to $x_{j}$ and $y_{j}$, respectively. Thanks to the commutative property of the derivative, given a pair $(\boldsymbol{k},\boldsymbol{l})$ one can compute  $\nabla^{x_j,y_j}_{\boldsymbol{k},\boldsymbol{l}}M$ by taking derivatives from any row, at any order, until completing the derivative orders specified in $(\boldsymbol{k},\boldsymbol{l})$. We refer to each of these possible choices as a derivative path, and define those paths that only involve regular simple shifts, i.e. after each derivative there are not two equal rows, as \textbf{regular derivative paths}. The number of regular derivative paths is denoted by $C_{\boldsymbol{k},\boldsymbol{l}}(M)$.
\end{defn}

The following lemma provides an expression for the derivatives of the determinant of an interpolation matrix in terms of a weighted sum of determinants of other interpolation 
 matrices. 

\begin{lem}
\label{lem:regular_permutations}
Let $\boldsymbol{k}\in[0:K-1]^{KL}$, $\boldsymbol{l}\in[0:L-1]^{KL}$ and $\alpha_{1}=\sum_{i=1}^{KL}\boldsymbol{k}(i)$
and $\alpha_{2}=\sum_{i=1}^{KL}\boldsymbol{l}(i)$. Then, we have\textup{ 
\begin{equation}
\frac{\partial^{\alpha_{1}+\alpha_{2}}}{\partial x_j^{\alpha_{1}}\partial y_j^{\alpha_{2}}}\det(M)\biggl|_{x_{j}=x_{i},y_{j}=y_{i}}=\sum_{
\left(\boldsymbol{k,l}\right)\in \mathcal{R}_M(\alpha_1,\alpha_2)}C_{\boldsymbol{k},\boldsymbol{l}}(M)\det\left(\nabla^{x_j,y_j}_{\boldsymbol{k},\boldsymbol{l}}M\right)\biggl|_{x_{j}=x_{i},y_{j}=y_{i}}\label{eq:derivatives_regular_j_l}
\end{equation}
}
where $\mathcal{R}_M(\alpha_1,\alpha_2)$ is the set of  $(\boldsymbol{k},\boldsymbol{l})$ pairs satisfying  $C_{\boldsymbol{k},\boldsymbol{l}}(M)\neq 0$, i.e., there is at least one derivative path for which $\nabla^{x_j,y_j}_{\boldsymbol{k},\boldsymbol{l}}$ can be applied by using only regular simple shifts. 

\end{lem}
We defer the Proof of \lemref{regular_permutations} to \appref{proof_regular_permutations}. 



\subsection{Choosing an $(\alpha_1,\alpha_2)$ pair in a coalescence}
Recall that our objective is to find an $(\alpha_1,\alpha_2)$ pair for each step in the successive coalescence procedure. \lemref{regular_permutations}  is an important step in this direction as it allows us to  express $D_{\alpha_{1},\alpha_{2}}(\tilde{Z})$ in terms of a sum of determinants of interpolation matrices. 
However, it still does not provide us a clear clue on how to choose $(\alpha_1,\alpha_2)$, so that $D_{\alpha_{1},\alpha_{2}}(\tilde{Z})\neq 0$. Next, we define a structure over the derivative sets of the interpolation matrices, similar to the ones defined for the computation orders in \subsecref{almost_regular_schemes}, which will eventually help us to define the 
\textbf{quasi-unique shift} pairs $(\alpha_1,\alpha_2)$, which satisfy the conditions needed for completing a coalescence procedure successfully.

\color{black}
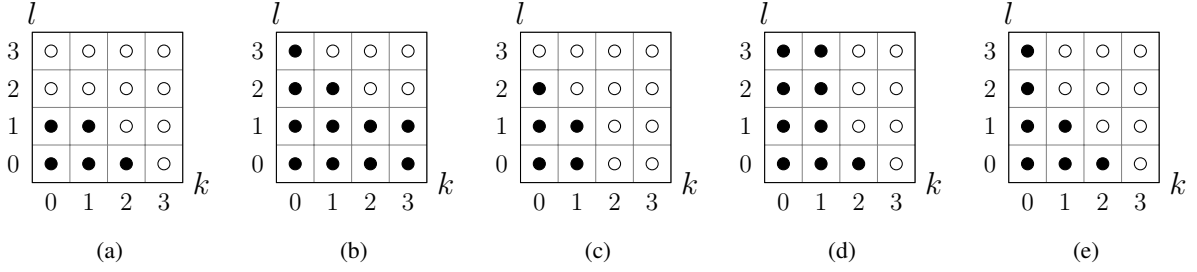
\begin{figure}
\centering

\subfloat[\label{fig:n-zig-zag-1}]{
\begin{tikzpicture}

\draw [help lines,  step=0.5cm] (-3.5, -3.5) node (v19) {} grid (-1.5,-1.5); 
\node[scale=0.8] at (-3.25,-3.75) {$0$};  
\node[scale=1] at (-1.25,-3.5) {$k$};  
\node[scale=1] at (-3.5,-1.25) {$l$};         
\node[scale=0.8] at (-1.75,-3.75) {$3$};   
\node[scale=0.8] at (-2.25,-3.75) {$2$};   
\node[scale=0.8] at (-2.75,-3.75) {$1$};   
\node[scale=0.8] at (-3.75,-3.25) {$0$};              
\node[scale=0.8] at (-3.75,-1.75) {$3$};  
\node[scale=0.8] at (-3.75,-2.25) {$2$};  
\node[scale=0.8] at (-3.75,-2.75) {$1$};

\draw[fill, color=black] (-3.25,-3.25) node (v1) {} circle (.08); 
\draw[fill,color=black] (-2.75,-3.25) node (v2) {} circle (.08); 

\draw[fill, color=black] (-2.25,-3.25) node (v3) {} circle (.08);  
\draw[fill, color=black] (-3.25,-2.75) node (v4) {} circle (.08);  
\draw[fill, color=black] (-2.75,-2.75) node (v5) {} circle (.08);  
\draw[color=black] (-2.25,-2.75) node (v6) {} circle (.08);  
\draw[color=black] (-3.25,-2.25) node (v7) {} circle (.08);  
\draw[color=black] (-2.75,-2.25) node (v8) {} circle (.08);  
\draw[color=black] (-2.25,-2.25) node (v9) {} circle (.08);  
\draw[color=black] (-3.25,-1.75) node (v10) {} circle (.08);  
\draw[color=black] (-2.75,-1.75) node (v11) {} circle (.08);  
\draw[color=black] (-2.25,-1.75) node (v12) {} circle (.08);              
\draw[color=black] (-1.75,-3.25) node (v19) {} circle (.08);      
\draw[color=black] (-1.75,-2.75) node (v22) {} circle (.08);      
\draw[color=black] (-1.75,-2.25) node (v25) {} circle (.08);      
\draw[color=black] (-1.75,-1.75) node (v28) {} circle (.08);                 

\draw  (-3.5,-1.5) rectangle (-1.5,-3.5);
\end{tikzpicture}}
\subfloat[\label{fig:n-zig-zag-2}]{
\begin{tikzpicture}

\draw [help lines,  step=0.5cm] (-3.5, -3.5) node (v19) {} grid (-1.5,-1.5); 
\node[scale=0.8] at (-3.25,-3.75) {$0$};  
\node[scale=1] at (-1.25,-3.5) {$k$};  
\node[scale=1] at (-3.5,-1.25) {$l$};         
\node[scale=0.8] at (-1.75,-3.75) {$3$};   
\node[scale=0.8] at (-2.25,-3.75) {$2$};   
\node[scale=0.8] at (-2.75,-3.75) {$1$};   
\node[scale=0.8] at (-3.75,-3.25) {$0$};              
\node[scale=0.8] at (-3.75,-1.75) {$3$};  
\node[scale=0.8] at (-3.75,-2.25) {$2$};  
\node[scale=0.8] at (-3.75,-2.75) {$1$};

\draw[fill, color=black] (-3.25,-3.25) node (v1) {} circle (.08); 
\draw[fill,color=black] (-2.75,-3.25) node (v2) {} circle (.08); 

\draw[fill, color=black] (-2.25,-3.25) node (v3) {} circle (.08);  
\draw[fill, color=black] (-3.25,-2.75) node (v4) {} circle (.08);  
\draw[fill, color=black] (-2.75,-2.75) node (v5) {} circle (.08);  
\draw[fill, color=black] (-2.25,-2.75) node (v6) {} circle (.08);  
\draw[fill, color=black] (-3.25,-2.25) node (v7) {} circle (.08);  
\draw[fill, color=black] (-2.75,-2.25) node (v8) {} circle (.08);  
\draw[color=black] (-2.25,-2.25) node (v9) {} circle (.08);  
\draw[fill, color=black] (-3.25,-1.75) node (v10) {} circle (.08);  
\draw[color=black] (-2.75,-1.75) node (v11) {} circle (.08);  
\draw[color=black] (-2.25,-1.75) node (v12) {} circle (.08);              
\draw[fill, color=black] (-1.75,-3.25) node (v19) {} circle (.08);      
\draw[fill, color=black] (-1.75,-2.75) node (v22) {} circle (.08);      
\draw[color=black] (-1.75,-2.25) node (v25) {} circle (.08);      
\draw[color=black] (-1.75,-1.75) node (v28) {} circle (.08);                 

\draw  (-3.5,-1.5) rectangle (-1.5,-3.5);
\end{tikzpicture}}
\subfloat[\label{fig:z-zig-zag-1}]{
\begin{tikzpicture}

\draw [help lines,  step=0.5cm] (-3.5, -3.5) node (v19) {} grid (-1.5,-1.5); 
\node[scale=0.8] at (-3.25,-3.75) {$0$};  
\node[scale=1] at (-1.25,-3.5) {$k$};  
\node[scale=1] at (-3.5,-1.25) {$l$};         
\node[scale=0.8] at (-1.75,-3.75) {$3$};   
\node[scale=0.8] at (-2.25,-3.75) {$2$};   
\node[scale=0.8] at (-2.75,-3.75) {$1$};   
\node[scale=0.8] at (-3.75,-3.25) {$0$};              
\node[scale=0.8] at (-3.75,-1.75) {$3$};  
\node[scale=0.8] at (-3.75,-2.25) {$2$};  
\node[scale=0.8] at (-3.75,-2.75) {$1$};

\draw[fill, color=black] (-3.25,-3.25) node (v1) {} circle (.08); 
\draw[fill,color=black] (-2.75,-3.25) node (v2) {} circle (.08); 

\draw[color=black] (-2.25,-3.25) node (v3) {} circle (.08);  
\draw[fill, color=black] (-3.25,-2.75) node (v4) {} circle (.08);  
\draw[fill, color=black] (-2.75,-2.75) node (v5) {} circle (.08);  
\draw[color=black] (-2.25,-2.75) node (v6) {} circle (.08);  
\draw[fill, color=black] (-3.25,-2.25) node (v7) {} circle (.08);  
\draw[color=black] (-2.75,-2.25) node (v8) {} circle (.08);  
\draw[color=black] (-2.25,-2.25) node (v9) {} circle (.08);  
\draw[color=black] (-3.25,-1.75) node (v10) {} circle (.08);  
\draw[color=black] (-2.75,-1.75) node (v11) {} circle (.08);  
\draw[color=black] (-2.25,-1.75) node (v12) {} circle (.08);              
\draw[color=black] (-1.75,-3.25) node (v19) {} circle (.08);      
\draw[color=black] (-1.75,-2.75) node (v22) {} circle (.08);      
\draw[color=black] (-1.75,-2.25) node (v25) {} circle (.08);      
\draw[color=black] (-1.75,-1.75) node (v28) {} circle (.08);                 

\draw  (-3.5,-1.5) rectangle (-1.5,-3.5);
\end{tikzpicture}}
\subfloat[\label{fig:z-zig-zag-2}]{
\begin{tikzpicture}

\draw [help lines,  step=0.5cm] (-3.5, -3.5) node (v19) {} grid (-1.5,-1.5); 
\node[scale=0.8] at (-3.25,-3.75) {$0$};  
\node[scale=1] at (-1.25,-3.5) {$k$};  
\node[scale=1] at (-3.5,-1.25) {$l$};         
\node[scale=0.8] at (-1.75,-3.75) {$3$};   
\node[scale=0.8] at (-2.25,-3.75) {$2$};   
\node[scale=0.8] at (-2.75,-3.75) {$1$};   
\node[scale=0.8] at (-3.75,-3.25) {$0$};              
\node[scale=0.8] at (-3.75,-1.75) {$3$};  
\node[scale=0.8] at (-3.75,-2.25) {$2$};  
\node[scale=0.8] at (-3.75,-2.75) {$1$};

\draw[fill, color=black] (-3.25,-3.25) node (v1) {} circle (.08); 
\draw[fill,color=black] (-2.75,-3.25) node (v2) {} circle (.08); 

\draw[fill, color=black] (-2.25,-3.25) node (v3) {} circle (.08);  
\draw[fill, color=black] (-3.25,-2.75) node (v4) {} circle (.08);  
\draw[fill, color=black] (-2.75,-2.75) node (v5) {} circle (.08);  
\draw[color=black] (-2.25,-2.75) node (v6) {} circle (.08);  
\draw[fill, color=black] (-3.25,-2.25) node (v7) {} circle (.08);  
\draw[fill, color=black] (-2.75,-2.25) node (v8) {} circle (.08);  
\draw[color=black] (-2.25,-2.25) node (v9) {} circle (.08);  
\draw[fill, color=black] (-3.25,-1.75) node (v10) {} circle (.08);  
\draw[fill, color=black] (-2.75,-1.75) node (v11) {} circle (.08);  
\draw[color=black] (-2.25,-1.75) node (v12) {} circle (.08);              
\draw[color=black] (-1.75,-3.25) node (v19) {} circle (.08);      
\draw[color=black] (-1.75,-2.75) node (v22) {} circle (.08);      
\draw[color=black] (-1.75,-2.25) node (v25) {} circle (.08);      
\draw[color=black] (-1.75,-1.75) node (v28) {} circle (.08);                 

\draw  (-3.5,-1.5) rectangle (-1.5,-3.5);
\end{tikzpicture}}
\subfloat[\label{fig:non1}]{
\begin{tikzpicture}

\draw [help lines,  step=0.5cm] (-3.5, -3.5) node (v19) {} grid (-1.5,-1.5); 
\node[scale=0.8] at (-3.25,-3.75) {$0$};  
\node[scale=1] at (-1.25,-3.5) {$k$};  
\node[scale=1] at (-3.5,-1.25) {$l$};         
\node[scale=0.8] at (-1.75,-3.75) {$3$};   
\node[scale=0.8] at (-2.25,-3.75) {$2$};   
\node[scale=0.8] at (-2.75,-3.75) {$1$};   
\node[scale=0.8] at (-3.75,-3.25) {$0$};              
\node[scale=0.8] at (-3.75,-1.75) {$3$};  
\node[scale=0.8] at (-3.75,-2.25) {$2$};  
\node[scale=0.8] at (-3.75,-2.75) {$1$};

\draw[fill, color=black] (-3.25,-3.25) node (v1) {} circle (.08); 
\draw[fill,color=black] (-2.75,-3.25) node (v2) {} circle (.08); 

\draw[fill, color=black] (-2.25,-3.25) node (v3) {} circle (.08);  
\draw[fill, color=black] (-3.25,-2.75) node (v4) {} circle (.08);  
\draw[fill, color=black] (-2.75,-2.75) node (v5) {} circle (.08);  
\draw[color=black] (-2.25,-2.75) node (v6) {} circle (.08);  
\draw[fill, color=black] (-3.25,-2.25) node (v7) {} circle (.08);  
\draw[color=black] (-2.75,-2.25) node (v8) {} circle (.08);  
\draw[color=black] (-2.25,-2.25) node (v9) {} circle (.08);  
\draw[fill, color=black] (-3.25,-1.75) node (v10) {} circle (.08);  
\draw[color=black] (-2.75,-1.75) node (v11) {} circle (.08);  
\draw[color=black] (-2.25,-1.75) node (v12) {} circle (.08);              
\draw[color=black] (-1.75,-3.25) node (v19) {} circle (.08);      
\draw[color=black] (-1.75,-2.75) node (v22) {} circle (.08);      
\draw[color=black] (-1.75,-2.25) node (v25) {} circle (.08);      
\draw[color=black] (-1.75,-1.75) node (v28) {} circle (.08);                 

\draw  (-3.5,-1.5) rectangle (-1.5,-3.5);
\end{tikzpicture}}
\caption{Example sets of N-zig-zag ordered (a,b), Z-zig-zag ordered (c,d) and
neither N-zig-zag ordered nor Z-zig-zag ordered (e).}\label{fig:example_N_zig_zag}
\end{figure}

\begin{defn}
A derivative set $\mathcal{U}_{z,M}$ is said to be \textbf{N-zig-zag ordered} with parameter $\mu_{B}$ if $(i,j)\in \mathcal{U}_{z,M}$ implies that all the derivatives with order $(k,l)$
such that $S_{\mathcal{N}}(1,1)\geq S_{\mathcal{N}}(k+1,l+1)\geq S_{\mathcal{N}}(i+1,j+1)$
are also in $\mathcal{U}_{z,M}$, where \color{\revisioncolor} $S_{\mathcal{N}}(k,l)=(K-1)\mu_{B}\left(\left\lceil \frac{l}{\mu_{B}}\right\rceil -1\right)+\mu_{B}(k-1)+l$ as in \subsecref{almost_regular_schemes}. \color{black} Similarly, $\mathcal{U}_{z,M}$ is \textbf{Z-zig-zag
ordered} with parameter $\mu_{A}$ if $(i,j)\in \mathcal{U}_{z,M}$ implies
that all $(k,l)$ such that $S_{\mathcal{Z}}(1,1)\geq S_{\mathcal{Z}}(k+1,l+1)\geq S_{\mathcal{Z}}(i+1,j+1)$
are in $\mathcal{U}_{z,M}$, \color{\revisioncolor} where $S_{\mathcal{Z}}(k,l)=(L-1)\mu_{A}\left(\left\lceil \frac{k}{\mu_{A}}\right\rceil -1\right)+\mu_{A}(l-1)+k$ as in \subsecref{almost_regular_schemes}. \color{black}
\end{defn}
\begin{example}
\label{Ex:natural_order_visualization}Consider the derivative sets
with $K=L=4$. The derivative sets
illustrated in \figref{n-zig-zag-1} and \figref{n-zig-zag-2} are N-zig-zag
ordered for $\mu_{B}=2$, and the sets in \figref{z-zig-zag-1}
and \figref{z-zig-zag-2} are Z-zig-zag ordered for $\mu_{A}=2$.
The set in \figref{non1} is neither N-zig-zag nor
Z-zig-zag ordered.
\end{example}
\color{\revisioncolor}

Hereafter, for brevity, we stick to the N-zig-zag order. This will allow us to prove the part $a$ of \thmref{main_thm}. The proof of part $b$ follows similarly using the Z-zig-zag order instead, and thus we omit it here. 

\color{black}

\color{\revisioncolor}

\begin{defn}
Consider $(x_{j},y_{j})$ is the variable node and $(x_{i},y_{i})$
is the pivot node.  Suppose that $\mathcal{U}_{z_j,M}$ obeys the N-zig-zag order, and define $M^*\triangleq\nabla^{x_j,y_j}_{\boldsymbol{k}^*,\boldsymbol{l}^*}M\bigl|_{x_{j}=x_{i},y_{j}=y_{i}}$. If there is only one $(\boldsymbol{k}^*,\boldsymbol{l}^*)\in \mathcal{R}_M(\alpha_1,\alpha_2)$ such that $\mathcal{U}_{z_j,M^*}$ obeys the N-zig-zag order, then $(\alpha_1,\alpha_2)$ is called
\textbf{quasi-unique}. 
\end{defn}

\color{black}
\begin{figure}
\centering\subfloat[\label{fig:quasi-unique-1-a}]{
\begin{tikzpicture} 
\draw [help lines,  step=0.5cm] (-3.5, -3.5) node (v19) {} grid (-2, -0.5); 
\node[scale=0.8] at (-3.25,-3.75) {$0$};  
\node[scale=1] at (-1.75,-3.5) {$k$};  
\node[scale=1] at (-3.5,-0.25) {$l$};      
\node[scale=0.8] at (-2.25,-3.75) {$2$};   
\node[scale=0.8] at (-2.75,-3.75) {$1$};   
\node[scale=0.8] at (-3.75,-3.25) {$0$};     
\node[scale=0.8] at (-3.75,-0.75) {$5$};  
\node[scale=0.8] at (-3.75,-1.25) {$4$};  
\node[scale=0.8] at (-3.75,-1.75) {$3$};  
\node[scale=0.8] at (-3.75,-2.25) {$2$};  
\node[scale=0.8] at (-3.75,-2.75) {$1$};        
\draw[fill, color=black] (-3.25,-2.25) node (v1) {} circle (.08);  
\draw[fill, color=black] (-2.75,-3.25) node (v3) {} circle (.08);      
\draw[fill, color=black] (-3.25,-3.25) node (v2) {} circle (.08); 
\draw[fill, color=black] (-3.25,-2.75) node (v2) {} circle (.08); 
\draw  (-3.5,-2) rectangle (-2,-3.5); 
\draw  (-3.5,-0.5) rectangle (-2,-2); \end{tikzpicture}}
\subfloat[\label{fig:quasi-unique-1-b}]{
\begin{tikzpicture}

\draw [help lines,  step=0.5cm] (-3.5, -3.5) node (v19) {} grid (-2, -0.5); 
\node[scale=0.8] at (-3.25,-3.75) {$0$};  
\node[scale=1] at (-1.75,-3.5) {$k$};  
\node[scale=1] at (-3.5,-0.25) {$l$};   
\node[scale=0.8] at (-2.25,-3.75) {$2$};   
\node[scale=0.8] at (-2.75,-3.75) {$1$};   
\node[scale=0.8] at (-3.75,-3.25) {$0$};   
    
\node[scale=0.8] at (-3.75,-0.75) {$5$};  
\node[scale=0.8] at (-3.75,-1.25) {$4$};  
\node[scale=0.8] at (-3.75,-1.75) {$3$};  
\node[scale=0.8] at (-3.75,-2.25) {$2$};  
\node[scale=0.8] at (-3.75,-2.75) {$1$};       

\draw  (-3.5,-2) rectangle (-2,-3.5); 
\draw  (-3.5,-0.5) rectangle (-2,-2);

\node at (-3.25,-3.25) {$1$}; 
\node at (-3.25,-2.75) {$2$}; \end{tikzpicture}}
\subfloat[\label{fig:quasi-unique-1-c}]{
\begin{tikzpicture} 
\draw [help lines,  step=0.5cm] (-3.5, -3.5) node (v19) {} grid (-2, -0.5); 
\node[scale=0.8] at (-3.25,-3.75) {$0$};  
\node[scale=1] at (-1.75,-3.5) {$k$};  
\node[scale=1] at (-3.5,-0.25) {$l$};      
\node[scale=0.8] at (-2.25,-3.75) {$2$};   
\node[scale=0.8] at (-2.75,-3.75) {$1$};   
\node[scale=0.8] at (-3.75,-3.25) {$0$};     
\node[scale=0.8] at (-3.75,-0.75) {$5$};  
\node[scale=0.8] at (-3.75,-1.25) {$4$};  
\node[scale=0.8] at (-3.75,-1.75) {$3$};  
\node[scale=0.8] at (-3.75,-2.25) {$2$};  
\node[scale=0.8] at (-3.75,-2.75) {$1$};        
\draw[fill, color=black] (-3.25,-2.25) node (v1) {} circle (.08);  
\draw[fill, color=black] (-2.75,-3.25) node (v3) {} circle (.08);      
\draw[fill, color=black] (-3.25,-3.25) node (v2) {} circle (.08); 
\draw[fill, color=black] (-3.25,-2.75) node (v2) {} circle (.08); 
\draw  (-3.5,-2) rectangle (-2,-3.5); 
\draw  (-3.5,-0.5) rectangle (-2,-2);

\node at (-2.75,-2.75) {$1$}; 
\node at (-2.75,-2.25) {$2$}; \end{tikzpicture}}
\subfloat[\label{fig:quasi-unique-1-d}]{
\begin{tikzpicture} 
\draw [help lines,  step=0.5cm] (-3.5, -3.5) node (v19) {} grid (-2, -0.5); 
\node[scale=0.8] at (-3.25,-3.75) {$0$};  
\node[scale=1] at (-1.75,-3.5) {$k$};  
\node[scale=1] at (-3.5,-0.25) {$l$};      
\node[scale=0.8] at (-2.25,-3.75) {$2$};   
\node[scale=0.8] at (-2.75,-3.75) {$1$};   
\node[scale=0.8] at (-3.75,-3.25) {$0$};     
\node[scale=0.8] at (-3.75,-0.75) {$5$};  
\node[scale=0.8] at (-3.75,-1.25) {$4$};  
\node[scale=0.8] at (-3.75,-1.75) {$3$};  
\node[scale=0.8] at (-3.75,-2.25) {$2$};  
\node[scale=0.8] at (-3.75,-2.75) {$1$};        
\draw[fill, color=black] (-3.25,-2.25) node (v1) {} circle (.08);  
\draw[fill, color=black] (-2.75,-3.25) node (v3) {} circle (.08);      
\draw[fill, color=black] (-3.25,-3.25) node (v2) {} circle (.08); 
\draw[fill, color=black] (-3.25,-2.75) node (v2) {} circle (.08); 
\draw  (-3.5,-2) rectangle (-2,-3.5); 
\draw  (-3.5,-0.5) rectangle (-2,-2);

\node at (-2.25,-3.25) {$1$}; 
\node at (-3.25,-1.75) {$2$}; \end{tikzpicture}}
\caption{Depictions of derivative sets in \exaref{quasi-unique}.}
\end{figure}
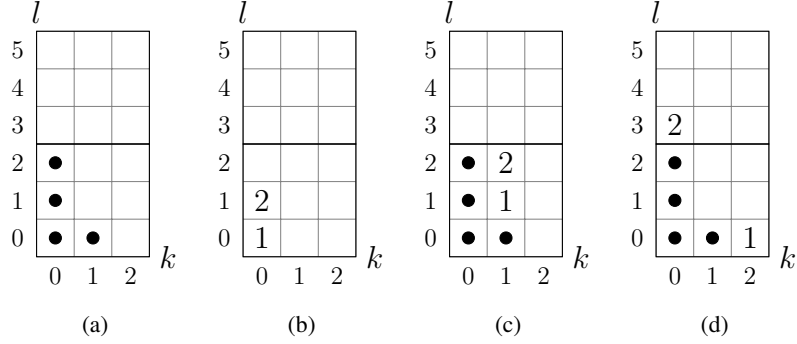
\color{\revisioncolor}

\begin{example}
\label{exa:quasi-unique}Let $(\alpha_1,\alpha_2)=(2,2)$ and $K=3, L=6$, $\mu_{B}=3$. We assume
the derivative sets of the pivot and variable nodes are as depicted in the derivative order space in \figref{quasi-unique-1-a} and \figref{quasi-unique-1-b}, respectively. We observe that the interpolation matrix has two rows that depend on the variable node and four rows that depend on the pivot node. Without loss of generality, we assume rows 1 and 2 depend on the variable node. In this example, we stick to the definition in \eqref{d_a1_a2_y_first}, i.e., we take derivatives of the interpolation matrix first with respect to the
$y$ component of the variable node and then the $x$ component. Therefore, $\mathcal{R}_M(2,2)=\{\left([1,1,\mathbf{0}_{KL-2}],[1,1,\mathbf{0}_{KL-2}]\right),\left([2,0,\mathbf{0}_{KL-2}],[0,2,\mathbf{0}_{KL-2}]\right)\}$, where $\mathbf{0}_{KL-2}$ is the all-zero vector with dimension $KL-2$. Note that there is no other $(\boldsymbol{k},\boldsymbol{l})$ pair such that $\nabla^{x_j,y_j}_{\boldsymbol{k},\boldsymbol{l}}$ can be applied by using only regular simple shifts. When we apply $\nabla^{x_j,y_j}_{\boldsymbol{k},\boldsymbol{l}}$ with $(\boldsymbol{k},\boldsymbol{l})=([1,1,\mathbf{0}_{KL-2}],[1,1,\mathbf{0}_{KL-2}])$, we obtain a derivative set as depicted in \figref{quasi-unique-1-c}, and obtain the one in \figref{quasi-unique-1-d} with $(\boldsymbol{k},\boldsymbol{l})=([2,0,\mathbf{0}_{KL-2}],[0,2,\mathbf{0}_{KL-2}])$. Note that the derivative set in \figref{quasi-unique-1-c} obeys the N-zig-zag order while the one in \figref{quasi-unique-1-d} does not. Since there is only one $(\boldsymbol{k},\boldsymbol{l})$ pair resulting in an N-zig-zag ordered derivative set, $(\alpha_1,\alpha_2)$ is quasi-unique.
\end{example}

Next we describe, in detail, the first two iterations of the recursive coalescence
procedure, and then generalize the result to any iteration.  Without loss of generality, we choose $(x_{n},y_{n})$ as the pivot node for all the
coalescences in the recursion, and coalesce it with the variable node $z_{i}$ in
the $i^{th}$ coalescence from $i=1$ to $n-1$. Let us define the set
of remaining nodes before applying the $j^{th}$ coalescence as $Z_{j}\triangleq\{(x_{i},y_{i})\mid i\in[j:n]\}$. For the first coalescence, let $M_{1}=M$, and suppose we find a quasi-unique
shift for order $(\alpha_{1}^{*},\alpha_{2}^{*})$. 
We denote by $M_{2}\triangleq C_{\boldsymbol{k}^{*},\boldsymbol{l}^{*}}(M_1)\nabla_{\boldsymbol{k}^{*},\boldsymbol{l}^{*}}^{x_{1},y_{1}}M_1\bigl|_{x_{1}=x_{n},y_{1}=y_{n}}$
the unique matrix such that $\mathcal{U}_{z_{n},M_{2}}$ satisfies the N-zig-zag order, and define the set of matrices containing the rest of the interpolation matrices as
\begin{equation}
\Phi_{2}\triangleq\left\{ C_{\boldsymbol{k},\boldsymbol{l}}(M_1)\nabla_{\boldsymbol{k},\boldsymbol{l}}^{x_{1},y_{1}}M_1\bigl|_{x_{1}=x_{n},y_{1}=y_{n}}|(\boldsymbol{k},\boldsymbol{l})\in\mathcal{R}_{M_1}(\alpha_{1}^{*},\alpha_{2}^{*})\setminus(\boldsymbol{k}^{*},\boldsymbol{l}^{*}))\right\}.     
\end{equation}
Then, from \eqref{derivatives_regular_j_l}, we can write
\begin{equation}
D_{2}(Z_{2})=\frac{\partial^{\alpha_{1}^*+\alpha_{2}^*}}{\partial x_{1}^{\alpha_{1}^*}\partial y_{1}^{\alpha_{2}^*}}\det(M_{1})\biggl|_{x_{1}=x_{n},y_{1}=y_{n}}=\det\left(M_{2}\right)+\sum_{\bar{M}\in\Phi_{2}}\det\left(\bar{M}\right).
\end{equation}
For the second coalescence, taking $(x_{n},y_{n})$ as the pivot node
and $(x_{2},y_{2})$ as the variable node, we write the Taylor series
expansion of $D_{2}(Z_{2})$ as
\begin{equation}
D_{2}(Z_{2})=\sum_{(\alpha_{1},\alpha_{2})\in\mathbb{N}^{2}}\frac{1}{\alpha_{1}!\alpha_{2}!}(x_{2}-x_{n})^{\alpha_{1}}(y_{2}-y_{n})^{\alpha_{2}}D_{\alpha_{1},\alpha_{2}}(Z_{3})\label{eq:d_i_z_i_taylor-1}
\end{equation}
where 
\begin{equation}
D_{\alpha_{1},\alpha_{2}}(Z_{3})=\frac{\partial^{\alpha_{1}+\alpha_{2}}}{\partial x_{2}^{\alpha_{1}}\partial y_{2}^{\alpha_{2}}}\det(M_{2})\biggl|_{x_{2}=x_{n},y_{2}=y_{n}}+\sum_{\bar{M}\in\Phi_{2}}\frac{\partial^{\alpha_{1}+\alpha_{2}}}{\partial x_{2}^{\alpha_{1}}\partial y_{2}^{\alpha_{2}}}\det\left(\bar{M}\right)\biggl|_{x_{2}=x_{n},y_{2}=y_{n}}.\label{eq:d_i_z_i_taylor-2}
\end{equation}
Next, we apply \eqref{derivatives_regular_j_l} to \eqref{d_i_z_i_taylor-2}. This time, we
find a quasi-unique shift $(\alpha_1^*,\alpha_2^*)$ by only considering matrix $M_{2}$. 
Note  that, the $(\alpha_{1}^{*},\alpha_{2}^{*})$
pair is different for each recursion but for a clearer notation, we omit the recursion index. Since the choice of  $(\alpha_{1}^{*},\alpha_{2}^{*})$ only considers $M_2$, it does not imply the existence of quasi-unique shifts for all the other matrices in $\Phi_{2}$. We denote by $M_{3}\triangleq C_{\boldsymbol{k}^{*},\boldsymbol{l}^{*}}(M_2)\nabla_{\boldsymbol{k}^{*},\boldsymbol{l}^{*}}^{x_{1},y_{1}}M_{2}\bigl|_{x_{2}=x_{n},y_{2}=y_{n}}$
the unique matrix satisfying that $\mathcal{U}_{z_{n},M_{3}}$ follows the N-zig-zag
order, and define the set of matrices containing the rest
of weighted interpolation matrices, originated from $M_{2}$ or from
$\bar{M}\in\Phi_{2}$, as
\begin{align}
\Phi_{3} \triangleq & \left\{ C_{\boldsymbol{k},\boldsymbol{l}}(M_{2})\nabla_{\boldsymbol{k},\boldsymbol{l}}^{x_{2},y_{2}}M_{2}\bigl|_{x_{2}=x_{n},y_{2}=y_{n}}|(\boldsymbol{k},\boldsymbol{l})\in\mathcal{R}_{M_{2}}(\alpha_{1}^{*},\alpha_{2}^{*})\setminus(\boldsymbol{k}^{*},\boldsymbol{l}^{*}))\right\} \nonumber \\
 & \cup\left\{ C_{\boldsymbol{k},\boldsymbol{l}}(\bar{M})\nabla_{\boldsymbol{k},\boldsymbol{l}}^{x_{2},y_{2}}\bar{M}\bigl|_{x_{2}=x_{n},y_{2}=y_{n}}|(\boldsymbol{k},\boldsymbol{l})\in\mathcal{R}_{M_{2}}(\alpha_{1}^{*},\alpha_{2}^{*}),\bar{M}\in\Phi_{2}\right\}. 
\end{align}
Then, we can write 
\begin{equation}
D_{3}(Z_{3})=D_{\alpha_{1}^{*},\alpha_{2}^{*}}(Z_{3})=\det\left(M_{3}\right)+\sum_{\bar{M}\in\Phi_{3}}\det\left(\bar{M}\right).
\end{equation}

We follow the same procedure until all nodes are coalesced with the
pivot node and we reach $D_{n}(Z_{n})$. In general, the expressions
in this procedure are defined recursively as follows.
\begin{equation}
M_{i+1}\triangleq C_{\boldsymbol{k}^{*},\boldsymbol{l}^{*}}(M_{i})\nabla_{\boldsymbol{k}^{*},\boldsymbol{l}^{*}}^{x_{i},y_{i}}M_{i}\bigl|_{x_{i}=x_{n},y_{i}=y_{n}},i\in[1:n-1]\label{eq:M_i}
\end{equation}

\begin{equation}
\label{eq:d_i_z_i}
D_{i}(Z_{i})\triangleq D_{\alpha_{1}^*,\alpha_{2}^*}(Z_{i})=\det(M_i)+\sum_{\bar{M}\in \Phi_i}\det(\bar{M}).
\end{equation}

\begin{equation}
\label{eq:d_i_z_i_taylor}
D_i(Z_i)=\sum_{(\alpha_1,\alpha_2)\in\mathbb{N}^{2}}\frac{1}{\alpha_{1}!\alpha_{2}!}(x_{i}-x_{n})^{\alpha_{1}}(y_{i}-y_{n})^{\alpha_{2}}D_{\alpha_{1},\alpha_{2}}(Z_{i+1}).
\end{equation}

\begin{align}
\Phi_{i+1}\triangleq & \left\{ C_{\boldsymbol{k},\boldsymbol{l}}(M_{i})\nabla_{\boldsymbol{k},\boldsymbol{l}}^{x_{i},y_{i}}M_{i}\bigl|_{x_{i}=x_{n},y_{i}=y_{n}}:(\boldsymbol{k},\boldsymbol{l})\in\mathcal{R}_{M_{i}}(\alpha_{1}^*,\alpha_{2}^*)\setminus(\boldsymbol{k}^{*},\boldsymbol{l}^{*})\right\} \nonumber \\
 & \cup\left\{ C_{\boldsymbol{k},\boldsymbol{l}}(\bar{M})\nabla_{\boldsymbol{k},\boldsymbol{l}}^{x_{i},y_{i}}\bar{M}\bigl|_{x_{i}=x_{n},y_{i}=y_{n}}:(\boldsymbol{k},\boldsymbol{l})\in\mathcal{R}_{M_{i}}(\alpha_{1}^*,\alpha_{2}^*),\bar{M}\in\Phi_{i}\right\}.\label{eq:phi_i}
\end{align}

\begin{lem}
\label{lem:quasi-unique}Consider the recursive Taylor series expansion procedure on a fixed
pivot, $(x_n,y_n)$. If, for every step $i\in[1:n]$, we can find a quasi-unique
shift $(\alpha_1^*,\alpha_2^*)$ for $M_i$ in \eqref{M_i} , then
\begin{equation}
\label{eq:quasi-unique-lem}
D_n(Z_n)=\det(M_n), 
\end{equation}
where 
$M_n$  depends only on $Z_n=\{(x_n,y_n)\}$. Therefore, the associated interpolation matrix $M_n$, and hence, $M_1$ are invertible for almost all choices of evaluation points. 
\end{lem}

The proof of \lemref{quasi-unique} is given in \appref{proof-quasi-unique}.


In the next lemma, we present a set of situations for which a quasi-unique shift exits in a coalescence step between a pivot node and a variable node. These are not the only situations for which quasi-unique shifts exit but are sufficient to derive the recovery threshold presented in \thmref{main_thm}, as we show in the next subsection.

\begin{lem}
\label{lem:quasi-unique-conds}
Assume that in the $i^{th}$ coalescence step we have the variable node $z_i=(x_i,y_i)$ and the pivot node $z_n=(x_n,y_n)$. Define $r_f\triangleq |\mathcal{U}_{z_i,M_i}| (\mod \mu_B$) and $l_e\triangleq \mu_B - |\mathcal{U}_{z_n,M_i}| (\mod \mu_B)$. That is, when depicted in the derivative order space, $r_f$ is the number of elements in the rightmost partially-occupied column of the derivative set of the variable node, and $l_e$ is the number of empty places in the rightmost partially-occupied column of the derivative set of the pivot node. Then, if $|\mathcal{U}_{z_i,M_i}|+|\mathcal{U}_{z_n,M_i}|>\mu_BK$ and one of the following conditions is satisfied:
\begin{enumerate}
    \item $r_f=0$,
    \item $r_f=l_e$,
    \item $l_e=0$, 
\end{enumerate}
or 
\begin{enumerate}
   \setcounter{enumi}{3}
    \item $|\mathcal{U}_{z_i,M_i}|+|\mathcal{U}_{z_n,M_i}|\leq \mu_BK$,
\end{enumerate}
then there exists a quasi-unique shift for the coalescence of these nodes. 
\end{lem}

The proof of \lemref{quasi-unique-conds} is given in \appref{proof-quasi-unique-conds}.

\subsection{Derivation of the Recovery Threshold Expression}

The existence of a quasi-unique shift depends on the joint structure of the derivative sets of the pivot and the variable nodes. If the derivative sets of the pivot node and the variable node satisfy the conditions in \lemref{quasi-unique-conds}, then, in a coalescence step, i.e., recursive Taylor series expansion, it is possible to find a quasi-unique shift for this recursive step and we can proceed to the next recursion. Otherwise, by simply ignoring specific computations provided by the worker whose evaluation point corresponds to the variable node under consideration, we can have the structure of the remaining computations satisfy the conditions in \lemref{quasi-unique-conds}. This adds an overhead of ignored computations to the recovery threshold expression. In the following lemma, we provide an upper bound on the total number of computations we may need to ignore throughout the whole recursion process by analysing the worst-case scenario.

\begin{lem}
Assume that the conditions of \lemref{quasi-unique-conds} hold in none of the coalescences in the recursive Taylor series expansion process. Then, in the worst case, by ignoring at most $(\mu_{B}-2)(\frac{L}{\mu_{B}}-1)$ computations throughout all the recursion steps suffices to guarantee decodability for almost all choices of evaluation points.
\end{lem}

\begin{IEEEproof}
Assume that
none of the conditions of \lemref{quasi-unique-conds} hold. If $r_f>l_e$,
we can satisfy condition 2, i.e., $r_f=l_e$, in \lemref{quasi-unique-conds} by ignoring $r_f-l_e$ computations received from the worker whose evaluation point is the variable node. Thus,
in the worst case, we ignore $\max(r_f-l_e)=(\mu_{B}-1)-1=\mu_{B}-2$
computations. Note that the minimum value of $l_e$ is 1. Otherwise, condition 3 in \lemref{quasi-unique-conds} would be satisfied. Moreover, the maximum value of $r_f$ is $\mu_B-1$. Otherwise, condition 1 in \lemref{quasi-unique-conds} would be satisfied. On the other hand, if $r_f<l_e$, we can ignore $r_f$ computations
and satisfy the condition 1 in \lemref{quasi-unique-conds}. Since $r_f<l_e<\mu_{B}$,
in the worst case, we need to ignore $\max(r_f)=\mu_{B}-2$ computations.
Thus, in either case, the maximum number of computations we ignore
is $\mu_{B}-2$. Observe that generating a new block in the derivative order space as a result of a coalescence and not satisfying any of the conditions in \lemref{quasi-unique-conds} are possible only if when $|\mathcal{U}_{z_i,M_i}|+|\mathcal{U}_{z_n,M_i}|>\mu_BK$. Given that there are $L/\mu_{B}$
blocks in the whole derivative order space, the maximum number of coalescences
for which $|\mathcal{U}_{z_i,M_i}|+|\mathcal{U}_{z_n,M_i}|>\mu_BK$ is at most $(L/\mu_{B}-1)$. 
Thus, in the worst-case, the total number of ignored computations
is $(\mu_{B}-2)(\frac{L}{\mu_{B}}-1)$. 
\end{IEEEproof}

Since the polynomial we need to interpolate, $A(x)B(y)$, has $KL$ coefficients, in the worst case the recovery threshold becomes $KL+(\mu_{B}-2)(\frac{L}{\mu_{B}}-1)$. Since this number guarantees the existence of a quasi-unique shift in every recursive Taylor series expansion, by \lemref{quasi-unique}, we can
conclude that our original interpolation matrix is invertible for
almost all choices of the evaluation points. This completes the proof
of \thmref{main_thm}.
\color{black}

\section{\label{sec:Conclusion}Conclusion}

In this work, we studied the memory-efficient exploitation of stragglers
in distributed matrix multiplication where workers are allowed to have
heterogeneous computation and storage capacities. We proposed bivariate
polynomial coding schemes allowing efficient use of workers' memories.
Bivariate polynomial coding poses the problem of invertibility
of an interpolation matrix, which is highly non-trivial, unlike univariate polynomial codes. We first proposed a coding scheme based on
the fact that the interpolation matrix of bivariate interpolation
is always invertible if the evaluation points form a rectangular grid.
However, in this scheme, some computations received by the master
may not be useful since the information they provide is already obtained
from previous responses. In order to tackle this problem, we
showed that as long as the workers follow a specific computation order, the interpolation
matrix is invertible for almost every choice of the interpolation points. Based on this, we proposed BPC-VO and BPC-HO
solving the problem of redundant computations. However, the constraints
imposed by the computation orders in BPC-VO and BPC-HO harm the average
computation time when the storage capacities of the workers are limited.
To overcome this, in BPC-NZO and BPC-ZZO, we relax these constraints
by allowing a few redundant computations, which are still much less
than those of B-PROC. The ability of the proposed schemes to exploit the workers' \color{\revisioncolor} storage \color{black} capacities is close to the optimal. For different storage capacities, we numerically
showed that in terms of the average
computation time, the proposed schemes in the paper outperform existing schemes in the literature.

The proof of the almost regularity of bivariate polynomial coding
schemes is itself a theoretically interesting one, and it may guide
proofs of other multivariate interpolation schemes for distributed
matrix multiplication in more general situations. Another interesting
line of work is the application of bivariate polynomial coding to private matrix multiplication.

\ifCLASSOPTIONcaptionsoff
  \newpage
\fi

\appendices
\color{\revisioncolor}
\section{Proof of \lemref{regular_permutations}}
\label{app:proof_regular_permutations}
\begin{IEEEproof}
Given an interpolation matrix
$M$, we first prove
\begin{equation}
\label{eq:det_simple_derivative}
\frac{\partial}{\partial x_j}\det\left(M\right)=\sum_{i=1}^{KL}\det(\partial_{i,x_j}M),
\end{equation}
and
\begin{equation}
\label{eq:det_simple_derivative_y}
\frac{\partial}{\partial y_j}\det\left(M\right)=\sum_{i=1}^{KL}\det(\partial_{i,y_j}M).
\end{equation}

They follow directly from the chain rule. Let $m_{i,j}$'s denote
the elements of $M$, and $S_{KL}$ the set of all permutations
of the columns of $M$. We use the fact $\det(M)=\sum_{\pi\in S_{KL}}\text{sgn}(\pi)\prod_{i=1}^{KL}m_{i,\pi(i)}$
\cite[Definition 7.4]{liesen_linear_2015-1}, where $\pi$ is a permutation, and $\text{sgn}(\pi)$ is its parity. Then, 
\begin{align}
\frac{\partial}{\partial x_j}\det(M)=\sum_{\pi\in S_{KL}}\text{sgn}(\pi)\frac{\partial}{\partial x_j}\prod_{i=1}^{KL}m_{i,\pi(i)}=\sum_{\pi\in S_{KL}}\text{sgn}(\pi)\sum_{i=1}^{KL}\left(\frac{\partial}{\partial x_j}m_{i,\pi(i)}\right)\prod_{j\in[1:KL]\setminus\{i\}}m_{j,\pi(j)}\nonumber\\
=\sum_{i=1}^{KL}\sum_{\pi\in S_{n}}\text{sgn}(\pi)\left(\frac{\partial}{\partial x_j}m_{i,\pi(i)}\right)\prod_{j\in[1:KL]\setminus\{i\}}m_{j,\pi(j)}=\sum_{i=1}^{KL}\det(\partial_{i,x_j}M).
\end{align}
The proof of \eqref{det_simple_derivative_y} can be done similarly. Next, consider part of a derivative path $s\triangleq\partial_{i_{l},y_j}\cdots\partial_{i_{2},y_j}\partial_{i_{1},y_j}$
of length $l<\alpha_{2}$ such that it has two identical rows or at
least one zero row, resulting in $\det(sM)=0$. Now let us consider
the other sequences having $s$ as the suffix. Applying \eqref{det_simple_derivative_y}
$m\leq\alpha_{2}-l$ times, 
\begin{equation}
\frac{\partial}{\partial y_j^{m}}\det(sM)=\sum_{i_{l+m}=1}^{KL}\cdots\sum_{i_{l+1}=1}^{KL}\det(\partial_{i_{l+m},y_j}\cdots\partial_{i_{l+1},y_j}sM).
\end{equation}

However, $\frac{\partial}{\partial y_j^{m}}\det(sM)=0$
since $\det(sM)=0$. The same applies to $x$ directional derivatives. That is, while taking the derivatives of $\det(M)$, i.e.,
applying $\nabla^{x_j,y_j}_{\boldsymbol{k},\boldsymbol{l}}$, if we encounter
a sub-sequence $s$ such that $\det(sM)=0$, then the sum of determinants
of all matrices having $sM$ as suffix, i.e., $\partial_{i_{l+m},y_j}\cdots\partial_{i_{l+1},y_j}sM$,
adds up to zero. Thus, only the sequences in which all simple shifts are regular contribute to \eqref{derivatives_regular_j_l}, while applying $\nabla^{x_j,y_j}_{\boldsymbol{k},\boldsymbol{l}}$. Given a $(\boldsymbol{k},\boldsymbol{l})$ pair, if $C_{\boldsymbol{k},\boldsymbol{l}}$ denotes the number of sequences composed of only regular simple shifts, we obtain \eqref{derivatives_regular_j_l}.
\end{IEEEproof}

\section{Proof of \lemref{quasi-unique}}
\label{app:proof-quasi-unique}
We first present another lemma that will be useful in the proof. 
\begin{lem}
\label{lem:phi_i_nozigzag}
No derivative set corresponding to the evaluation points $z_i$ in $\Phi_{i+1},\forall i \in[1:N-1]$, defined in \eqref{phi_i} obeys N-zig-zag order.
\end{lem}
\begin{IEEEproof}
From the definition of quasi-unique shift, it is clear that no elements of the first set in \eqref{phi_i} obeys N-zig-zag order. To show the same for the second set, consider the elements of the variable node at the coalescence step $i$. $(\alpha_1^*,\alpha_2^*)$ is chosen such that there is only one $(\boldsymbol{k}^*,\boldsymbol{l}^*)$ such that when the elements of the variable node are shifted according to $(\boldsymbol{k}^*,\boldsymbol{l}^*)$, and evaluated at $(x_n,y_n)$, the resulting derivative set obeys the N-zig-zag order. Assume now that no element in $\Phi_i$ obeys the N-zig-zag order. Take any $\bar{M}\in \Phi_i$ and apply $\nabla^{x_i,y_i}_{\boldsymbol{k},\boldsymbol{l}}\bar{M}$ for some $(\boldsymbol{k},\boldsymbol{l})$. If $(\boldsymbol{k},\boldsymbol{l})\neq (\boldsymbol{k}^*,\boldsymbol{l}^*)$, then at least one of the elements of the variable node will be placed to a location whose priority score is larger than those of all the locations to which the elements of the variable node would be placed if $(\boldsymbol{k}^*,\boldsymbol{l}^*)$ were applied. This is because $\exists j$ such that $\boldsymbol{k}(j)>\boldsymbol{k}^*$ or $\exists j$ such that $\boldsymbol{l}(j)>\boldsymbol{l}^*(j)$. In $\nabla^{x_i,y_i}_{\boldsymbol{k},\boldsymbol{l}}\bar{M}|_{x_i=x_n,y_i=y_n}$, if the derivative set of $(x_n,y_n)$ obeyed the N-zig-zag order, the variable node would have to have more elements than it originally had since the largest priority score whose corresponding location is occupied is larger for $\nabla^{x_i,y_i}_{\boldsymbol{k},\boldsymbol{l}}\bar{M}|_{x_i=x_n,y_i=y_n}$ than $\nabla^{x_i,y_i}_{\boldsymbol{k}^*,\boldsymbol{l}^*}M_i|_{x_i=x_n,y_i=y_n}$. Therefore, it is not possible that the derivative set of the pivot node for $\nabla^{x_i,y_i}_{\boldsymbol{k},\boldsymbol{l}}\bar{M}|_{x_i=x_n,y_i=y_n}$ obeys the N-zig-zag order when $(\boldsymbol{k},\boldsymbol{l})\neq (\boldsymbol{k}^*,\boldsymbol{l}^*)$. On the other hand, if $(\boldsymbol{k},\boldsymbol{l})= (\boldsymbol{k}^*,\boldsymbol{l}^*)$, since we assume that no element of $\Phi_i$ obeys the N-zig-zag order, the derivative set of the pivot node for $\nabla^{x_i,y_i}_{\boldsymbol{k}^*,\boldsymbol{l}^*}\bar{M}|_{x_i=x_n,y_i=y_n}$ does not obey the N-zig-zag order. Since we know that no element in $\Phi_2$ obeys the N-zig-zag order by definition, by induction, we conclude that none of the elements in $\Phi_{i+1},\forall i \in[2:N]$ obeys the N-zig-zag order. 
\end{IEEEproof}

The rows of $M_n$ only depend on the pivot node $(x_n,y_n)$, and thus, the derivative set associated to $(x_n,y_n)$ has $KL$ elements and satisfies the N-zig-zag order. 
Similarly, for all matrices in $\Phi_n$, the derivative sets of the pivot node have $KL$ elements. However, as \lemref{phi_i_nozigzag} suggests, in this case, no elements of $\Phi_n$ satisfies the N-zig-zag order. 
This implies that all matrices in $\Phi_n$ have at least one duplicate row, or a zero row. Therefore, $\sum_{\bar{M}\in\Phi_n}\det(\bar{M})|_{x_{n-1}=x_n,y_{n-1}=y_n}=0$. This proves \eqref{quasi-unique-lem}. The proof of $\det(M_n)\neq 0$ follows, directly, from the fact that, for $M_n$, the derivative set of the pivot node obeys the N-zig-zag order. This means that each row of $M_n$ corresponds to $\partial_k A(x_n)\partial_l B(y_n)$, $\forall k\in[0:K-1]$, $\forall l \in [0:L-1]$. Therefore, $M_n$ can be written as an upper triangular matrix, and therefore, invertible, implying $D_n(Z_n)\neq 0$. Remember that $D_{i+1}(Z_{i+1})\neq0$ implies $D_i(Z_i)\neq 0$ for all $i\in[0:n-1]$ due to the linear independence between $(x_i-x_n)^{\alpha_1}(y_i-y_n)^{\alpha_2}$ for different $(\alpha_1,\alpha_2)$ pairs. Thus, $D_n(Z_n)\neq 0$ implies $D_1(Z_1)\neq0$ recursively, and thus, $M_1$ is invertible. This proves the claim of the lemma. \hfill $\blacksquare$

\section{Proof of \lemref{quasi-unique-conds}} \label{app:proof-quasi-unique-conds}

First, note that the existence of a quasi-unique shift is only related to the structure of the uppermost blocks of the pivot and variable nodes' derivative sets. Therefore, even if the derivative sets of the pivot and variable nodes occupy more than one block, in the derivative order space, it is sufficient to consider only the uppermost blocks since the fully occupied blocks can be handled only by additional $y$-directional derivatives. Thus, we proceed as if there exist only the uppermost blocks of the derivative sets of the pivot and variable nodes.

Our proof is based on determining some sufficient conditions for the existence of a quasi-unique shift, which will reduce to the conditions claimed in the lemma. We first state our problem visually in the derivative order space in terms of the derivative sets and the derivatives of the evaluations, then we find the sufficient conditions on this visual problem statement.

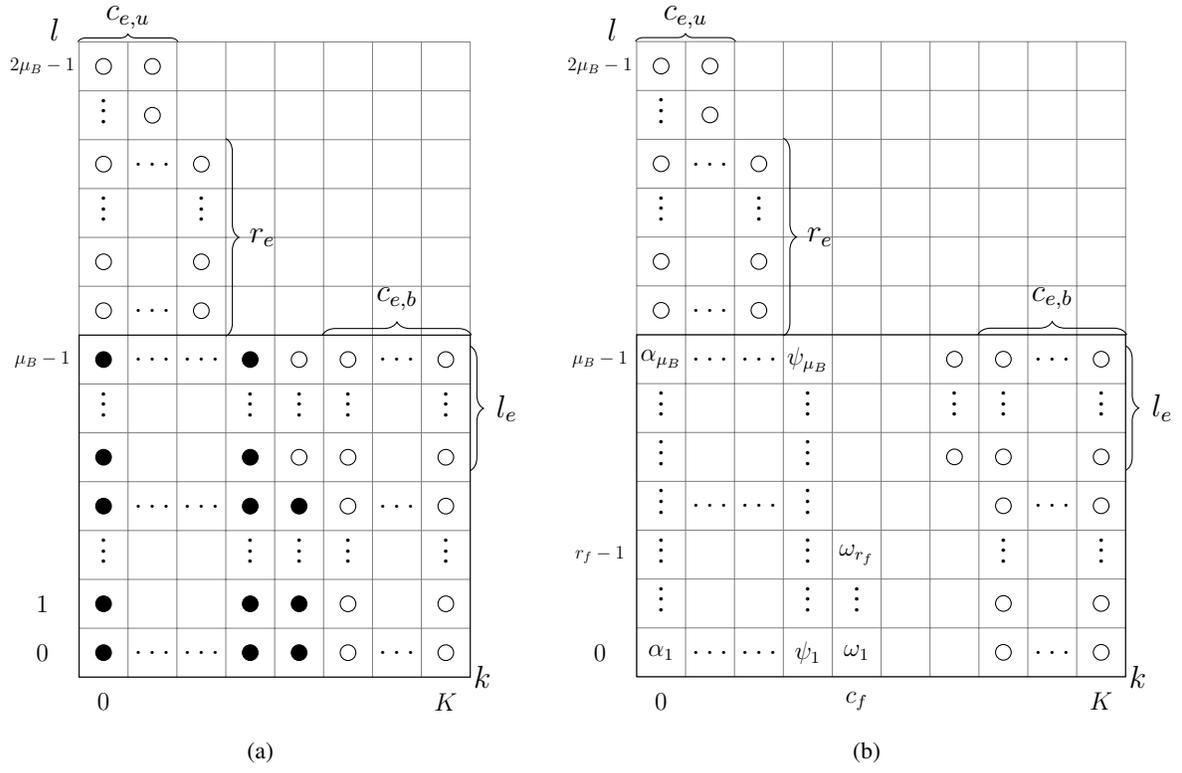
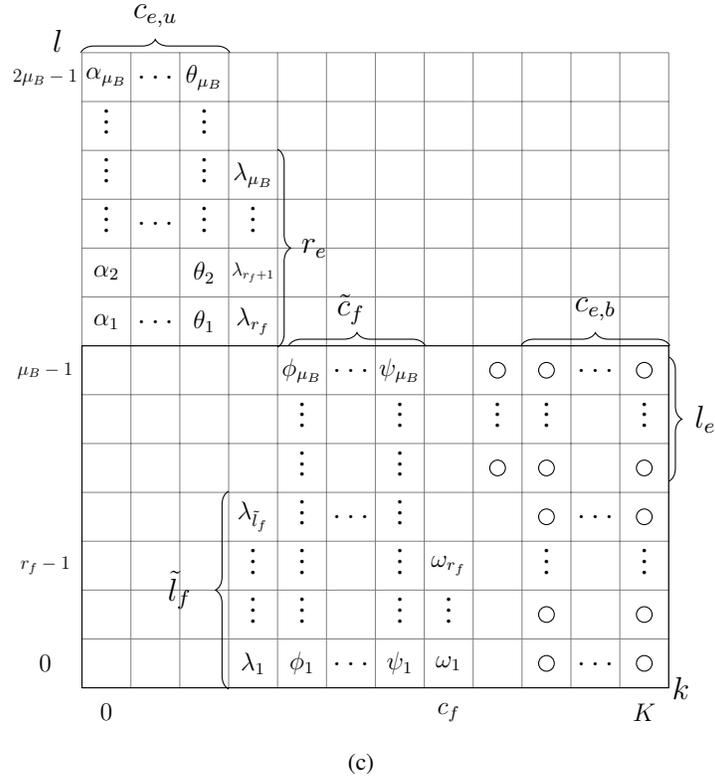
\begin{figure}
\centering
\subfloat[\label{fig:y-dir-proof-1}]{
\usetikzlibrary{decorations.pathreplacing}
\begin{tikzpicture}[scale=1.3]

\draw [help lines,  step=0.5cm] (-3.5, -3.5) node (v19) {} grid (0.5,3) node (v2) {}; 

\node[scale=0.8] at (-3.25,-3.75) {$0$};  
\node[scale=1] at (0.625,-3.5) {$k$};  
\node[scale=1] (v1) at (-3.75,3.125) {$l$};   
\node[scale=0.8] at (0.25,-3.75) {$K$};

\node[scale=0.8] at (-3.875,-3.25) {$0$};    
  
\node[scale=0.6] at (-3.875,-0.25) {$\mu_B-1$};
\node[scale=0.6] at (-3.875,2.75) {$2\mu_B-1$};

\node[scale=0.8] at (-3.875,-2.75) {$1$};  
 
\draw[fill, color=black] (-1.75,-1.75) circle (.08);  
\draw[fill, color=black] (-3.25,-1.75) circle (.08);  
\draw[fill, color=black] (-1.75,-2.75) circle (.08);  
\draw[fill, color=black] (-3.25,-2.75) circle (.08);  
\draw[fill, color=black] (-1.25,-3.25) circle (.08);  
\draw[fill, color=black] (-1.25,-1.75) circle (.08); 
\draw[fill, color=black] (-1.25,-2.75) circle (.08); 
\draw[fill, color=black] (-1.75,-3.25) circle (.08); 
\draw[fill, color=black] (-3.25,-3.25) circle (.08); 
\draw[fill, color=black] (-1.75,-0.25) circle (.08); 
\draw[fill, color=black] (-3.25,-0.25) circle (.08); 
\draw[fill, color=black] (-1.75,-1.25) circle (.08); 
\draw[fill, color=black] (-3.25,-1.25) circle (.08);

\draw  (-3.5,0) rectangle (0.5,-3.5); 

\draw[color=black] (-1.25,-1.25) circle (.08);  
\draw[color=black] (0.25,-0.25) circle (.08);  
\draw[color=black] (0.25,-1.75) circle (.08);  
\draw[color=black] (0.25,-1.25) circle (.08);  
\draw[color=black] (0.25,-2.75) circle (.08);  
\draw[color=black] (-0.75,-3.25) circle (.08);  
\draw[color=black] (-0.75,-2.75) circle (.08);  
\draw[color=black] (-0.75,-1.75) circle (.08); 
\draw[color=black] (0.25,-3.25) circle (.08); 
\draw[color=black] (-1.25,-0.25) circle (.08); 
\draw[color=black] (-0.75,-1.25) circle (.08); 
\draw[color=black] (-0.75,-0.25) circle (.08);  
\draw[color=black] (-2.25,1.75) circle (.08); 
\draw[color=black] (-2.25,0.75) circle (.08); 
\draw[color=black] (-2.25,0.25) circle (.08); 
\draw[color=black] (-3.25,0.75) circle (.08); 
\draw[color=black] (-3.25,0.25) circle (.08); 
\draw[color=black] (-3.25,1.75) circle (.08); 
\draw[color=black] (-3.25,2.75) circle (.08); 
\draw[color=black] (-2.75,2.75) circle (.08); 
\draw[color=black] (-2.75,2.25) circle (.08); 

\node at (-1.25,-2.125) {$\vdots$};
\node at (-1.75,-2.125) {$\vdots$};
\node at (-3.25,-2.125) {$\vdots$};
\node at (-1.75,-0.625) {$\vdots$};
\node at (-3.25,-0.625) {$\vdots$};
\node at (-3.25,2.375) {$\vdots$};
\node at (-3.25,1.375) {$\vdots$};
\node at (-2.25,1.375) {$\vdots$};
\node at (-1.25,-0.625) {$\vdots$};
\node at (-0.75,-2.125) {$\vdots$};
\node at (-0.75,-0.625) {$\vdots$};
\node at (0.25,-2.125) {$\vdots$};
\node at (0.25,-0.625) {$\vdots$};
\node at (-2.75,-1.75) {$\ldots$};
\node at (-2.75,-0.25) {$\ldots$};
\node at (-2.75,1.75) {$\ldots$};
\node at (-2.75,0.25) {$\ldots$};
\node at (-2.25,-3.25) {$\ldots$};
\node at (-2.25,-1.75) {$\ldots$};
\node at (-2.25,-0.25) {$\ldots$};
\node at (-2.75,-3.25) {$\ldots$};
\node at (-0.25,-0.25) {$\ldots$};
\node at (-0.25,-1.75) {$\ldots$};
\node at (-0.25,-3.25) {$\ldots$};

\node (v3) at (-3.625,3) {};
\node (v4) at (-2.375,3) {};
\draw [decorate, decoration={brace, amplitude=3pt}] (v3) -- (v4);
\node at (-3,3.25) {$c_{e,u}$};
\node (v5) at (-2,2.125) {};
\node (v6) at (-2,-0.125) {};
\draw [decorate, decoration={brace, amplitude=5pt}] (v5) -- (v6);
\node at (-1.625,1) {$r_e$};
\node (v7) at (-1.125,0) {};
\node (v8) at (0.625,0) {};
\draw [decorate, decoration={brace, amplitude=5pt}] (v7) -- (v8);
\node at (-0.25,0.375) {$c_{e,b}$};
\node (v9) at (0.5,0) {};
\node (v10) at (0.5,-1.5) {};
\draw [decorate, decoration={brace, amplitude=5pt}] (v9) -- (v10);
\node at (0.875,-0.75) {$l_e$};

\end{tikzpicture}}
\subfloat[\label{fig:y-dir-proof-2}]{
\usetikzlibrary{decorations.pathreplacing}
\begin{tikzpicture}[scale=1.3]

\draw [help lines,  step=0.5cm] (-3.5, -3.5) node (v19) {} grid (1.5,3) node (v2) {}; 

\node[scale=0.8] at (-3.25,-3.75) {$0$};  
\node[scale=1] at (1.625,-3.5) {$k$};  
\node[scale=1] (v1) at (-3.75,3.125) {$l$};   
\node[scale=0.8] at (1.25,-3.75) {$K$};   
   
\node[scale=0.8] at (-1.25,-3.75) {$c_f$};   
\node[scale=0.8] at (-3.875,-3.25) {$0$};    
  
\node[scale=0.6] at (-3.875,-0.25) {$\mu_B-1$};
\node[scale=0.6] at (-3.875,2.75) {$2\mu_B-1$};  
\node[scale=0.6] at (-3.875,-2.25) {$r_f-1$};  

\draw  (-3.5,0) rectangle (1.5,-3.5); 

\draw[color=black] (-0.25,-1.25) circle (.08);  
\draw[color=black] (1.25,-0.25) circle (.08);  
\draw[color=black] (1.25,-1.75) circle (.08);  
\draw[color=black] (1.25,-1.25) circle (.08);  
\draw[color=black] (1.25,-2.75) circle (.08);  
\draw[color=black] (0.25,-3.25) circle (.08);  
\draw[color=black] (0.25,-2.75) circle (.08);  
\draw[color=black] (0.25,-1.75) circle (.08); 
\draw[color=black] (1.25,-3.25) circle (.08); 
\draw[color=black] (-0.25,-0.25) circle (.08); 
\draw[color=black] (0.25,-1.25) circle (.08); 
\draw[color=black] (0.25,-0.25) circle (.08);  
\draw[color=black] (-2.25,1.75) circle (.08); 
\draw[color=black] (-2.25,0.75) circle (.08); 
\draw[color=black] (-2.25,0.25) circle (.08); 
\draw[color=black] (-3.25,0.75) circle (.08); 
\draw[color=black] (-3.25,0.25) circle (.08); 
\draw[color=black] (-3.25,1.75) circle (.08); 
\draw[color=black] (-3.25,2.75) circle (.08); 
\draw[color=black] (-2.75,2.75) circle (.08); 
\draw[color=black] (-2.75,2.25) circle (.08); 

\node at (-1.25,-2.625) {$\vdots$};
\node at (-1.75,-2.125) {$\vdots$};
\node at (-1.75,-2.625) {$\vdots$};
\node at (-3.25,-2.125) {$\vdots$};
\node at (-3.25,-2.625) {$\vdots$};
\node at (-1.75,-1.125) {$\vdots$};
\node at (-1.75,-1.625) {$\vdots$};
\node at (-1.75,-0.625) {$\vdots$};
\node at (-3.25,-1.125) {$\vdots$};
\node at (-3.25,-1.625) {$\vdots$};
\node at (-3.25,-0.625) {$\vdots$};
\node at (-3.25,2.375) {$\vdots$};
\node at (-3.25,1.375) {$\vdots$};
\node at (-2.25,1.375) {$\vdots$};
\node at (-0.25,-0.625) {$\vdots$};
\node at (0.25,-2.125) {$\vdots$};
\node at (0.25,-0.625) {$\vdots$};
\node at (1.25,-2.125) {$\vdots$};
\node at (1.25,-0.625) {$\vdots$};
\node at (-2.75,-1.75) {$\ldots$};
\node at (-2.75,-0.25) {$\ldots$};
\node at (-2.75,1.75) {$\ldots$};
\node at (-2.75,0.25) {$\ldots$};
\node at (-2.25,-3.25) {$\ldots$};
\node at (-2.25,-1.75) {$\ldots$};
\node at (-2.25,-0.25) {$\ldots$};
\node at (-2.75,-3.25) {$\ldots$};
\node at (0.75,-0.25) {$\ldots$};
\node at (0.75,-1.75) {$\ldots$};
\node at (0.75,-3.25) {$\ldots$};

\node (v3) at (-3.625,3) {};
\node (v4) at (-2.375,3) {};
\draw [decorate, decoration={brace, amplitude=3pt}] (v3) -- (v4);
\node at (-3,3.25) {$c_{e,u}$};
\node (v5) at (-2,2.125) {};
\node (v6) at (-2,-0.125) {};
\draw [decorate, decoration={brace, amplitude=5pt}] (v5) -- (v6);
\node at (-1.625,1) {$r_e$};
\node (v7) at (-0.125,0) {};
\node (v8) at (1.625,0) {};
\draw [decorate, decoration={brace, amplitude=5pt}] (v7) -- (v8);
\node at (0.75,0.375) {$c_{e,b}$};
\node (v9) at (1.5,0) {};
\node (v10) at (1.5,-1.5) {};
\draw [decorate, decoration={brace, amplitude=5pt}] (v9) -- (v10);
\node at (1.875,-0.75) {$l_e$};

\node[scale=0.8] at (-3.25,-3.25) {$\alpha_1$};
\node[scale=0.8] at (-3.25,-0.25) {$\alpha_{\mu_B}$};

\node[scale=0.8] at (-1.75,-3.25) {$\psi_1$};

\node[scale=0.8] at (-1.75,-0.25) {$\psi_{\mu_B}$};

\node[scale=0.8] at (-1.25,-3.25) {$\omega_1$};
\node[scale=0.8] at (-1.25,-2.25) {$\omega_{r_f}$};
\end{tikzpicture}}\\
\subfloat[\label{fig:y-dir-proof-3}]{
\usetikzlibrary{decorations.pathreplacing}
\begin{tikzpicture}[scale=1.3]

\draw [help lines,  step=0.5cm] (-3.5, -3.5) node (v19) {} grid (2.5,3) node (v2) {}; 

\node[scale=0.8] at (-3.25,-3.75) {$0$};  
\node[scale=1] at (2.625,-3.5) {$k$};  
\node[scale=1] (v1) at (-3.75,3.125) {$l$};   
\node[scale=0.8] at (2.25,-3.75) {$K$};   
   
\node[scale=0.8] at (0.25,-3.75) {$c_f$};   
\node[scale=0.8] at (-3.875,-3.25) {$0$};    
  
\node[scale=0.6] at (-3.875,-0.25) {$\mu_B-1$};
\node[scale=0.6] at (-3.875,2.75) {$2\mu_B-1$};  
\node[scale=0.6] at (-3.875,-2.25) {$r_f-1$};

\draw  (-3.5,0) rectangle (2.5,-3.5); 

\draw[color=black] (0.75,-1.25) circle (.08);  
\draw[color=black] (2.25,-0.25) circle (.08);  
\draw[color=black] (2.25,-1.75) circle (.08);  
\draw[color=black] (2.25,-1.25) circle (.08);  
\draw[color=black] (2.25,-2.75) circle (.08);  
\draw[color=black] (1.25,-3.25) circle (.08);  
\draw[color=black] (1.25,-2.75) circle (.08);  
\draw[color=black] (1.25,-1.75) circle (.08); 
\draw[color=black] (2.25,-3.25) circle (.08); 
\draw[color=black] (0.75,-0.25) circle (.08); 
\draw[color=black] (1.25,-1.25) circle (.08); 
\draw[color=black] (1.25,-0.25) circle (.08);

\node at (0.25,-2.625) {$\vdots$};
\node at (-0.25,-2.125) {$\vdots$};
\node at (-0.25,-2.625) {$\vdots$};
\node at (-3.25,1.375) {$\vdots$};
\node at (-0.25,-1.125) {$\vdots$};
\node at (-0.25,-1.625) {$\vdots$};
\node at (-0.25,-0.625) {$\vdots$};
\node at (-3.25,2.375) {$\vdots$};
\node at (-3.25,1.875) {$\vdots$};
\node at (-2.25,1.375) {$\vdots$};
\node at (-2.25,1.875) {$\vdots$};
\node at (-2.25,2.375) {$\vdots$};

\node at (-1.25,-2.625) {$\vdots$};
\node at (-1.25,-2.125) {$\vdots$};
\node at (-1.25,-1.625) {$\vdots$};
\node at (-1.25,-1.125) {$\vdots$};
\node at (-1.25,-0.625) {$\vdots$};

\node at (-1.75,1.375) {$\vdots$};
\node at (-1.75,-2.625) {$\vdots$};
\node at (-1.75,-2.125) {$\vdots$};
\node at (0.75,-0.625) {$\vdots$};
\node at (1.25,-2.125) {$\vdots$};
\node at (1.25,-0.625) {$\vdots$};
\node at (2.25,-2.125) {$\vdots$};
\node at (2.25,-0.625) {$\vdots$};
\node at (-2.75,1.25) {$\ldots$};
\node at (-2.75,2.75) {$\ldots$};
\node at (-0.75,-3.25) {$\ldots$};
\node at (-0.75,-1.75) {$\ldots$};
\node at (-0.75,-0.25) {$\ldots$};

\node at (-2.75,0.25) {$\ldots$};
\node at (1.75,-0.25) {$\ldots$};
\node at (1.75,-1.75) {$\ldots$};
\node at (1.75,-3.25) {$\ldots$};

\node (v7) at (0.875,0) {};
\node (v8) at (2.625,0) {};
\draw [decorate, decoration={brace, amplitude=5pt}] (v7) -- (v8);
\node at (1.75,0.375) {$c_{e,b}$};
\node (v9) at (2.5,0) {};
\node (v10) at (2.5,-1.5) {};
\draw [decorate, decoration={brace, amplitude=5pt}] (v9) -- (v10);
\node at (2.875,-0.75) {$l_e$};

\node[scale=0.8] at (-3.25,0.25) {$\alpha_1$};
\node[scale=0.8] at (-3.25,0.75) {$\alpha_2$};
\node[scale=0.8] at (-3.25,2.75) {$\alpha_{\mu_B}$};

\node[scale=0.8] at (-0.25,-3.25) {$\psi_1$};
\node[scale=0.8] at (-0.25,-0.25) {$\psi_{\mu_B}$};

\node[scale=0.8] at (0.25,-3.25) {$\omega_1$};
\node[scale=0.8] at (0.25,-2.25) {$\omega_{r_f}$};

\node[scale=0.8] at (-2.25,0.25) {$\theta_1$};
\node[scale=0.8] at (-2.25,0.75) {$\theta_2$};
\node[scale=0.8] at (-2.25,2.75) {$\theta_{\mu_B}$};

\node[scale=0.8] at (-1.75,0.25) {$\lambda_{r_f}$};
\node[scale=0.6] at (-1.75,0.75) {$\lambda_{r_f+1}$};
\node[scale=0.8] at (-1.75,1.75) {$\lambda_{\mu_B}$};

\node[scale=0.8] at (-1.75,-3.25) {$\lambda_{1}$};
\node[scale=0.8] at (-1.75,-1.75) {$\lambda_{\tilde{l}_f}$};

\node[scale=0.8] at (-1.25,-3.25) {$\phi_1$};
\node[scale=0.8] at (-1.25,-0.25) {$\phi_{\mu_B}$};

\node (v3) at (-2,-1.375) {};
\node (v4) at (-2,-3.625) {};

\draw [decorate, decoration={brace, amplitude=5pt}] (v4) -- (v3);
\node at (-2.5,-2.5) {$\tilde{l}_f$};
\node (v5) at (-3.625,3) {};
\node (v6) at (-1.875,3) {};
\draw [decorate, decoration={brace, amplitude=5pt}] (v5) -- (v6);
\node at (-2.75,3.375) {$c_{e,u}$};
\node (v11) at (-1.5,2.125) {};
\node (v12) at (-1.5,-0.125) {};
\draw [decorate, decoration={brace, amplitude=5pt}] (v11) -- (v12);
\node at (-1.125,1) {$r_e$};
\node (v13) at (-1.5,0) {};
\node (v14) at (0.125,0) {};
\draw [decorate, decoration={brace, amplitude=5pt}] (v13) -- (v14);
\node at (-0.75,0.375) {$\tilde{c}_f$};
\end{tikzpicture}}
\caption{Visualization of the derivative sets of the pivot and variable nodes.}\label{fig:y-dir-proof}
\end{figure}

In this part of the proof, we take all the $y$-directional derivatives before the $x$-directional ones. We depict the elements in the derivative set of the pivot node, $z_n=(x_n,y_n)$, in the derivative order space by filled circles in \figref{y-dir-proof-1}. Since the sum of the elements in the derivative sets of the pivot and variable nodes is larger than the size of one block, i.e., $|\mathcal{U}_{z_n,M_i}|+|\mathcal{U}_{z_i,M_i}|>\mu_BK$, the coalescence generates a new block. The unfilled circles in \figref{y-dir-proof-1} represent the locations of the elements of the variable node to be coalesced with the pivot node after the coalescence. Their locations are determined such that, after the coalescence, the resulting derivative set obeys the N-zig-zag order. Therefore, from the structure in the figure, we write $|\mathcal{U}_{z_i,M_i}|=l_e+(c_{e,b}+c_{e,u})\mu_B+r_e$.

Since, after determining the locations to which the elements of the variable node are placed, we no longer need the elements of the pivot node. Therefore, in \figref{y-dir-proof-2}, we remove the elements of the pivot node from the picture, and, instead, we depict the elements of the variable node in their original places such that they obey N-zig-zag order. Note that in this proof, our goal is to find a quasi-unique shift $(\alpha_1^*,\alpha_2^*)$ such that there is only one unique placement, characterized by $(\boldsymbol{k}^*,\boldsymbol{l}^*)$, of the elements of the variable node along with the elements of the pivot node. Therefore, we need to track the final location of each element of the variable node and make sure that to the location each element is placed, it is not possible to place another element from the variable node. Therefore, we denote the elements of the variable node by Greek letters and their subscripts. Note that the letters used for this purpose should not be mixed with the other uses of the Greek letters throughout the paper. 
\color{black}

\color{\revisioncolor}
Given the depictions in \figref{y-dir-proof-2}, the next step is to determine $y$-directional shifts such that all elements of the variable node are placed to the correct row in the derivative order space. Since, according to \lemref{regular_permutations}, only regular simple shifts are considered, while taking $y$-directional derivatives, the sequence of the elements having the same $x$-directional derivative order cannot change. Therefore, for example, $\alpha_{\mu_B}$ stays always above the elements denoted by $\alpha_i,i\in[1:\mu_B-1]$. Thanks to this property, filling the locations determined to be filled in the new block is straightforward. Shifting the block composed of the variable node's elements with the same shape as the locations to be filled towards $y$-direction uniquely determines the elements to be moved to the new block. The remaining $y$-directional shifts will be of the remaining elements of the variable node in the lower block. In \figref{y-dir-proof-3}, we depict the shifted elements to the upper new block and the remaining elements together. To have $y$-directional shifts which generate quasi-unique shifts, whenever we fill a row in the locations determined to be filled in the lower block, the elements to be placed there must be uniquely determined. For example, while filling the top $l_e$ rows, for each row, there must be exactly $c_{e,b}+1$ columns among the elements of the variable node that are available to provide their top-most element. After filling top $l_e$ rows, in the remaining rows, there must be exactly $c_{e,b}$ columns of the elements of the variable node that can provide their top-most element. Therefore, to guarantee this, a sufficient condition is that the shape of the remaining elements of the variable node and the shape of the remaining empty locations match. That is, $\tilde{c}_f=c_{e,b}$ and the remaining elements of the variable node have only one partially-occupied column with $l_e$ elements. There might be several structures satisfying this condition. One of them is when $r_f=0$ since this implies $l_e+(c_{e,b}+c_{e,u})\mu_B+r_e \equiv 0 \mod \mu_B$. Therefore, $\tilde{l}_f=l_e$. This proves condition 1 of the lemma. Another structure satisfying the sufficient condition is that $r_f=l_e$. When this is the case, $l_e+(c_{e,b}+c_{e,u})\mu_B+r_e=r_f+(c_{e,b}+c_{e,u})\mu_B+r_e=r_f+c_f\mu_B$, implying $\tilde{l}_f=0$. This proves condition 2 of the lemma. For completeness, note that after the elements are aligned with their final rows via $y$-directional derivatives, necessary $x$-directional shifts can be easily applied such that the elements of the variable node are finally placed to their intended locations. Again, due to \lemref{regular_permutations}, we consider only regular simple shifts and therefore, while taking $x$-directional derivatives, the sequence of the elements having the same $y$-directional derivative order cannot change.

In the remaining of the proof, we take all $x$-directional derivatives before $y$-directional derivatives. In this case, \figref{y-dir-proof-1} and \figref{y-dir-proof-2} are still valid. However, since we are taking $x$-directional derivatives first, we first align all the elements of the variable node that are to stay in the lower block with their intended columns. We start with the rightmost column of the lower block, which is column $K$. When $|\mathcal{U}_{z_n,M_i}|+|\mathcal{U}_{z_i,M_i}|\leq \mu_BK$, this column are not intended to be fully occupied, let us say only $\tilde{l}_e$ of them will be filled, but the empty locations start from the bottom and they are consecutive until the end. Therefore, the rows of the elements of the variable node that will provide elements to these locations are uniquely determined, namely the rows $[0:\tilde{l}_e-1]$ of the elements of the variable node from the bottom. Note that, if $|\mathcal{U}_{z_n,M_i}|+|\mathcal{U}_{z_i,M_i}|> \mu_BK$, then $\tilde{l}_e=\mu_B$, which does not break our argument. After the rightmost elements from the rows $[0:\tilde{l}_e-1]$ of the elements of the variable node are shifted to the $K^{th}$ column via $x$-directional shifts, next, we fill the columns starting from column $K-1$ to column $K-c_{e,b}-1$. Note that since each of these columns are intended to be fully occupied, they are directly filled with the rightmost elements of each row via $x$-directional shifts. Finally, we fill the column $K-c_{e,b}$, which has $l_e$ locations intended to be occupied after the coalescence. If $|\mathcal{U}_{z_n,M_i}|+|\mathcal{U}_{z_i,M_i}|\leq \mu_BK$, then the upper block is not generated and the number of remaining elements of the variable node is equal to $l_e$, each on different rows. Thanks to the property that the sequence of the elements having the same $y$-directional derivative orders cannot change by $x$-directional shifts, the elements to be placed to the $l_e$ empty locations are uniquely determined. This proves condition 4 of the lemma. On the other hand, when $|\mathcal{U}_{z_n,M_i}|+|\mathcal{U}_{z_i,M_i}|> \mu_BK$, a new block is generated, so there will be always more than $l_e$ remaining elements of the variable node. Therefore, to have a unique shift, in this case, we need $l_e=0$, which proves condition 3 of the lemma. \hfill $\blacksquare$

\color{black}

\bibliographystyle{IEEEtran}
\bibliography{references}


\section*{Supplementary Material}
\section*{Alternative Formulation of Almost Regular Interpolation Schemes} \label{app:equivalency-proof}

In this section, we discuss an alternative formulation of almost regular interpolation schemes based on the interpolation of $A(x)B(y)$ from only its evaluations, as done in B-PROC. In such an approach, for worker $i$, $A(x)$ would be evaluated at the distinct evaluation points $\{x_{i,k}:k\in[0:m_{A,i}-1]\}$ and $B(y)$ would be evaluated at the distinct evaluation points $\{y_{i,l}:l\in[0:m_{B,i}-1]\}$. In this case, computations assigned to worker $i$ would be $A(x_{i,k})B(y_{i,l})$. Remember that in all almost regular interpolation schemes, we have a priority score which determines the order in which the computations will be carried out by each worker. Each priority score is a function of the computation index, which is $(k,l)$. For our alternative formulation, we can use the same priority scores defined for Hermite interpolation-based schemes such that $(k,l)$ is the index for the computation $A(x_{i,k})B(y_{i,l})$. Then vertical, horizontal, N-zig-zag and Z-zig-zag order definitions follow. Thus, two formulations are equivalent to each other, under the almost regularity condition, and Theorem 1 and Corollary 1 are also valid for this case. Before giving a proof for this claim, we first present two useful lemmas. Note that we only provide the proof for N-zig-zag order. The proof can be trivally extended to Z-zig-zag order case.
\setcounter{lem}{5}
\begin{lem}
\label{Lemma:y-dir-der}Assume all the responses from the workers
obey N-zig-zag order. Consider the nodes,
i.e., evaluation points of the received responses at the master, $z_{1,i}=(x_{i},y_{1})$
and $z_{2,i}=(x_{i},y)$ for $i\in[1:p]$ for $p\in\mathbb{Z}^{+}$,
where the interpolation matrix depends on all of them. That is, all
nodes $z_{2,i}$ share the same $y$ coordinate. Let $\mathcal{U}_{z_{1},M}=\{(0,0),(0,1),\cdots,(0,l-1)\}$
for any $1\leq l\leq L-1$, and $\mathcal{U}_{z_{2},M}=\{(0,0)\},\forall i\in[1:p]$.
Consider $y_{1}$ as the pivot and $y$ as a variable. Then $(\alpha_{1},\alpha_{2})=(0,lp)$
is a quasi-unique shift and $|\mathcal{R}_M(\alpha_1,\alpha_2)|=1$. That is 
\[
\frac{\partial^{lp}}{\partial y^{lp}}\det(M(y))\big|_{y=y_{1}}=C_{\boldsymbol{\tilde{k},\tilde{l}}}(M)\det(\nabla_{\boldsymbol{\tilde{k},\tilde{l}}}^yM)\big|_{y=y_{1}}
\]
by Definition 7, where $\boldsymbol{\tilde{k}=0}$ and
$\boldsymbol{\tilde{l}}(i)=l,\forall i\in[1:p]$. After such a coalescence,
we obtain $\mathcal{U}_{z_{1},\tilde{M}}=\{(0,0),(0,1),\cdots,(0,l-1),(0,l)\}$, where $\tilde{M}=\nabla_{\boldsymbol{\tilde{k},\tilde{l}}}^yM\big|_{y=y_{1}}$. 
\end{lem}
\begin{IEEEproof}
For any $i\in[1:p]$, consider two nodes $(x_{i},y_{1})$ and $(x_{i},y)$.
While taking the derivative of $M(y)$ with respect to $y$, the minimum
derivative order to be applied to the $i^{\text{th}}$ row is $l$
since $\mathcal{U}_{z_{1},M}=\{(0,0),(0,1),\cdots,(0,l-1)\}$ has all the orders
up to $l$. Otherwise, we would get two identical rows in $\frac{\partial^{lp}}{\partial y^{lp}}\det(M(y))|_{y=y_{1}}$.
Since we have $p$ rows depending on $y$, and $\alpha_{2}=lp$, we
must have $\boldsymbol{\tilde{l}}(i)=l,\forall i\in[1:p]$. This is
the only possible $\boldsymbol{\tilde{l}}$ and proves the claim.
\end{IEEEproof}
\begin{lem}
\label{Lemma:x-dir-der}Assume all the responses from the workers
obey N-zig-zag order. Let $z_{1}=(x_{1},y_{1})$ and $z=(x,y_{1})$,
i.e., they share the same $y$ coordinate. Let $\mathcal{U}_{z_{1},M}=\{(i,j):i\in[0:k-1],j\in[0:m-1]\}$,
that is, it contains $k$ columns with exactly $m\leq L$ elements,
and $\mathcal{U}_{z,M}=\{(0,0),(0,1),\cdots,(0,l-1)\}$, for any $l\in[0:m-1]$,
i.e., one column with $l$ elements. Consider $x$ as a variable and
$x_{1}$ as the pivot. Then, $(\alpha_{1},\alpha_{2})=(lk,0)$ is a quasi-unique shift with $|\mathcal{R}_M(\alpha_1,\alpha_2)|=1$. That is,
\[
\frac{\partial^{lk}}{\partial x^{lk}}\det(M(x))\big|_{x=x_{1}}=C_{\tilde{\textbf{\ensuremath{\boldsymbol{k}}}},\boldsymbol{\tilde{l}}}(M)\det(\nabla_{\boldsymbol{\tilde{k},\tilde{l}}}^xM)\big|_{x=x_{1}}
\]
by Definition 7, where $\tilde{\textbf{\ensuremath{\boldsymbol{k}}}}(i)=k$,
$\forall i\in[1:l]$ and $\boldsymbol{\tilde{l}=0}.$ After such a
coalescence, we obtain $\mathcal{U}_{z_{1},\tilde{M}}=\{(k,0),(k,1),\cdots,(k,l-1)\}$ where $\tilde{M}=\nabla_{\boldsymbol{\tilde{k},\tilde{l}}}^xM\big|_{x=x_{1}}$.
\end{lem}
\begin{IEEEproof}
While taking the $kl$-th order derivative of $M(x)$ with respect
to $x$ ,we need to allocate $kl$ shifts into the rows of $M(x)$
depending on $x$, each of which corresponds to one element in the
$\mathcal{U}_{z,M}$. We have $\mathcal{U}_{z_{1},M}=\{(i,j):i\in[0:k-1],j\in[0:m-1]\}$.
After the shift with the order $(\alpha_{1},\alpha_{2})=(kl,0)$,
there should not be any duplicate element in the derivative set of
the pivot node. Thus, to each element of $\tilde{U}_{z,M}$, we will assign
$k$ derivative order, i.e., $\tilde{\textbf{\ensuremath{\boldsymbol{k}}}}(i)=k$,
$\forall i\in[1:l]$. All other allocations would generate duplicate
elements in $\mathcal{U}_{z_{1},M}$ after the coalescence. 
\end{IEEEproof}

Next, we state the equivalency between the alternative formulation we give in this section and the formulation in Section VI.

\begin{lem}
\label{lem:conversion-to-hermite}
Assuming the alternative formula we have introduced is employed, if the set of evaluation points,
or nodes, assigned to a worker has N-zig-zag order, then by a series
of unique shifts, it can be reduced to a single node whose derivative
set has also N-zig-zag order. 
\end{lem}
\begin{IEEEproof}
Assume the master receive $\tilde{k}_{i}\tilde{l_{i}}$
evaluations of $A(x)B(y)$ from worker $i$. That is, the master receives
the evaluations $\{A(x_{i,k})B(y_{i,l}):k\in[1:\tilde{k}_{i}],l\in[1:\tilde{l}_{i}]\}$
where $\tilde{k_{i}}\leq m_{A,i}$ and $\tilde{l_{i}}\leq m_{B,i}$
such that they are in accordance with the constraints imposed by N-zig-zag
order. Thus, we have derivative sets $\tilde{U}_{(x_{i,k},y_{i,l}),M}=\{(0,0)\}$,
$\forall k\in[1:\tilde{k}_{i}],\forall l\in[1:\tilde{l}_{i}]$, $\forall i\in[1:N]$
assuming each worker sent at least one response. Then if we apply
the coalescence procedure described in Section VI, and if we take
$y$-directional derivatives first, according to Lemma \ref{Lemma:y-dir-der},
we can always find unique shifts in the coalescence procedure. Assume,
without losing generality, during the $y$-directional derivatives,
$y_{i,1}$ is taken as the pivot. After taking all $y$- directional
derivatives, the derivative sets become $\mathcal{U}_{(x_{i,k},y_{i,1}),M_2}=\{(0,l):l\in[1:\tilde{l}_{i}]\}$,
$\forall i\in[1:N]$, $\forall k\in[1:\tilde{k}_{i}]$. Then, according
to Lemma \ref{Lemma:x-dir-der}, by only taking $x$-directional derivatives,
we can find unique shifts in every coalescence step. Without losing
generality, assuming $x_{i,1}$ is taken as pivot for the $x$-directional
derivatives, we end up with $\mathcal{U}_{(x_{i,1},y_{i,1}),M_3}=\{(k,l):k\in[\tilde{k}_{i}],l\in[\tilde{l}_{i}]\}$,
$\forall i\in[1:N]$, as claimed by the lemma.
\end{IEEEproof}

Observe that this derivatives sets are equivalent
to the derivative sets obtained as a result of Hermite interpolation-based
bivariate polynomial codes. This proves the equivalency under the almost
regularity condition.




%








\end{document}